\documentclass[11pt,preprintnumbers,pra,aps,
nofootinbib,tightenlines,floatfix, twocolumn,superscriptaddress,longbibliography]{revtex4-2}

\usepackage[colorlinks=true]{hyperref}
\hypersetup{
  linkcolor  = TealBlue!90!Black,
  citecolor  = YellowGreen!90!Black,
  urlcolor   = Peach!90!Black,
  colorlinks = true,
}
\usepackage[dvipsnames]{xcolor}
\usepackage[all]{hypcap}
\usepackage{amsmath,amssymb}
\usepackage{graphicx}
\usepackage{balance}
\usepackage[utf8]{inputenc}
\usepackage{csquotes}
\usepackage{empheq}
\usepackage{multirow}
\usepackage[section]{placeins}
\usepackage{comment}
\usepackage[customcolors]{hf-tikz}
\usepackage[top=1in,bottom=1in,left=1.8cm,right=1.6cm]{geometry}

\allowdisplaybreaks
\makeatletter
\renewcommand\onecolumngrid{
\do@columngrid{one}{\@ne}
\def\set@footnotewidth{\onecolumngrid}
\def\footnoterule{\kern-6pt\hrule width 1.5in\kern6pt}%
}
\makeatother

\newcommand{\dif}{\ensuremath{\text{d}}}
\newcommand\scalemath[2]{\scalebox{#1}{\mbox{\ensuremath{\displaystyle #2}}}}
\newcommand{\Tr}{\ensuremath{\text{Tr}}}
\def\Nslash{{ {\cal N}\hskip-0.55em /}}

\newcount\hour \newcount\hourminute \newcount\minute
\hour=\time \divide \hour by 60
\hourminute=\hour \multiply \hourminute by 60
\minute=\time \advance \minute by -\hourminute
\newcommand{\mydate}{\ \today \ - \number\hour :\number\minute}

\begin{document}

\title{\texorpdfstring{Entanglement Structures in Quantum Field Theories II: \\ Distortions of Vacuum Correlations Through the Lens of Local Observers}{Entanglement Structures in Quantum Field Theories II: Distortions of Vacuum Correlations Through the Lens of Local Observers}}
\author{Natalie Klco}
\email{natalie.klco@duke.edu}
\affiliation{{Duke Quantum Center and Department of Physics, Duke University, Durham, NC 27708, USA}}
\author{D.~H.~Beck}
\email{dhbeck@illinois.edu}
\affiliation{Department of Physics and Illinois Quantum Information Science and Technology (IQUIST) Center, University of Illinois at Urbana-Champaign,  Urbana, IL 61801, USA}

\begin{abstract}
When observing a quantum field via detectors with access to only the mixed states of spatially separated, local regions---a ubiquitous experimental design---the capacity to access the full extent of distributed entanglement can be limited, shrouded by classical correlations.
By performing projective measurements of the field external to two detection patches and classically communicating the results, underlying pure states may be identified for which entanglement quantification is clear.
In the Gaussian continuous-variable states of the free scalar field vacuum, this protocol uncovers a disparity between the spacelike entanglement established within the field and that which is locally detectable.
This discrepancy is found to grow exponentially with the separation between observation regions.
The protocol developed herein offers insight and practical guidance for clarifying the unavoidable distortion of quantum field correlations when viewed from the vantage of a pair of local observers.
\end{abstract}
\date{\mydate}
\maketitle

{
\footnotesize
\tableofcontents
}
\section{Introduction}

Modern understanding of quantum correlations recognizes their unique capacity to improve the efficiency of natural information processing.
Beyond being a valuable resource in quantum-based simulation~\cite{Feynman1982,Banuls:2019bmf,Klco:2021lap,Bauer:2022hpo}, communications security~\cite{bennet1984quantum,PhysRevLett.67.661}, and sensing~\cite{2017RvMP...89c5002D} protocols, quantum entanglement throughout nature is increasingly recognized as a key element in guiding symmetries, structure, and time evolution~\cite{Kharzeev:2017qzs,Baker:2017wtt,Cervera-Lierta:2017tdt,Berges:2018cny,Beane:2018oxh,GortonJohnson2019a,Beane:2019loz,Tu:2019ouv,Beane:2020wjl,Beane:2021zvo,Kharzeev:2021yyf,Robin:2020aeh,Low:2021ufv,Gong:2021bcp,Roggero:2021asb,Mueller:2021gxd,Liu:2022grf}.
Both separately and together, the rapid experimental progress in the development of quantum technologies~\cite{Altman:2019vbv} and the goals of basic sciences in quantum many-body physics motivate continued research to clarify the fundamental interconnectivity of quantum systems---in particular, for quantum fields describing the Standard Model of particle physics or governing large-scale quantum computational frameworks.

The observation that spacelike entanglement is distributed throughout the vacuum state of quantum fields has grown from abstract algebraic origins~\cite{Reeh1961,summers1985vacuum,summers1987bell1,summers1987bell2,Halvorson:1999pz,Witten:2018zxz} toward operational protocols~\cite{VALENTINI1991321,Reznik:2002fz,Reznik:2003mnx,Retzker_2005,Klco:2021cxq}.
Complementing progress in quantifying entanglement entropies for bipartitions of global pure states~\cite{Calabrese:2004eu,Casini:2008wt,Calabrese:2009qy,Casini:2009sr,Calabrese:2009ez,Coser_2017,Ruggiero:2018hyl}, quantifications of the mixed-state entanglement between disjoint patches of the field are emerging~\cite{Marcovitch:2008sxc,Calabrese:2012ew,Calabrese:2012nk,Klco:2020rga,Klco:2021biu,Klco:2021cxq}.
Though currently analytically out-of-reach due to sign problems or challenging analytic continuations, even for free scalar fields, such calculations have addressed an important physical property: the amount of entanglement that is experimentally accessible (organized into two-mode entangled pairs~\cite{Klco:2021cxq}) when provided access to two spacelike-separated local detection patches of the field~\cite{Marcovitch:2008sxc,Klco:2020rga,Klco:2021biu}.
To refine understanding of such results, the present manuscript focuses upon a different, but closely related, physical property:  the amount of entanglement that can be theoretically identified between two spacelike-separated detection patches leveraging the additional resource of local operations and classical communications (LOCC) informed by measurement of the external field volume.
Because LOCC cannot increase the entanglement between the field patches~\cite{Plenio:2005cwa,chitambar2014everything}, this added resource will aid in clarifying the entanglement structure present throughout the field vacuum, relevant both fundamentally and for application in quantum simulation design.

\begin{figure}
\includegraphics[width = 0.9\columnwidth]{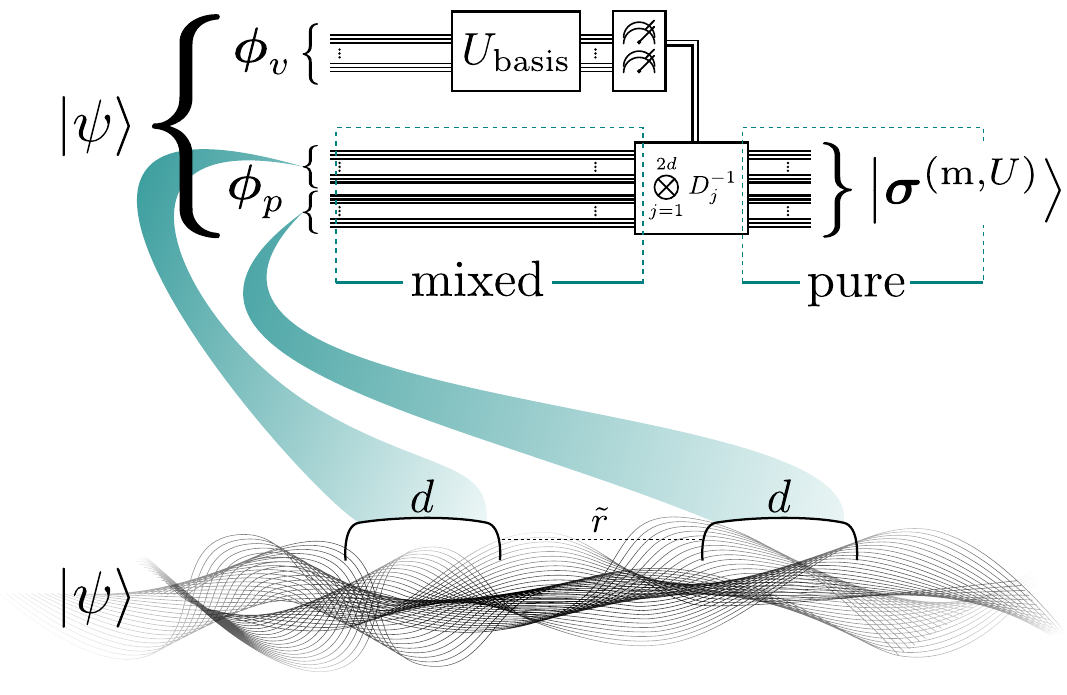}
\caption{Schematic representation of the thought-experiment guiding the present calculations for a non-interacting scalar field vacuum in the infinite volume of one spatial dimension, $|\psi\rangle$.
When latticized, the free scalar field vacuum is a Gaussian continuous-variable (CV) system with one CV quantum degree of freedom per lattice site, each represented by a horizontal line in the quantum circuit.
For local observation, two patches of the field vacuum, each with $d$ lattice sites, are extracted and compose the Hilbert space of $\boldsymbol{\phi}_p$, while the remaining external volume is represented by $\boldsymbol{\phi}_v$.
The locally-observed mixed state of the patch pair results from tracing the external volume.
However, a measurement of the external volume can
be classically communicated to inform local unitary displacement operators capable of
unifying the mixed convex decomposition into a single measurement-basis-dependent underlying pure state, $|\boldsymbol{\sigma}^{(\mathrm{m},U)}\rangle$.
In particular, measurement of the infinite volume in the field and conjugate momentum bases, producing $|\boldsymbol{\sigma}^{({\rm m}, \phi)}\rangle$ and $|\boldsymbol{\sigma}^{({\rm m}, \pi)}\rangle$ respectively, are at the center of this work.
}
\label{fig:protocolcircuit}
\end{figure}

The distinction between the entanglement truly distributed by the field at spacelike separations and that accessible by local observations is here explored through calculation of the thought experiments depicted in Fig.~\ref{fig:protocolcircuit}.
By propagating classical communications regarding the configuration of the entirety of the external field volume,  local unitaries can be designed that align all quantum states in the convex decomposition of the patch-patch mixed state in order to reveal an underlying entangled pure state (for which entanglement quantification is direct) between the two regions of the field.
Using this technique, some important properties of the locally accessible entanglement between regions of the field vacuum---e.g., its decay exponentially more rapid than the two-point correlators and abrupt termination at long distances in the presence of a UV truncation---become more rigorously understood as a response of the field to the action of local observation.

Leveraging the well-developed language of Gaussian quantum information~\cite{Simon:2000zz,Duan_2000,giedke2001distillability,2001PhRvL..87p7904G,2003PhRvA..67e2311B,Braunstein:2005zz,2012RvMP...84..621W,2014arXiv1401.4679A,2016arXiv161205215L,Serafini2017}, this manuscript explores the entanglement in the vacuum of the simplest quantum field theory---the non-interacting lattice scalar field vacuum in one spatial dimension---as a leading-order approximation of more complicated theories of interest.
While the formalisms developed in this work for analyzing the protocols of Fig.~\ref{fig:protocolcircuit} are not mathematically complex, the physical interpretation of resulting entanglement calculations is important.
By contrasting the features of locally-observed entanglement with that fundamentally present in the field, aspects of observable quantum information resources are identified that specifically arise as a result of classical correlations generated by the loss of entangled portions of the vacuum.
Similar phenomena of fragmented quantum correlations producing a noisy lens through which experimental observations must be interpreted may be responsible for apparent thermalization of local observables in, for example, extended quantum simulations~\cite{2016Sci...353..794K} and  heavy ion collisions~\cite{Ho:2015rga}.

After discussing the Gaussian representation of the free lattice scalar field vacuum and establishing the covariance matrix (CM) of regions within the infinite volume, section~\ref{sec:latticeScalarFieldVacuum} develops the analysis techniques to perform the thought experiment of Fig.~\ref{fig:protocolcircuit} and to systematically identify entangled pure states between the disjoint field regions.
In section~\ref{sec:consequences}, this protocol is directly applied to the scalar field vacuum to reveal several consequences of local entanglement observation---from parametric deviations in the scaling of the accessible entanglement, both in the functional form and range, to the short-distance scale dependence at long-distance separations and the delocalization of structured entanglement.
From entanglement-inspired optimizations of quantum simulations for quantum field theories to  the basic quantum information properties of fields, the protocols of Fig.~\ref{fig:protocolcircuit} offer insight for the contextualized decoding of observations
from the disjoint local detection regions that commonly constitute our experimental measurements of the quantum fields of nature.

\section{Lattice Scalar Field Vacuum}
\label{sec:latticeScalarFieldVacuum}
When spatially latticized with local continuous fields, the Hamiltonian of the lattice scalar field is
\begin{equation}
  \hat{H} = \frac{1}{2} \sum_\mathbf{x} \Big( \hat{\pi}(\mathbf{x})^2 + m^2 \hat{\phi}(\mathbf{x})^2 + \left(\nabla_a \hat{\phi}(\mathbf{x}) \right)^2\Big) \ \ \ ,
  \label{eq:hamiltonian}
\end{equation}
where field redefinitions have been performed to produce dimensionless quantities for both the field, $\hat{\phi}$, its conjugate momentum, $\hat{\pi}$, and the mass, $m$.
\subsection{Gaussian Formalism}
Governed by a quadratic Hamiltonian, the vacuum of the free scalar field is a Gaussian continuous-variable (CV) quantum state.
The vacuum wavefunction of this systematic quantum many-body discretization of the continuum field may be analytically determined to be
\begin{equation}
|\psi\rangle = \frac{\det \mathbf{K}^{\frac{1}{4}}}{\pi^{\frac{N}{4}}} \int \dif \boldsymbol{\phi} \ e^{-\frac{1}{2} \boldsymbol{\phi}^T \mathbf{K} \boldsymbol{\phi}}  \ |\boldsymbol{\phi}\rangle \ \ \ ,
\label{eq:vacuumWF}
\end{equation}
where $\boldsymbol{\phi}$ is a vector containing the field values distributed across the lattice sites, $N$ is the total number of sites composing the lattice, and $\mathbf{K}$ is the inter-site correlation matrix.
For a one-dimensional field with periodic boundary conditions, the correlation matrix elements may be calculated analytically as $K_{ij} = \frac{1}{N} \sum_{k = 0}^{N-1} \cos\left(\frac{2 \pi k}{N}(i-j)\right) \sqrt{m^2 + \hat{k}^2}$, with lattice momentum $\hat{k} = 2 \sin\left(\frac{\pi k}{N}\right)$ in a finite-difference representation of the gradient operator.
Matrix elements of this correlation matrix can be calculated analytically in the thermodynamic limit of infinite volume where momentum modes become continuous within their $2\pi$-Brillouin zones.   In this regime, the matrix elements are
\begin{equation}
K_{\hat{r}}^{\infty} = \sqrt{4 +m^2} \  {}_3\bar{F}_{2} \left(\begin{matrix} -\frac{1}{2}, \frac{1}{2}, 1 \\ 1 - \hat{r}, 1+\hat{r}\ \end{matrix}; \frac{4}{4 + m^2 } \right) \ \ \ ,
\label{eq:KrhatTDL}
\end{equation}
where $\hat{r}$ is the lattice separation between the two sites defining the matrix element and ${}_3\bar{F}_{2}\left( \begin{matrix} a, b, c \\ d, e \end{matrix}; z\right) = {}_3F_{2}\left( \begin{matrix} a, b, c \\ d, e \end{matrix}; z\right) /\Gamma(d) \Gamma(e) $ is the regularized hypergeometric function.
In one dimension, the mass serves as an important IR regulator of the theory and cannot be analytically set to zero.
However, the asymptotic decays of the correlation matrix elements are found, as expected, to be polynomial, $\sim \frac{4}{\pi - 4 \pi \hat{r}^2}$, and exponential, $\sim-\sqrt{\frac{m}{2\pi r^3}} e^{-m r}$, in the massless and massive regimes, respectively.  With explicit scale, $m$, the massive regime has been expressed in the continuum, with $r$ the continuous separation.

The inverse of this correlation matrix, connected to the two-point field-space correlation functions, may be similarly calculated as $\left(\mathbf{K}^{-1}\right)_{ij} = \frac{1}{N} \sum_{k = 0}^{N-1} \cos\left(\frac{2 \pi k}{N}(i-j)\right) \left(m^2 + \hat{k}^2\right)^{-\frac{1}{2}}$.  In the infinite volume limit, matrix elements of this inverse become
\begin{equation}
\left(K^{-1}\right)_{\hat{r}}^{\infty} = \frac{{}_3\bar{F}_2 \left( \begin{matrix}
\frac{1}{2}, \frac{1}{2}, 1 \\
1 - \hat{r}, 1+\hat{r}
\end{matrix}; \frac{4}{4+m^2} \right)}{\sqrt{4 + m^2}} \ \ \ .
\end{equation}
From the correlation matrix and its inverse, the two-point correlation functions in the pure vacuum state may be directly expressed as
\begin{align}
  G_{ij} &= \langle \psi|(\hat{\phi}_i - \bar{\phi}_i) (\hat{\phi}_j-\bar{\phi}_j) |\psi\rangle = \frac{1}{2} \left(\mathbf{K}^{-1}\right)_{ij} \label{eq:phiphiG}\\
  H_{ij} &=  \langle \psi| (\hat{\pi}_i-\bar{\pi}_i)( \hat{\pi}_j-\bar{\pi}_j)|\psi\rangle = \frac{1}{2} K_{ij} \ \ .
\end{align}
Note trivially that $\left(\mathbf{K}^{-1}\right)_{x,y} \neq \left( \mathbf{K}_{x,y}\right)^{-1}$, where $x, y$ may be individual indices or lists indexing the matrix of correlators among portions of the latticized field vacuum.
The asymptotic decays of the inverse correlation matrix elements, connected to the two-point field correlators, are found to be logarithmic, $\sim -\left( \ln(mr)+\gamma-\ln[2]\right)/\pi$, and exponential, $\sim \frac{e^{-mr}}{\sqrt{2\pi m r}}$, in the massless and massive regimes, respectively.  For the massless regime, the remaining mass factor regulates an IR divergence  and $\gamma$ is the Euler-Mascheroni constant.

Together, these collections of quadrature expectation values completely define the covariance matrix (CM) of a Gaussian CV state, commonly structured modewise in the form
\begin{equation}
  \sigma_{ij} = \langle \psi| \left\{ \hat{q}_i-\bar{q}_i, \hat{q}_j-\bar{q}_j \right\}_+ |\psi\rangle \ \ \ ,
\end{equation}
with $\mathbf{q} = \left\{ \hat{\phi}_1, \hat{\pi}_1, \hat{\phi}_2, \hat{\pi}_2, \cdots, \hat{\phi}_N, \hat{\pi}_N\right\}$.
The pure vacuum state has vanishing first-moment displacements $\langle \hat{\pi}\rangle = \bar{\pi} = 0 =  \bar{\phi} = \langle \hat{\phi}\rangle $.
The CM is thus a riffled form of $\mathbf{G}$ and $\mathbf{H}$ with an additional factor of 2.
In this formalism,
the uncertainty relations may be expressed as the matrix inequality $\boldsymbol{\sigma} - i \boldsymbol{\Omega} \geq 0$, where the symplectic matrix encoding the commutation relations is
$ \left[ \mathbf{q}, \mathbf{q}^T \right] = i \boldsymbol{\Omega}_N $
with $\boldsymbol{\Omega}_N = \bigoplus_{j = 1}^N i \boldsymbol{\tau}_y$ and $\boldsymbol{\tau}_y$ the second Pauli matrix.
The purity, $1/\sqrt{\det \boldsymbol{\sigma}}$, of full lattice CMs indicates that they represent closed quantum systems with no further volume beyond.
By choreographing an increase in the volume and a decrease in the lattice spacing, a series of lattices provide a pure-state trajectory to the continuum limit~\footnote{Note that this differs from an alternate perspective of the lattice producing a discretized sampling of the continuous vacuum, in which case the lattice state would be mixed.
The distinction is similar to that discussed in the context of vacuum state preparation in Refs.~\cite{Klco:2019yrb,Klco:2020aud}, where preparing the vacuum state of a particular lattice is distinct from preparing a state that samples the continuum vacuum state.
These two perspectives differ in UV completions, but importantly share the same continuum limit.
}.

In the Gaussian language,
the experimentally accessible entanglement between field patches can be directly quantified~\cite{Klco:2021cxq} by the logarithmic negativity~\cite{Horodecki:1996nc,Simon:2000zz,Vidal:2002zz,Plenio:2005cwa}
\begin{equation}
  \mathcal{N}_{A|B} = - \sum_{j = 1}^N \log_2 \min \left( \nu_i^\Gamma,1\right) \ \ \ ,
\end{equation}
where $\Gamma$ indicates the partial transpose (PT), a positive transformation $\hat{\pi} \rightarrow -\hat{\pi}$ acting locally in the $A$ or $B$ spaces~\cite{Simon:2000zz}, and the PT symplectic spectrum may be calculated as
\begin{equation}
\boldsymbol{\nu}^\Gamma = 2 \text{spec}\sqrt{\mathbf{G} \mathbf{H}^\Gamma} \ \ \ .
\label{eq:specGHgamma}
\end{equation}
Note that only PT symplectic eigenvalues between 0 and 1 contribute to the logarithmic negativity.
It is using this language that Refs.~\cite{Marcovitch:2008sxc,Klco:2020rga,Klco:2021biu,Klco:2021cxq} have addressed  the negativity between disjoint patches of the free scalar field vacuum.

Physical interpretation of the negativity entanglement measure is generically limited to an upper bound of the distillable entanglement between the $A,B$ Hilbert spaces~\cite{Vidal:2002zz}.
However, the additional structure of the free scalar vacuum allows this entanglement measure to directly quantify the two-mode accessible entanglement between field patches in a single instance of the quantum vacuum.
More specifically, the entirety of the negativity may be consolidated via local unitaries into $(1_A \times 1_B)$ mode pairs, generically in the pure states identified by volume measurement~\cite{2003PhRvA..67e2311B,2003quant.ph..1038G,2004PhRvA..70e2329B}, $\boldsymbol{\sigma}^{({\rm m})}$, as well as for the mixed states, $\boldsymbol{\sigma}^{({\rm t})}$, representing an isolated pair of disjoint field patches~\cite{Klco:2021cxq}.

\subsection{Extraction of Disjoint Field Patches}

When quantifying the entanglement between two patches of the field, modes associated with field sites external to the patches may be traced or measured.
The act of measurement is commonly acknowledged to rely upon interactions that modify the observed quantum systems, incorporating the observer as an active part of the resulting quantum state.  Similarly, the absence of measurement is an action that leads to physical consequences.
Omission of entangled quantum degrees of freedom generates classical correlations throughout the remaining Hilbert space.
In a Bayesian sense, the pure/mixed final state density matrix after partial measurement/omission characterizes an operationally relevant state of knowledge specific to an observer's retention/loss of information.

Tracing the volume external to a set of field patches represents the typical information processing that occurs in standard experimental design, where interaction with quantum fields is mediated through spatially localized detectors.
In the covariance matrix formalism, this external volume tracing may be performed by extracting intra- and inter-patch
matrix elements, as these observables are unchanged by volume integration.
Explicitly, for indices $i, j \in \{p\}$ within the two patches, the matrix of two-point field correlators upon tracing of the external volume is
\begin{equation}
G^{\rm (t)}_{ij} = \Tr_p \left[ \hat{\phi}_i \hat{\phi}_j \rho_p\right] = \Tr \left[ \hat{\phi}_i \hat{\phi}_j \rho\right]  \ \ \ ,
\end{equation}
where the $v, p$ subscripts indicate that operators reside in the external volume~\footnote{Note: though it is common terminology to utilize \enquote{volume} to indicate the entirety of a latticized field, it will be used presently to indicate the entirety of the latticized field \emph{external to the detection patches of interest}.} or in the patches of interest, $\Tr[\cdot] = \Tr_p[\Tr_v[\cdot]]$, and $\rho_p = \Tr_v[\rho]$.
As a result,
\begin{align}
  \mathbf{G}^{\rm (t)} = \mathbf{G}_{pp} &= \frac{1}{2} \left(\mathbf{K}^{-1}\right)_{pp} \label{eq:GHtrace} \\ \mathbf{H}^{\rm (t)} = \mathbf{H}_{pp} &= \frac{1}{2} \mathbf{K}_{pp} \ \ \ ,
\end{align}
with superscript, $(\rm t)$, indicating tracing of the external volume.
This convenient feature of the Gaussian formalism
has allowed numerical calculations of continuum limit extrapolations of the infinite-volume entanglement structure in 1-, 2-, and 3-dimensions~\cite{Marcovitch:2008sxc,Klco:2020rga,Klco:2021biu}, despite
strong cancellations in determination of the PT symplectic spectrum.

\subsection{External Volume Measurement}

Beyond a description of local detector observations, retaining information of measured volume configurations allows for exploration of the underlying entanglement distributed in the field.
This measurement produces an index that distinguishes distinct pure state ensembles, allowing recovery of a pure state in the Hilbert space of the patches, as expressed diagrammatically in Fig.~\ref{fig:protocolcircuit} (see Appendix~\ref{app:underlying} for a simple discrete-variable example).
In this way, the patches can be analyzed in terms of a physical decomposition of pure states arising from the infinite-volume vacuum rather than as a mixed state of unknown origin or composition~\footnote{Note: in general, mixed states are known to not have unique pure-state convex decompositions.}.

When a globally pure vacuum state, $|\psi\rangle$, is partially projected onto a particular configuration $|\boldsymbol{\phi}_v\rangle$ in the basis of field eigenstates exterior to the patches, the remaining pure state in the patch Hilbert space is
\begin{multline}
|\psi_p\rangle_{\boldsymbol{\phi}_v} \propto \langle \boldsymbol{\phi}_v | \psi\rangle =  \frac{\det \mathbf{K}^{\frac{1}{4}}}{\pi^{\frac{N}{4}}} e^{-\frac{1}{2} \boldsymbol{\phi}_v^T \mathbf{K}_{vv} \boldsymbol{\phi}_v} \\ \int \dif \boldsymbol{\phi}_p \ e^{-\frac{1}{2} \boldsymbol{\phi}_p^T \mathbf{K}_{pp} \boldsymbol{\phi}_p - \boldsymbol{u}^T\boldsymbol{\phi}_p }|\boldsymbol{\phi}_p\rangle \ \ \ ,
\end{multline}
with $\mathbf{u} = \mathbf{K}_{pv} \boldsymbol{\phi}_v$.
In this application, the union of $\boldsymbol{\phi}_p$ and $\boldsymbol{\phi}_v$ captures CV modes throughout the entirety of the field volume in which the vacuum is a pure state, i.e., $[p]+[v] = N$.
Such partial projections are composed of the pure state after measurement and the associated amplitude,
$\langle \boldsymbol{\phi}_v | \psi\rangle \equiv \mathcal{A}_{\boldsymbol{\phi}_v} |\psi_p\rangle_{\boldsymbol{\phi}_v}$.
Identifying the normalized pure state in the patch Hilbert space after this projection leads to
\begin{multline}
  |\psi_p\rangle_{\boldsymbol{\phi}_v} = \frac{\det \mathbf{K}_{pp}^{\frac{1}{4}}}{\pi^{\frac{[p]}{4}}} e^{-\frac{1}{2} \mathbf{u}^T (\mathbf{K}_{pp})^{-1} \mathbf{u} } \\ \int \dif \boldsymbol{\phi}_p \ \exp\left[-\frac{1}{2} \boldsymbol{\phi}_p^T \mathbf{K}_{pp} \boldsymbol{\phi}_p - \boldsymbol{u}^T \boldsymbol{\phi}_p \right] |\boldsymbol{\phi}_p\rangle  \ \ \ ,
  \label{eq:purestatepostmeasurement}
\end{multline}
employing the standard Gaussian integral
\begin{equation}
\int d^n \mathbf{x} \ e^{- \mathbf{x}^T \mathbf{A} \mathbf{x} - 2 \mathbf{B}^T \mathbf{x}} = \sqrt{\frac{\pi^n}{\det \mathbf{A}}} e^{\mathbf{B}^T \mathbf{A}^{-1} \mathbf{B}} \ \ \ .
\end{equation}
The associated
$\boldsymbol{\phi}_v$-configuration-dependent amplitude is thus determined to be
\begin{equation}
  \mathcal{A}_{\boldsymbol{\phi}_v} = \frac{1}{\pi^{\frac{[v]}{4}}}  \frac{\det \mathbf{K}^{\frac{1}{4}}}{\det \mathbf{K}_{pp}^{\frac{1}{4}} } \frac{e^{-\frac{1}{2} \boldsymbol{\phi}_v^T \mathbf{K}_{vv} \boldsymbol{\phi}_v}}{e^{-\frac{1}{2} \mathbf{u}^T (\mathbf{K}_{pp})^{-1} \mathbf{u} }} \ \ \ .
\end{equation}
If the volume external to the patches were to be traced (i.e., loss of measurement information),
this weighted set of pure-states would form an explicit convex decomposition of the patch density matrix as
\begin{align}
  \rho_p &=  \int  \dif \boldsymbol{\phi}_v \ \langle \boldsymbol{\phi}_v |\psi\rangle \langle \psi | \boldsymbol{\phi}_v \rangle
  \\
  &= \int \dif \boldsymbol{\phi}_v \ |\mathcal{A}_{\boldsymbol{\phi}_v}|^2 \ \rho\Big( |\psi_p\rangle_{\boldsymbol{\phi}_v} \Big)\ \ \ ,
\end{align}
with $\rho(|\chi\rangle) = |\chi\rangle \langle \chi|$ a pure state density matrix.

Note from equation~\eqref{eq:purestatepostmeasurement}
that the pure-state ensembles distinguished by $\boldsymbol{\phi}_v$-configuration index are simply displacements of the same Gaussian state, i.e., the $\boldsymbol{\phi}_v$-content of the state appears only in the vector of first moment displacements, $\mathbf{u}^T$.
Therefore, each pure state shares the same CM, and thus the same entanglement.
In the following, this feature will allow
a description in terms of a single CM per measurement basis, rather than per measurement configuration.

As usual for pure Gaussian states, the volume-projected patch CMs, those of  Eq.~\eqref{eq:purestatepostmeasurement}, are independent of the specific $\boldsymbol{\phi}_v$-configuration measured.
As such, the mixed state CM representing $\rho_p$ upon tracing of the volume may be described by a convex decomposition of  displaced Gaussian states each with the same pure CM characterized by components,
\begin{equation}
\mathbf{G}^{\rm (m, \phi)} = \frac{1}{2} \left(\mathbf{K}_{pp}\right)^{-1} \ \ , \ \  \mathbf{H}^{\rm (m, \phi)} = \frac{1}{2} \mathbf{K}_{pp} \ \ \ ,
\label{eq:GHphimeasurement}
\end{equation}
with the superscript, $({\rm m}, \phi)$, indicating measurement of the external volume in the $\phi$-basis and classical communication of the result as shown in Fig.~\ref{fig:protocolcircuit}.
Note that, upon projecting the volume into the field eigenbasis, no displacements have been generated in conjugate momentum space and $\mathbf{H}^{\rm (m, \phi)} = \mathbf{H}^{\rm (t)}$.
The field correlators are found to differ by an order of operations, with volume coordinate removal occurring before(after) inversion for sites that are measured(traced), see respectively Eqs.~\eqref{eq:GHphimeasurement} and~\eqref{eq:GHtrace}.  This is consistent with the relation $\mathbf{G}^{\rm{(pure)}} = \frac{1}{4} \left(\mathbf{H}^{\rm{(pure)}}\right)^{-1}$, producing a pure-state CM with unit determinant.

Expressing the fact that volume measurement naturally leads to a particular pure-state convex decomposition of the patch-patch density matrix,
the traced CM can be written as
\begin{equation}
  \boldsymbol{\sigma}^{\rm (t)} = \boldsymbol{\sigma}^{\rm (m)} + \mathbf{Y} \ \ \ ,
\label{eq:sigmaplusYrelation}
\end{equation}
where $\mathbf{Y} \ge 0$ is a matrix that may be interpreted physically as governing a classical distribution of first-moment displacements.
Upon tracing of the field volume external to the patches, the measurement information distinguishing displaced versions of the $\boldsymbol{\sigma}^{({\rm m})}$ pure state is lost, generating a mixed-state ensemble with CM $\boldsymbol{\sigma}^{({\rm t})}$ (see Appendix~\ref{app:cmmixedformalism} for further discussion).
As noted above, no displacements are generated in conjugate momentum space and thus the associated matrix elements of $\mathbf{Y}$ vanish, $\mathbf{Y}^{\rm (m, \phi)}_\pi = \mathbf{0}$.
The field operator elements of $\mathbf Y$ may be expressed as
\begin{equation}
\scalemath{0.98}{
  \left(\mathbf{Y}^{\rm (m,\phi)}_\phi\right)_{ij} = 2 \int \dif \boldsymbol{\phi}_v \ \left|\mathcal{A}_{\boldsymbol{\phi}_v} \right|^2 \ \langle \hat{\phi}_{p,i} \rangle_{\boldsymbol{\phi}_v}  \langle \hat{\phi}_{p,j} \rangle_{\boldsymbol{\phi}_v} } \  ,
  \label{eq:Yexpression}
\end{equation}
where $\mathbf{G}^{\rm (t)} = \mathbf{G}^{\rm (m, \phi)} + \frac{1}{2} \mathbf{Y}_\phi^{\rm (m, \phi)}$.
The displacements in field space are calculated (e.g., by standard source-variation techniques of partial differentiation with respect to $\mathbf{u}$ incorporated into the  Eq.~\eqref{eq:purestatepostmeasurement} integrand) to be
\begin{equation}
  \langle \hat{\boldsymbol{\phi}}_p \rangle_{\boldsymbol{\phi}_v} ={}_{\boldsymbol{\phi}_v}\langle\psi_p| \boldsymbol{\hat{\phi}}_p |\psi_p\rangle_{\boldsymbol{\phi}_v} = -\left(\mathbf{K}_{pp}\right)^{-1} \mathbf{u} \ \ \ ,
\label{eq:displacementsfromvolumeconfig}
\end{equation}
as a function of the projected volume configuration.

The integration of Eq.~\eqref{eq:Yexpression} can be performed with the external-volume-configuration-dependent displacements of Eq.~\eqref{eq:displacementsfromvolumeconfig}.
However, it may be convenient to express the classical noise, $\mathbf{Y}_\phi^{({\rm m}, \phi)}$, directly in terms of the matrix elements of $\mathbf{K}$, the site-wise correlation matrix.
To do so, the standard partial Gaussian integration,
\begin{equation}
\int \cdots \int \text{d} \bar{\mathbf{x}}\  e^{-\frac{1}{2} \mathbf{x}^T \mathbf{A} \mathbf{x}} = \frac{(2\pi)^{\frac{\left[\bar{\mathbf{x}}\right]}{2}}}{\sqrt{\det \bar{\mathbf{A}}}} e^{-\frac{1}{2} \mathbf{x}_0^T \mathbf{F} \mathbf{x}_0}
\end{equation}
with
\begin{equation}
\mathbf{A} \mathbf{x} = \begin{pmatrix}
\mathbf{A}_0 & \mathbf{B} \\
\mathbf{B}^T & \bar{\mathbf{A}}
\end{pmatrix} \begin{pmatrix}
\mathbf{x}_0 \\ \bar{\mathbf{x}}
\end{pmatrix} \ \ , \ \ \mathbf{F} = \mathbf{A}_0 - \mathbf{B} \bar{\mathbf{A}}^{-1} \mathbf{B}^T \ \ ,
\end{equation}
is applied to the scalar field vacuum density matrix.
Identifying $\mathbf{A}_0 = \mathbf{K}_{pp} \oplus \mathbf{K}_{pp}$, $\mathbf{B} = \begin{pmatrix} \mathbf{K}_{pv} \\ \mathbf{K}_{pv} \end{pmatrix}$, and $\bar{\mathbf{A}} = 2 \mathbf{K}_{vv}$ for the traced integration of $\bar{\mathbf{x}} = \boldsymbol{\phi}_v$ yields the reduced density matrix of the patches as
\begin{multline}
\rho_p = \frac{\det \mathbf{K}^{\frac{1}{2}}}{\pi^\frac{[p]}{2}\sqrt{\det \mathbf{K}_{vv}}} \int \text{d} {\boldsymbol\phi}_p \text{d}{\boldsymbol\phi}_p' \\ \exp\left[-\frac{1}{2} \begin{pmatrix}
\boldsymbol{\phi}_p & \boldsymbol{\phi}_p'
\end{pmatrix}^T
\mathbf{F}
\begin{pmatrix}
\boldsymbol{\phi}_p \\ \boldsymbol{\phi}_p'
\end{pmatrix}\right] |\boldsymbol{\phi}_p\rangle \langle \boldsymbol{\phi}_p'|  \ \ .
\end{multline}
To determine the two-point field correlation functions, $\text{Tr}_p\left[ \hat{\phi}_i \hat{\phi}_j \rho_p\right]$, the relevant exponent becomes the sum over patch-dimension blocks, $\mathbf{F}_{\Sigma} =2\mathbf{K}_{pp} - 4\mathbf{K}_{pv}\left(2 \mathbf{K}_{vv}\right)^{-1} \mathbf{K}_{pv}^T$.
By inspection and connection to the weighted integrations leading to Eq.~\eqref{eq:phiphiG}, the field correlators may thus be read directly as
\begin{equation}
\scalemath{0.99}{
  \mathbf{G}^{\rm (t)} = \mathbf{F}_{\Sigma}^{-1} =\frac{1}{2}  \left( \mathbf{K}_{pp} - \mathbf{K}_{pv} \left(\mathbf{K}_{vv}\right)^{-1} \mathbf{K}_{pv}^T \right)^{-1} } \ \ .
\end{equation}
Expressing these traced expectation values as a deformation of those of the underlying pure state upon $\phi$-basis projective measurement of the external volume, Eq.~\eqref{eq:GHphimeasurement}, may be achieved through application of the following inverse relation~\cite{inversesumref}
\begin{equation}
\scalemath{0.98}{
  \left( \mathbf{C} + \mathbf{D}\right)^{-1} = \mathbf{C}^{-1} - \left( \mathbb{I} + \mathbf{C}^{-1} \mathbf{D} \right)^{-1} \mathbf{C}^{-1} \mathbf{D} \mathbf{C}^{-1}}  \ ,
\end{equation}
with $\mathbf{C} = \mathbf{K}_{pp}$.
As a result, the Gaussian noise corresponding to the first moment displacements may be written as
\begin{equation}
  \mathbf{Y}^{\rm (m, \phi)}_\phi = \left( \mathbb{I} - \bar{\mathbf{K}}\right)^{-1} \bar{\mathbf{K}} \left(\mathbf{K}_{pp}\right)^{-1} \ \ \ ,
\end{equation}
with $\bar{\mathbf{K}} = \left(\mathbf{K}_{pp}\right)^{-1} \mathbf{K}_{pv} \left(\mathbf{K}_{vv}\right)^{-1} \mathbf{K}_{pv}^T$.

A decomposition of the form in Eq.~\eqref{eq:sigmaplusYrelation} exists for each external-volume measurement basis (i.e., associated with $U$ in Fig.~\ref{fig:protocolcircuit}) and provides a physically meaningful pure-state convex decomposition of the quantum state in the field patches, leading to an infinite family of pure states $\boldsymbol{\sigma}^{\rm (m)} \leq \boldsymbol{\sigma}^{\rm (t)}$.
As one alternative to the $\phi$-basis, when
writing the vacuum state in the conjugate momentum basis,
\begin{equation}
|\psi\rangle = \frac{\det \mathbf{K}^{-\frac{1}{4}}}{\pi^{\frac{N}{4}}} \int \dif \boldsymbol{\pi} \  e^{-\frac{1}{2} \boldsymbol{\pi}^T \mathbf{K}^{-1} \boldsymbol{\pi}} |\boldsymbol{\pi}\rangle \ \ \ ,
\end{equation}
measurements may be considered that project onto conjugate-momentum-space eigenstates of the field external to the patches.
This representation may be derived from the field-space representation of Eq.~\eqref{eq:vacuumWF} through insertion of the conjugate-momentum-basis identity operator, $\int \dif \boldsymbol{\pi} \ |\boldsymbol{\pi} \rangle \langle \boldsymbol{\pi} | $, along with the plane-wave inner-product, $\langle \boldsymbol{\pi} |\boldsymbol{\phi} \rangle = \frac{1}{\sqrt{\left( 2 \pi\right)^N}} e^{-i \boldsymbol{\pi}^T \boldsymbol{\phi}}$.
Through direct application of the previous analysis, measurement of the volume in this basis yields the following patch-CM elements
\begin{align}
\mathbf{G}^{\rm (m, \pi)} &= \frac{1}{2} \left(\mathbf{K}^{-1}\right)_{pp} \\ \mathbf{H}^{\rm (m, \pi)} &= \frac{1}{2}\left( \left( \mathbf{K}^{-1}\right)_{pp}\right)^{-1} \ \ \  .
\end{align}
In this case, $ \mathbf{G}^{\rm (m, \pi)}= \mathbf{G}^{\rm (t)} $ and thus all displacement noise is relegated to the $\pi$-coordinates, $\mathbf{Y}^{\rm (m, \pi)}_\phi = \mathbf{0}$.

The fact that measured volume configurations index pure states distinguished only by first-moment displacements, Eq.~\eqref{eq:displacementsfromvolumeconfig}, allows the LOCC-creation of a single pure state per measurement basis, as shown in Fig.~\ref{fig:protocolcircuit}.
In the following, we analyze the entanglement of such pure states to quantify the underlying quantum correlations distributed in the field vacuum, significant portions of which are only visible when the classical noise generated by traced isolation of field patches is effectively removed through volume measurement.

\section{Detector-Observed Entanglement}
\label{sec:consequences}
Utilizing perspective gained by examining the pure quantum states of scalar field vacuum regions identified upon measurement of the external field volume, this section discusses entanglement properties that may be
specifically attributed to the classical correlations arising from local observation of a disjoint pair of detection patches.
In the following, four principle impacts are considered:
A) truncation of the spatial extent of quantum correlations,
B) parametric suppression of their decay with region separation,
C) mixing of information from disparate frequency regimes,
and
D) delocalization of entanglement information, challenging the understanding of many-body correlations in mixed quantum states.

\subsection{Truncated Range of Entanglement}
\label{sec:rtildeNslash}

\begin{figure*}
  \includegraphics[width=0.75\textwidth]{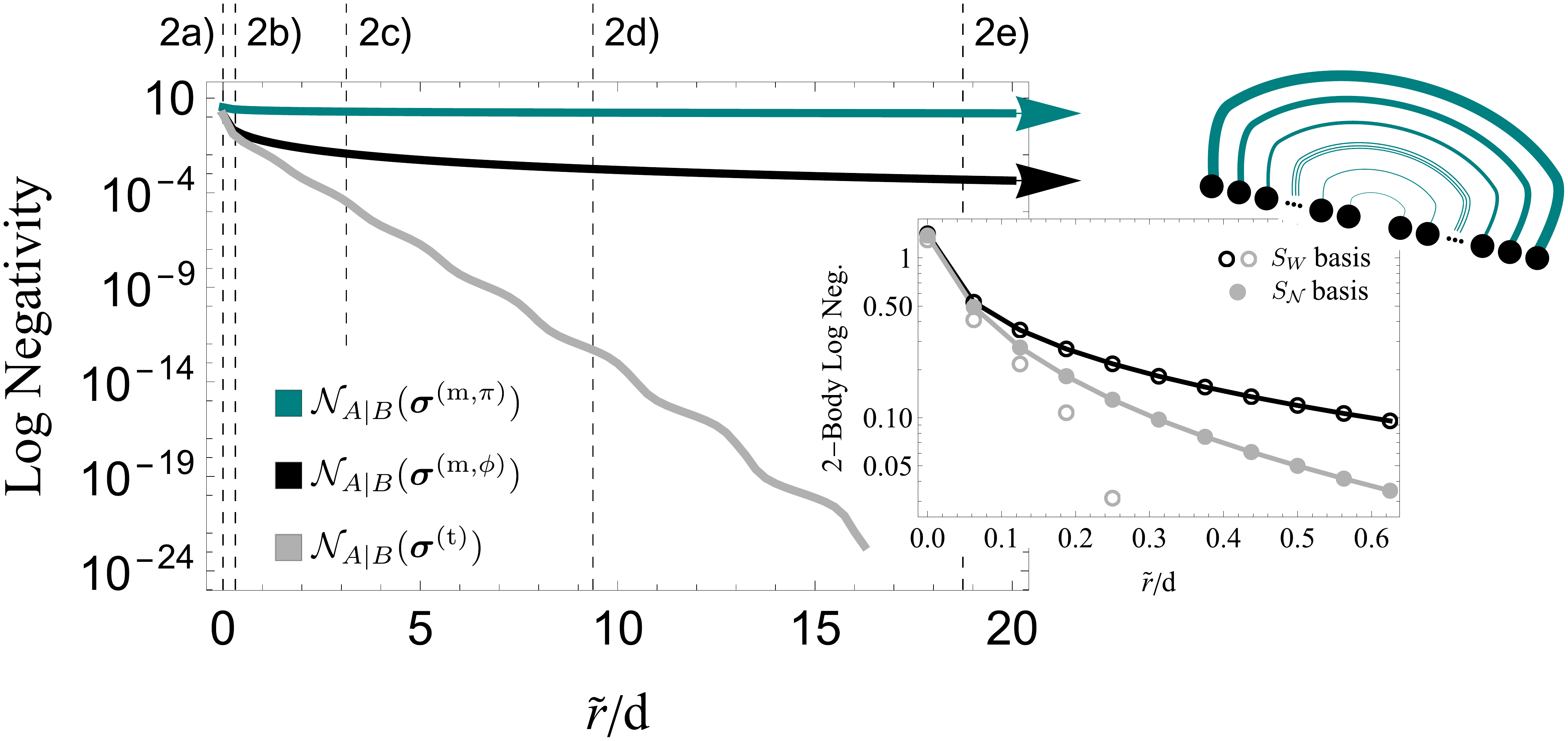}
  \caption{Underlying and locally accessible entanglement, quantified by the logarithmic negativities $\mathcal{N}_{A|B}(\boldsymbol\sigma^{({\rm m})})$ and $\mathcal{N}_{A|B}(\boldsymbol\sigma^{({\rm t})})$ respectively, between patches ($d = 16$ sites each) of the one-dimensional free lattice scalar field vacuum in the massless limit ($m = 10^{-10}$) as a function of dimensionless lattice separation, $\tilde{r}/d$, in the thermodynamic limit of infinite volume.  The classical noise generated upon tracing of the patch-external field volume produces a parametric suppression of the entanglement from poly/log to exponentially decaying, culminating in compatibility with a separable decomposition at finite dimensionless separation, $\tilde{r}_{\Nslash} = 264$. The vertical dashed lines are labeled in correspondence with the wavefunctions in the upper row of Fig.~\ref{fig:negGSwf}. \textbf{Inset:} Sum of two-body, $(1_A\times 1_B)$, contributions to the logarithmic negativity between volume-measured (black) and volume-traced (gray) field patches calculated in the local Williamson~\cite{Williamson1936,Serafini2017} and negativity-inspired~\cite{Klco:2021cxq} bases, as discussed in Sec.~\ref{sec:delocalizing}.  Local transformations to these bases (by $\mathbf{S}_W$ and $\mathbf{S}_{\mathcal{N}}$, respectively) are designed to maintain the parity symmetry of the pair of field patches, leading to the diagrammatically indicated two-body structure.
  }
  \label{fig:scalarUnderlyintEntanglement}
\end{figure*}

It has been long realized that the accessible entanglement between patches of UV-truncated fields vanishes at large spatial separations relative to the patch size~\cite{Audenaert:2002xfl,2004PhRvA..70e2329B,kofler2006entanglement,Marcovitch:2008sxc,Calabrese:2009ez,Zych:2010yk,Calabrese:2012ew,Calabrese:2012nk,MohammadiMozaffar:2017nri,Coser_2017,Klco:2019yrb,DiGiulio:2019cxv,Klco:2020rga,Klco:2021biu}.
Thus, in order to calculate the entanglement known to be present at all space-like separations in the continuum field~\cite{Reeh1961,summers1985vacuum,summers1987bell1,summers1987bell2}, a framework must be utilized that systematically controls the effects of the computationally-necessary UV truncation.
The present calculations utilize the systematic framework of lattice field theory, in which extrapolations may be performed with a series of lattices that pixelate the physical field patches with increasing resolution~\cite{Marcovitch:2008sxc,Klco:2020rga,Klco:2021biu}, i.e., with larger lattice diameter, $d$, of each field patch.
The separation at which the logarithmic negativity vanishes (and equivalently where patches of the field become separable~\cite{Klco:2021biu}), $\tilde{r}_{\Nslash}$, extends as the continuum limit is approached and has been found to scale as $\tilde{r}_{\Nslash}/d \sim (\gamma/a) d$, 
with $\gamma$ now a parameter scaling inversely with the spatial dimension~\cite{Klco:2021biu}.

To further understand latticized quantum field correlation structure, 
it is interesting to consider whether entanglement is truly present outside $\tilde{r}_{\Nslash}$, but has been buried by classical noise~\cite{PhysRevLett.80.2493,PhysRevA.58.826,Vidal:1998ch,PhysRevA.63.032306,Klco:2021cxq} as a result of the local entanglement detection schemes considered.
In particular, the reduced density matrix of field patches may become compatible with a separable convex decomposition despite the presence of entanglement in the contributing pure state ensembles.
By definition, such entanglement within a separable mixed state cannot be detected, neither theoretically nor experimentally, from the patch reduced density matrix alone.
However, it may be informed by the physical convex decompositions identified from the larger quantum state of the field.
To gain theoretical access to such quantities, Fig.~\ref{fig:protocolcircuit} indicates the present strategy
utilizing external volume measurement.
Such physical decompositions of the local mixed state isolate the classical noise added upon their local observation, conclusively establishing the underlying presence of entanglement distributed at all distances in the free lattice scalar field.

To illustrate this long-distance entanglement underlying the free lattice scalar vacuum, consider the simplest case $\boldsymbol{\phi}_p = \begin{pmatrix}
\phi_0 & \phi_{\tilde{r}+1}
\end{pmatrix}$, consisting of one site in each patch $(d = 1)$ separated by distance $\tilde{r}$ in lattice units.
If the volume external to these two sites is traced, the quantum state of the two sites becomes separable at separation $\tilde{r} = 1$ and beyond.
To characterize the pure states in the ensemble upon measurement of the volume in the local $\phi$-basis, the matrices of two-point correlation functions (Eq.~\eqref{eq:GHphimeasurement}) are
\begin{align}
\mathbf{G}^{({\rm m}, \phi)} &= \frac{1}{2} \begin{pmatrix}
K_{0,0} & K_{0, \tilde{r} + 1} \\
K_{0, \tilde{r}+1} & K_{0,0}
\end{pmatrix}^{-1} \\
\left(\mathbf{H}^{({\rm m}, \phi)}\right)^{\Gamma} &= \frac{1}{2} \begin{pmatrix}
K_{0,0} & - K_{0, \tilde{r} + 1} \\
-K_{0, \tilde{r}+1} & K_{0,0}
\end{pmatrix} \ \ \ .
\end{align}
Calculating the PT symplectic eigenvalues of $\boldsymbol{\sigma}^{({\rm m}, \phi)}$ from the eigenvalues of $2 \sqrt{\mathbf{G}\mathbf{H}^\Gamma}$ produces, $\nu_\pm = \frac{K_{0,0} \mp K_{0, \tilde{r}+1}}{\sqrt{K_{0,0}^2 - K_{0, \tilde{r}+1}^2}}$.
As off-diagonal elements of $K$ are negative (favoring positively correlated displacements of field lattice sites), the symplectic eigenvalue $\nu_-$ is the one that contributes to the patch-patch negativity.
The underlying negativity between two single-site patches is therefore
\begin{equation}
\mathcal{N}^{({\rm m}, \phi)}_{0|\tilde{r}+1} = -\log_2  \frac{K_{0,0} + K_{0, \tilde{r}+1}}{\sqrt{K_{0,0}^2 - K_{0, \tilde{r}+1}^2}}
\label{eq:d1underlyingnegativity}
\end{equation}
for measurements in the field eigenbasis.
Analysis for the case of $\pi$-basis measurement of the external volume results in underlying logarithmic negativity of the form
\begin{equation}
\mathcal{N}_{0|\tilde{r}+1}^{({\rm m}, \pi)} = -\log_2 \frac{\left(\mathbf{K}^{-1}\right)_{0,0} - \left(\mathbf{K}^{-1}\right)_{0,\tilde{r}+1}}{\sqrt{\left(\mathbf{K}^{-1}\right)^2_{0,0} - \left( \mathbf{K}^{-1}\right)^2_{0,\tilde{r}+1} }}
\label{eq:d1underlyingnegativityPI}
\end{equation}
Notably, these values are non-zero for all separations of the two single-site patches on the lattice.

For a system with larger $d$, and thus  a lattice further along the trajectory to the continuum, the main panel of Fig.~\ref{fig:scalarUnderlyintEntanglement} shows, in the massless regime, the decay of accessible entanglement quantified by the logarithmic negativity of the $\boldsymbol\sigma^{\rm (t)}$ mixed state and underlying entanglements quantified by the logarithmic negativity of the $\boldsymbol\sigma^{\rm (m, \phi)}$ and $\boldsymbol\sigma^{\rm (m, \pi)}$ pure states.
When measurements external to the patches are performed, either in the $\phi$- or $\pi$-basis, and communicated to
identify a common pure-state ensemble in the patch Hilbert space, entanglement between the patches is found to be present at all distances, as indicated by the arrows at right of the Fig.~\ref{fig:scalarUnderlyintEntanglement} main panel.
Therefore, entanglement can be accessed from the Hilbert space of the patches at separations beyond $\tilde{r}_{\Nslash}/d \sim 16$ if classical communications are provided from additional observations of the volume.

\subsection{Decay of Accessible Entanglement}
\label{sec:distillabledecay}

Though well-established, the phenomenon that accessible entanglement in the massless regime decays exponentially with the patch spatial separation (in units of the patch size), $\mathcal{N} \sim e^{-\beta \frac{\tilde{r}}{d}}$,  remains beyond current analytic control.
For massless scalar fields, numerical approaches have shown that $\beta$ scales
approximately linearly up to three spatial dimensions~\cite{Marcovitch:2008sxc,Calabrese:2012nk,Klco:2020rga,Klco:2021biu}, indicating a more rapid decay as the geometric phase space increases.
The feature limiting both analytic and numerical determinations of this entanglement structure is a sign problem, as poly/log correlation functions combine to produce the exponential decay of accessible entanglement.
The present analysis of underlying entanglement in physical convex decompositions of the patch-patch Hilbert space reveals that this decay is not an internal feature in the field, e.g., as would be the case for generation of an effectively massive quasi-particle propagation, but rather a
response pattern of the field to characterization by spatially local detectors alone.

\begin{figure*}
\includegraphics[width = 0.95\textwidth]{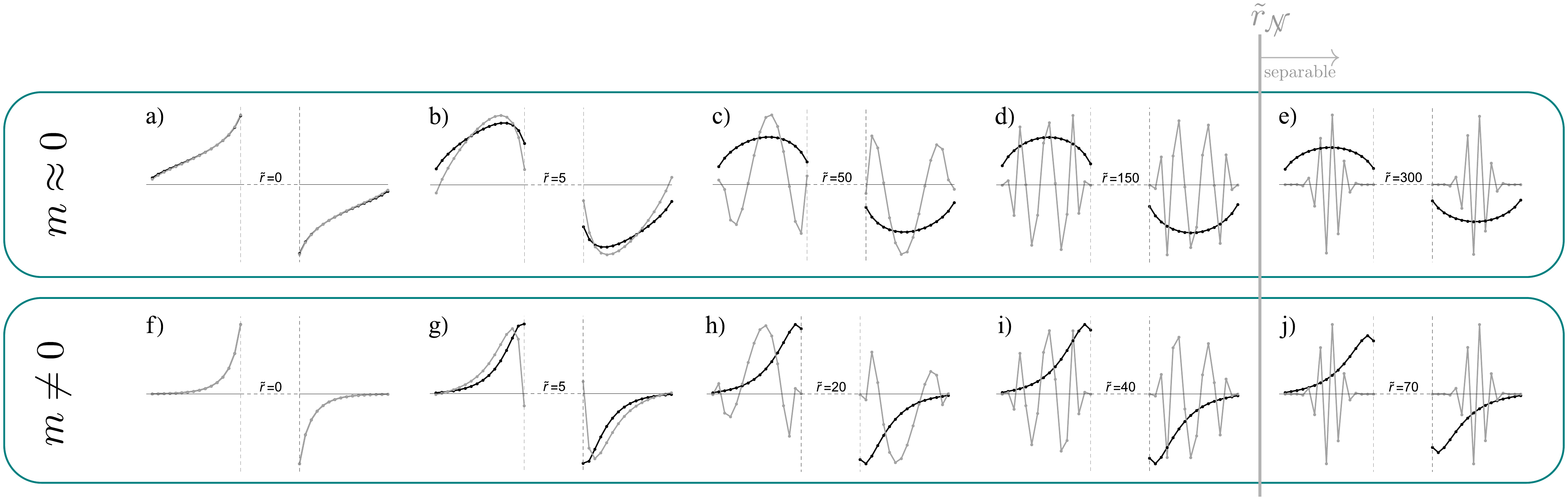}
  \caption{Spatial dependence of $\mathbf{G}\mathbf{H}^\Gamma$ ground state wavefunctions associated with the dominant contribution to the logarithmic negativity between field patches in the massless(top) regime with $d = 16, m = 10^{-10}$ ($\tilde{r}_{\Nslash} = 264$) and in the massive(bottom) regime with $d = 16, m = 0.3$ ($\tilde{r}_{\Nslash} = 59$).
In all cases, the wavefunction upon volume measurement (associated with $\boldsymbol\sigma^{({\rm m}, \phi)}$, black) remains predominantly in the IR, while the wavefunction upon volume tracing (associated with $\boldsymbol\sigma^{({\rm t})}$, gray)
is driven into the UV with increasing separation.
   }
  \label{fig:negGSwf}
\end{figure*}

Concretely, an asymptotic analysis of Eq.~\eqref{eq:d1underlyingnegativity} at large $\tilde{r}$ reveals the underlying entanglement to follow the envelope of the off-diagonal correlation matrix element, $ \mathcal{N}_{0|\tilde{r}+1}^{({\rm m}, \phi)} \rightarrow -\frac{K_{0, \tilde{r}+1}}{\ln[2] K_{0,0}}$, when the volume is measured in the $\phi$-basis.
In order to express the correlation structure of the massless field,
the IR regulation required for scalar fields in one spatial dimension may be rendered by inclusion of a mass chosen to have Compton wavelength significantly larger than the patch-pair spatial dimensions.
In this regime, $K_{0,\tilde{r}} \sim \frac{1}{\tilde{r}^2}$ and the underlying entanglement distributed in the field thus decays polynomially.
An analogous asymptotic analysis of Eq.~\eqref{eq:d1underlyingnegativityPI} for external volume measurement in the $\pi$-basis leads to logarithmic negativity that follows the envelope of off-diagonal inverse correlation matrix elements, $\mathcal{N}_{0|\tilde{r}+1}^{({\rm m}, \pi)} \rightarrow \frac{\left( \mathbf{K}^{-1} \right)_{0,\tilde{r}+1}}{\ln[2] \left( \mathbf{K}^{-1} \right)_{0,0}}$.
In the long-distance and massless regime, $\left(\mathbf{K}^{-1}\right)_{0,\tilde{r}} \sim \log (m\tilde{r})$ and the underlying entanglement distributed in the field thus decays logarithmically.
As illustrated in Fig.~\ref{fig:scalarUnderlyintEntanglement} for a higher resolution representation of the two field patches (larger $d$), when $\pi$-basis measurement results are provided from the volume, a pure state (CM $ \boldsymbol{\sigma}^{({\rm m}, \pi)}$) with parametrically greater underlying entanglement can be produced.

As is the case for the finite range of entanglement discussed in Sec.~\ref{sec:rtildeNslash}, the distinction clarified by the above examples is that the physically distillable entanglement is not falling exponentially in a massless theory because of a natural correlation structure in the field.
Rather, the underlying entanglement decays with the poly/log correlations and the act of observation via spatially local detectors produces classical noise to which the measurement of quantum correlations is sensitive.
This sensitivity translates into parametric suppression leading the accessible entanglement to fall exponentially.
The protocols of Fig.~\ref{fig:protocolcircuit} therefore allow conclusive determination that local detectors are capable of accessing only a small portion of the entanglement naturally distributed within the scalar field vacuum.

\subsection{Mixing of Separated Scales}

As indicated in Eq.~\eqref{eq:specGHgamma} for CV quantum states with vanishing $\langle\phi \pi\rangle$ correlators, the symplectic eigenvalues are calculable through the spectrum of $\mathbf{G} \mathbf{H}^{\Gamma}$.
The smallest eigenvalue, the ground state \enquote{energy} of $\mathbf{G}\mathbf{H}^{\Gamma}$, provides the dominant contribution to the logarithmic negativity.
Therefore, the ground state wavefunction of $\mathbf{G} \mathbf{H}^\Gamma$, shown in Fig.~\ref{fig:negGSwf} in a massive and massless regime for a range of patch separations, shows the leading-order in an entanglement hierarchy for the spatial structure of collective modes across the field patches~\footnote{Locally consolidating these collective modes into $(1_A \times 1_B)$ pairs of CV degrees of freedom~\cite{Klco:2021cxq} may then be physically regarded as a clear methodology for transferring entanglement to a quantum  detection system.}.
In both the massless and massive regimes, contributions to the logarithmic negativity organized by the symplectic spectrum exhibit an exponential hierarchy~\cite{Klco:2021biu,Klco:2021cxq}, and thus the leading-order reliably approximates the entanglement.

It has previously been shown that the collective modes allowing extraction of the dominant entanglement contribution between patches of the scalar vacuum contain frequencies that increase at larger spatial separation, potentially impacting the form of future effective theories aiming to capture non-local quantum properties of the field~\cite{Klco:2021biu}.
This relation can be seen in the ground-state wavefunctions of $\mathbf{G}\mathbf{H}^{\Gamma}$ when the volume external to the field patches is traced, depicted (gray) in Fig.~\ref{fig:negGSwf}.
In this case, the spatial wavefunctions associated with the $(1_A\times 1_B)$ mode pair of dominant logarithmic negativity contribution exhibit higher frequency spectra for longer distance patch separations.

The protocols of Fig.~\ref{fig:protocolcircuit} allow this UV/IR connection to be understood as a property of the field specific to the vantage of local observation by the pair of field patches.
As depicted (black) in  Fig.~\ref{fig:negGSwf}, when the volume external to the patches is measured and the result classically communicated to prevent classical mixing of 
ensembles indexed by volume configuration, the wavefunctions characterizing the dominant accessible entanglement do not exhibit UV/IR mixing.
Even at long distances, this leading-order wavefunction is a ground-state that retains an IR spectrum.
Calculated directly through the patch-pair PT symplectic eigenvectors, it is clear that the procedure of external volume tracing (gray) embeds both UV and IR information from the field volume, through classical correlations, into the reduced density matrix of the patches.
Extending this observation, connections between UV and IR energy scales may be expected to be a common feature of entanglement observables in quantum fields when probed by local detectors.

\subsection{Delocalization of Entanglement Information}
\label{sec:delocalizing}
Though classical correlations cannot generate additional entanglement, their effects on entanglement structure are far from neutral.
Beyond suppression, as discussed in Sections~\ref{sec:rtildeNslash} and~\ref{sec:distillabledecay}, a closely related consequence of classical correlations (e.g., accompanying local measurements) is to reorganize or delocalize information important to the observation of  entanglement.
In the following, this consequence will be demonstrated by the observation that classical noise may shift the basis in which a quantum system has $(1_A \times 1_B)^{\otimes n}$ entanglement structure.
This observable, the noise-sensitivity of the $(1\times 1)$ entanglement basis, is not only valuable from the perspective of local entanglement detection, but is also relevant in a variety of other contexts, e.g.,
initializing non-trivial quantum states, guiding quantifications of multipartite entanglement, and informing the implementation of ubiquitous computational and communications strategies centered around the pair-wise distribution of entanglement through noisy channels.

Though the phenomenon of classical correlations delocalizing entanglement information is well known, its importance motivates the discussion below in the specific context of
entanglement between disjoint patches of the scalar field vacuum.   To further aid intuition, Appendix~\ref{app:CCdelocalization} provides a brief example in a finite-dimensional system with correlated coherent errors.

There exist local unitary operators that transform the entanglement structure of scalar field patches into a set of $(1_A \times 1_B)$ mode pairs with a natural exponential entanglement hierarchy.
Consider first a common symplectic operator (unitary in the Hilbert space) acting locally and symmetrically, $\mathbf{S}_W = \mathbf{S}^{(A)}_W \oplus \mathbf{S}^{(B)}_W $, transforming the volume-measured CM to local Williamson normal form~\cite{Williamson1936} (see Appendix~\ref{app:williamson} for a constructive description) in each patch,
\begin{equation}
  \mathbf{S}_{W} \boldsymbol{\sigma}^{\rm (m)} \mathbf{S}_{W}^T = \boldsymbol{\sigma}^{({\rm m})}_W  \ \ \  .
\end{equation}
Due to the complementarity of Schmidt decompositions in pure states, defining this transformation requires only access to the patch-local blocks of the CM, independent of any patch-patch correlation functions.
For bipartite states with globally pure CMs, this transformation produces a set of $(1_A \times 1_B)_{\rm pure}$ entangled pairs connecting modes with matching local symplectic eigenvalues~\cite{2004PhRvA..70e2329B}.
Modulo degeneracies in the symplectic spectrum, the transformed pure CM, $\boldsymbol{\sigma}^{({\rm m})}_W$, becomes block diagonalized to form a tensor product set of $d$ entangled pairs spanning the patches.
As shown by the black open-circle points in the Fig.~\ref{fig:scalarUnderlyintEntanglement} inset, the additivity of the logarithmic negativity subsequently allows independent two-mode entanglement calculations to fully account for the underlying patch-patch entanglement for the pure CM $\boldsymbol{\sigma}^{({\rm m})}$ in the local Williamson basis.

To describe this same field, now
from the perspective of local detectors, the classical correlations from volume tracing, $\mathbf{Y}$ in Eq.~\eqref{eq:sigmaplusYrelation}, must also be included to generate the observed $\boldsymbol{\sigma}^{\rm (t)}$ mixed state.
Interestingly, the bases that block diagonalize $ \boldsymbol{\sigma}^{\rm (m)}$ and $\mathbf{Y}$ are not mutually compatible, and thus the ensemble of displacements remains classically correlated after local Williamson transformation.
While it is expected that classical correlations may reduce access to quantum correlations (separable mixed states generated with entangled ensembles being an extreme example),  the structure of displacements in the local Williamson basis, $\mathbf{Y}_W = \mathbf{S}_{W} \mathbf{Y} \mathbf{S}_{W}^T$, also eliminates  the entanglement additivity originally present in $\boldsymbol{\sigma}^{\rm (m)}_W$ described above.
Though the addition of these classical correlations leaves the separability among the $(1_A \times 1_B)$ mode pairs unchanged, the complete entanglement structure can no longer be ascertained by independent two-mode analyses---rather, the entirety of the $(d \times d)$ quantum state must be considered.
Loss of additivity in the local Williamson basis, $\boldsymbol{\sigma}_W^{\rm (t)}$ (gray open-circle points in the Fig.~\ref{fig:scalarUnderlyintEntanglement} inset), shows that the classical noise of volume tracing can cause information relevant for entanglement extraction to be dispersed beyond the entangled mode pair structure shared by every ensemble of the mixture.

Interestingly, it is possible to design a consolidating transformation for the isolated scalar field patches that does not disperse the entanglement information,
but allows additivity of $(1_A \times 1_B)$ negativity
in the presence of classically correlated $\mathbf{Y}$-displacements~\cite{Klco:2021cxq}.  Consider a second transformation, $\mathbf{S}_{\mathcal{N}} =\mathbf{S}^{(A)}_{\mathcal{N}}  \oplus \mathbf{S}^{(B)}_{\mathcal{N}} $, derived from the partial transpose of the traced CM
\begin{equation}
 \mathbf{S}_{\mathcal{N}} \boldsymbol{\sigma}^{\rm (t)} \mathbf{S}_{\mathcal{N}}^T = \boldsymbol{\sigma}^{\rm (t)}_{\mathcal{N}}  \ \ \  .
\end{equation}
The transformed CM, $\boldsymbol{\sigma}_{\mathcal{N}}^{({\rm t})}$ (gray solid points in the Fig.~\ref{fig:scalarUnderlyintEntanglement} inset), is not block diagonal but has the form $\boldsymbol{\sigma}^{\rm (t)}_{\mathcal{N}} = \boldsymbol{\sigma}_1 \oplus \boldsymbol{\sigma}_2 \oplus \cdots \oplus \boldsymbol{\sigma}_{n_-} \oplus \boldsymbol{\sigma}_h + \mathbf{Y}'$, where $\boldsymbol{\sigma}_i$ is a two-mode CM of one site from each field patch and $n_-$ is the number of PT symplectic eigenvalues contributing to the logarithmic negativity.
As discussed in detail in Ref.~\cite{Klco:2021cxq}, calculation of such a transformation may be achieved for the free scalar field through manipulation of the PT symplectic eigenvectors of $\boldsymbol{\sigma}^{\rm (t)}$, incorporating information from both intra- and inter-patch correlators.
With this transformation, the entanglement in $\boldsymbol{\sigma}^{\rm (t)}$ is entirely accessible through two-mode structure, with the logarithmic negativity of each $\boldsymbol{\sigma}_i$ equal to that of the corresponding PT symplectic eigenvalue.

This section has emphasized how classical correlations can delocalize information necessary to access quantum entanglement---requiring the use of larger many-body interactions than warranted by the entanglement structure of any particular ensemble of the underlying convex decomposition.
Though it is possible to control the dispersion of this information, as evidenced by the existence of alternate local transformation $\mathbf{S}_{\mathcal{N}}$, such localized procedures rely upon the PT eigenvectors of the full patch-pair CM, $\boldsymbol{\sigma}^{({\rm t})}$, beyond those local to each region.
Classical correlations are capable not only of simplifying quantum correlations via entanglement reduction, but are also capable of increasing the complexity of their structure via information delocalization from the perspective of local observers.

\section{Discussion}

With the methodology depicted in Fig.~\ref{fig:protocolcircuit}, the calculations above have explicitly shown that the free scalar field vacuum in fact distributes entanglement in a form compatible with the two-point correlation functions.
For the locally accessible entanglement, the parametrically suppressed exponential decay with spatial separation~(Sec.~\ref{sec:distillabledecay}) is thus an indirect property of the field---a response to the classical correlations generated upon tracing of the patch-external volume.
Furthermore, that the latticized field distributes entanglement at all spacelike separations, yet experiences a separability transition at finite separation when observed locally~(Sec.~\ref{sec:rtildeNslash}), indicates a subtle role of the lattice's finite bandwidth---resulting in long-distance entanglement that is systematically more vulnerable to uncertainty in the external field configuration.
This work contributes toward derivation of the
deformation parameters, $\beta$ and $\gamma$, by analyzing the field response to local observation, clarifying the physical mechanisms key to their origins.

For the scalar field Hamiltonian,  Eq.~\eqref{eq:hamiltonian}, the spacelike entanglement in the vacuum is entirely produced by the field-space gradient operator.
As a result, the $\langle \hat{\phi} \hat{\phi}\rangle$ correlation functions, decaying logarthmically with separation in the massless regime, are larger than their $\langle \hat{\pi} \hat{\pi}\rangle$ counterparts that decay polynomially.
This intuition can be extended to further understand the hierarchy in distributed logarithmic negativity, shown in Fig.~\ref{fig:scalarUnderlyintEntanglement}, upon measurement of the volume in these two bases.
By transferring quantum correlations to classical ones that may be removed by classically-controlled unitaries, as described by Fig.~\ref{fig:protocolcircuit}, the process of volume measurement in the $\phi$($\pi$)-basis yields remaining quantum correlations between field patches that scale as the $\pi$($\phi$)-correlators.
That the $\phi$-basis measurement requires entanglement to be identified within a stronger set of classical correlations further motivates the hierarchy, $\mathcal{N}_{A|B}(\boldsymbol{\sigma}^{({\rm m}, \phi)}) < \mathcal{N}_{A|B}(\boldsymbol{\sigma}^{({\rm m}, \pi)})$.
Though no proof has been achieved, the combination of this intuition and additional explorations of mixed-basis volume measurements suggest that these two bases, $\phi$ and $\pi$, may provide the bounds minimizing and maximizing, respectively, the distributed entanglement in the underlying pure states available from the full scalar field vacuum.

Though the suppression of accessible entanglement for local detectors may appear unfortunate from perspectives of utilizing quantum fields as entanglement resources naturally distributed at spacelike separations, the observed impact of classical correlations yields a potential for advantage in a reverse direction for computation and simulation.
In particular, field observables relevant to experiments with spatially localized detectors might be \emph{calculated} with significantly less entanglement than they could be fully \emph{simulated}.
In other words, calculating the results of local measurements could be performed by evolving the mixed-state detection regions directly (with non-unitary quantum channels arising from globally unitary dynamics) rather than simulating the entirety of a field to be locally projected at the end.
By leveraging the non-uniqueness of convex decompositions leading to locally indistinguishable reduced density matrices, abstracted deviations from the original vision of quantum computation---directly emulating physical quantum systems with precisely controlled quantum devices---may allow significant reductions in the amount of entanglement that must be established throughout physical arrays of quantum hardware.

At the core of this challenge in pursuing computational advantage is the fundamental question of the bound entanglement naturally present between regions of the scalar field, i.e., the difference between the accessible entanglement and that required to prepare the mixed state density matrix.
Unfortunately, this somewhat unphysical but potentially computationally impactful quantity is, in general, computationally challenging to optimize~\cite{Hiesmayr2021}.
For example, the present work does not rule out the possibility
of mixed state preparation protocols independent of the scalar field vacuum, which may require less distributed entanglement to create the reduced density matrix of the field patches.
Though the calculations presented may provide guidance for determining the entanglement of formation for the isolated mixed quantum states of local field patches, the main role of this work has been in establishing, via direct analysis of a thought-experiment measurement protocol, an exponential separation between the entanglement distributed between patches of the scalar field vacuum and the amount of entanglement that can be extracted from those patches by local detectors.

Consideration of the infinite-volume whole of simple quantum fields has revealed fundamental features of vacuum entanglement structure that naturally result from the perspective of spatially local observation regions.
In addition to illuminating physical origins for basic properties of field entanglement,
the established protocol highlights the essential impact of measurement locality on the perceived structure of quantum correlations
and offers additional guidance on the path toward beneficial synthesis of quantum field simulation and experimental design.

\vspace{0.2cm}
\begin{acknowledgments}
We would like to thank Aidan Murran, and Martin~J.~Savage for valuable discussions.
We have made extensive use of Wolfram Mathematica~\cite{Mathematica}.  DB is supported in part by NSF Nuclear Physics grant PHY-2111046.  In beginning stages of thus work, NK was supported in part by the Walter Burke Institute for Theoretical Physics, and by the U.S. Department of Energy Office of Science, Office of Advanced Scientific Computing Research, (DE-SC0020290), and
Office of High Energy Physics DE-ACO2-07CH11359.
\end{acknowledgments}

\bibliography{biblio}

\begin{thebibliography}{83}%
\makeatletter
\providecommand \@ifxundefined [1]{%
 \@ifx{#1\undefined}
}%
\providecommand \@ifnum [1]{%
 \ifnum #1\expandafter \@firstoftwo
 \else \expandafter \@secondoftwo
 \fi
}%
\providecommand \@ifx [1]{%
 \ifx #1\expandafter \@firstoftwo
 \else \expandafter \@secondoftwo
 \fi
}%
\providecommand \natexlab [1]{#1}%
\providecommand \enquote  [1]{``#1''}%
\providecommand \bibnamefont  [1]{#1}%
\providecommand \bibfnamefont [1]{#1}%
\providecommand \citenamefont [1]{#1}%
\providecommand \href@noop [0]{\@secondoftwo}%
\providecommand \href [0]{\begingroup \@sanitize@url \@href}%
\providecommand \@href[1]{\@@startlink{#1}\@@href}%
\providecommand \@@href[1]{\endgroup#1\@@endlink}%
\providecommand \@sanitize@url [0]{\catcode `\\12\catcode `\$12\catcode
  `\&12\catcode `\#12\catcode `\^12\catcode `\_12\catcode `\%12\relax}%
\providecommand \@@startlink[1]{}%
\providecommand \@@endlink[0]{}%
\providecommand \url  [0]{\begingroup\@sanitize@url \@url }%
\providecommand \@url [1]{\endgroup\@href {#1}{\urlprefix }}%
\providecommand \urlprefix  [0]{URL }%
\providecommand \Eprint [0]{\href }%
\providecommand \doibase [0]{https://doi.org/}%
\providecommand \selectlanguage [0]{\@gobble}%
\providecommand \bibinfo  [0]{\@secondoftwo}%
\providecommand \bibfield  [0]{\@secondoftwo}%
\providecommand \translation [1]{[#1]}%
\providecommand \BibitemOpen [0]{}%
\providecommand \bibitemStop [0]{}%
\providecommand \bibitemNoStop [0]{.\EOS\space}%
\providecommand \EOS [0]{\spacefactor3000\relax}%
\providecommand \BibitemShut  [1]{\csname bibitem#1\endcsname}%
\let\auto@bib@innerbib\@empty
\bibitem [{\citenamefont {Feynman}(1982)}]{Feynman1982}%
  \BibitemOpen
  \bibfield  {author} {\bibinfo {author} {\bibfnamefont {R.~P.}\ \bibnamefont
  {Feynman}},\ }\bibfield  {title} {\bibinfo {title} {Simulating physics with
  computers},\ }\href {https://doi.org/10.1007/BF02650179} {\bibfield
  {journal} {\bibinfo  {journal} {International Journal of Theoretical
  Physics}\ }\textbf {\bibinfo {volume} {21}},\ \bibinfo {pages} {467}
  (\bibinfo {year} {1982})}\BibitemShut {NoStop}%
\bibitem [{\citenamefont {Ba\~nuls}\ \emph {et~al.}(2020)\citenamefont
  {Ba\~nuls} \emph {et~al.}}]{Banuls:2019bmf}%
  \BibitemOpen
  \bibfield  {author} {\bibinfo {author} {\bibfnamefont {M.~C.}\ \bibnamefont
  {Ba\~nuls}} \emph {et~al.},\ }\bibfield  {title} {\bibinfo {title}
  {{Simulating Lattice Gauge Theories within Quantum Technologies}},\ }\href
  {https://doi.org/10.1140/epjd/e2020-100571-8} {\bibfield  {journal} {\bibinfo
   {journal} {Eur. Phys. J. D}\ }\textbf {\bibinfo {volume} {74}},\ \bibinfo
  {pages} {165} (\bibinfo {year} {2020})},\ \Eprint
  {https://arxiv.org/abs/1911.00003} {arXiv:1911.00003 [quant-ph]} \BibitemShut
  {NoStop}%
\bibitem [{\citenamefont {Klco}\ \emph {et~al.}(2022)\citenamefont {Klco},
  \citenamefont {Roggero},\ and\ \citenamefont {Savage}}]{Klco:2021lap}%
  \BibitemOpen
  \bibfield  {author} {\bibinfo {author} {\bibfnamefont {N.}~\bibnamefont
  {Klco}}, \bibinfo {author} {\bibfnamefont {A.}~\bibnamefont {Roggero}},\ and\
  \bibinfo {author} {\bibfnamefont {M.~J.}\ \bibnamefont {Savage}},\ }\bibfield
   {title} {\bibinfo {title} {{Standard model physics and the digital quantum
  revolution: thoughts about the interface}},\ }\href
  {https://doi.org/10.1088/1361-6633/ac58a4} {\bibfield  {journal} {\bibinfo
  {journal} {Rept. Prog. Phys.}\ }\textbf {\bibinfo {volume} {85}},\ \bibinfo
  {pages} {064301} (\bibinfo {year} {2022})},\ \Eprint
  {https://arxiv.org/abs/2107.04769} {arXiv:2107.04769 [quant-ph]} \BibitemShut
  {NoStop}%
\bibitem [{\citenamefont {Bauer}\ \emph {et~al.}(2022)\citenamefont {Bauer},
  \citenamefont {Davoudi}, \citenamefont {Balantekin}, \citenamefont
  {Bhattacharya}, \citenamefont {Carena}, \citenamefont {de~Jong},
  \citenamefont {Draper}, \citenamefont {El-Khadra}, \citenamefont {Gemelke},
  \citenamefont {Hanada}, \citenamefont {Kharzeev}, \citenamefont {Lamm},
  \citenamefont {Li}, \citenamefont {Liu}, \citenamefont {Lukin}, \citenamefont
  {Meurice}, \citenamefont {Monroe}, \citenamefont {Nachman}, \citenamefont
  {Pagano}, \citenamefont {Preskill}, \citenamefont {Rinaldi}, \citenamefont
  {Roggero}, \citenamefont {Santiago}, \citenamefont {Savage}, \citenamefont
  {Siddiqi}, \citenamefont {Siopsis}, \citenamefont {Van~Zanten}, \citenamefont
  {Wiebe}, \citenamefont {Yamauchi}, \citenamefont {Yeter-Aydeniz},\ and\
  \citenamefont {Zorzetti}}]{Bauer:2022hpo}%
  \BibitemOpen
  \bibfield  {author} {\bibinfo {author} {\bibfnamefont {C.~W.}\ \bibnamefont
  {Bauer}}, \bibinfo {author} {\bibfnamefont {Z.}~\bibnamefont {Davoudi}},
  \bibinfo {author} {\bibfnamefont {A.~B.}\ \bibnamefont {Balantekin}},
  \bibinfo {author} {\bibfnamefont {T.}~\bibnamefont {Bhattacharya}}, \bibinfo
  {author} {\bibfnamefont {M.}~\bibnamefont {Carena}}, \bibinfo {author}
  {\bibfnamefont {W.~A.}\ \bibnamefont {de~Jong}}, \bibinfo {author}
  {\bibfnamefont {P.}~\bibnamefont {Draper}}, \bibinfo {author} {\bibfnamefont
  {A.}~\bibnamefont {El-Khadra}}, \bibinfo {author} {\bibfnamefont
  {N.}~\bibnamefont {Gemelke}}, \bibinfo {author} {\bibfnamefont
  {M.}~\bibnamefont {Hanada}}, \bibinfo {author} {\bibfnamefont
  {D.}~\bibnamefont {Kharzeev}}, \bibinfo {author} {\bibfnamefont
  {H.}~\bibnamefont {Lamm}}, \bibinfo {author} {\bibfnamefont {Y.-Y.}\
  \bibnamefont {Li}}, \bibinfo {author} {\bibfnamefont {J.}~\bibnamefont
  {Liu}}, \bibinfo {author} {\bibfnamefont {M.}~\bibnamefont {Lukin}}, \bibinfo
  {author} {\bibfnamefont {Y.}~\bibnamefont {Meurice}}, \bibinfo {author}
  {\bibfnamefont {C.}~\bibnamefont {Monroe}}, \bibinfo {author} {\bibfnamefont
  {B.}~\bibnamefont {Nachman}}, \bibinfo {author} {\bibfnamefont
  {G.}~\bibnamefont {Pagano}}, \bibinfo {author} {\bibfnamefont
  {J.}~\bibnamefont {Preskill}}, \bibinfo {author} {\bibfnamefont
  {E.}~\bibnamefont {Rinaldi}}, \bibinfo {author} {\bibfnamefont
  {A.}~\bibnamefont {Roggero}}, \bibinfo {author} {\bibfnamefont {D.~I.}\
  \bibnamefont {Santiago}}, \bibinfo {author} {\bibfnamefont {M.~J.}\
  \bibnamefont {Savage}}, \bibinfo {author} {\bibfnamefont {I.}~\bibnamefont
  {Siddiqi}}, \bibinfo {author} {\bibfnamefont {G.}~\bibnamefont {Siopsis}},
  \bibinfo {author} {\bibfnamefont {D.}~\bibnamefont {Van~Zanten}}, \bibinfo
  {author} {\bibfnamefont {N.}~\bibnamefont {Wiebe}}, \bibinfo {author}
  {\bibfnamefont {Y.}~\bibnamefont {Yamauchi}}, \bibinfo {author}
  {\bibfnamefont {K.}~\bibnamefont {Yeter-Aydeniz}},\ and\ \bibinfo {author}
  {\bibfnamefont {S.}~\bibnamefont {Zorzetti}},\ }\href@noop {} {\bibinfo
  {title} {{Quantum Simulation for High Energy Physics}}} (\bibinfo {year}
  {2022}),\ \Eprint {https://arxiv.org/abs/2204.03381} {arXiv:2204.03381
  [quant-ph]} \BibitemShut {NoStop}%
\bibitem [{\citenamefont {{Bennett}}\ and\ \citenamefont
  {{Brassard}}(1984)}]{bennet1984quantum}%
  \BibitemOpen
  \bibfield  {author} {\bibinfo {author} {\bibfnamefont {C.~H.}\ \bibnamefont
  {{Bennett}}}\ and\ \bibinfo {author} {\bibfnamefont {G.}~\bibnamefont
  {{Brassard}}},\ }\bibfield  {title} {\bibinfo {title} {{Quantum cryptography:
  Public key distribution and coin tossing}},\ }in\ \href
  {https://doi.org/10.48550/arXiv.2003.06557} {\emph {\bibinfo {booktitle}
  {Proceedings of the IEEE International Conference on Computers, Systems, and
  Signal Processing, Bangalore, Dec. 1984}}}\ (\bibinfo {year} {1984})\ pp.\
  \bibinfo {pages} {175--179},\ \Eprint {https://arxiv.org/abs/2003.06557}
  {arXiv:2003.06557 [quant-ph]} \BibitemShut {NoStop}%
\bibitem [{\citenamefont {Ekert}(1991)}]{PhysRevLett.67.661}%
  \BibitemOpen
  \bibfield  {author} {\bibinfo {author} {\bibfnamefont {A.~K.}\ \bibnamefont
  {Ekert}},\ }\bibfield  {title} {\bibinfo {title} {{Quantum cryptography based
  on Bell's theorem}},\ }\href {https://doi.org/10.1103/PhysRevLett.67.661}
  {\bibfield  {journal} {\bibinfo  {journal} {Phys. Rev. Lett.}\ }\textbf
  {\bibinfo {volume} {67}},\ \bibinfo {pages} {661} (\bibinfo {year}
  {1991})}\BibitemShut {NoStop}%
\bibitem [{\citenamefont {{Degen}}\ \emph {et~al.}(2017)\citenamefont
  {{Degen}}, \citenamefont {{Reinhard}},\ and\ \citenamefont
  {{Cappellaro}}}]{2017RvMP...89c5002D}%
  \BibitemOpen
  \bibfield  {author} {\bibinfo {author} {\bibfnamefont {C.~L.}\ \bibnamefont
  {{Degen}}}, \bibinfo {author} {\bibfnamefont {F.}~\bibnamefont
  {{Reinhard}}},\ and\ \bibinfo {author} {\bibfnamefont {P.}~\bibnamefont
  {{Cappellaro}}},\ }\bibfield  {title} {\bibinfo {title} {{Quantum sensing}},\
  }\href {https://doi.org/10.1103/RevModPhys.89.035002} {\bibfield  {journal}
  {\bibinfo  {journal} {Reviews of Modern Physics}\ }\textbf {\bibinfo {volume}
  {89}},\ \bibinfo {eid} {035002} (\bibinfo {year} {2017})},\ \Eprint
  {https://arxiv.org/abs/1611.02427} {arXiv:1611.02427 [quant-ph]} \BibitemShut
  {NoStop}%
\bibitem [{\citenamefont {Kharzeev}\ and\ \citenamefont
  {Levin}(2017)}]{Kharzeev:2017qzs}%
  \BibitemOpen
  \bibfield  {author} {\bibinfo {author} {\bibfnamefont {D.~E.}\ \bibnamefont
  {Kharzeev}}\ and\ \bibinfo {author} {\bibfnamefont {E.~M.}\ \bibnamefont
  {Levin}},\ }\bibfield  {title} {\bibinfo {title} {{Deep inelastic scattering
  as a probe of entanglement}},\ }\href
  {https://doi.org/10.1103/PhysRevD.95.114008} {\bibfield  {journal} {\bibinfo
  {journal} {Phys. Rev. D}\ }\textbf {\bibinfo {volume} {95}},\ \bibinfo
  {pages} {114008} (\bibinfo {year} {2017})},\ \Eprint
  {https://arxiv.org/abs/1702.03489} {arXiv:1702.03489 [hep-ph]} \BibitemShut
  {NoStop}%
\bibitem [{\citenamefont {Baker}\ and\ \citenamefont
  {Kharzeev}(2018)}]{Baker:2017wtt}%
  \BibitemOpen
  \bibfield  {author} {\bibinfo {author} {\bibfnamefont {O.~K.}\ \bibnamefont
  {Baker}}\ and\ \bibinfo {author} {\bibfnamefont {D.~E.}\ \bibnamefont
  {Kharzeev}},\ }\bibfield  {title} {\bibinfo {title} {{Thermal radiation and
  entanglement in proton-proton collisions at energies available at the CERN
  Large Hadron Collider}},\ }\href {https://doi.org/10.1103/PhysRevD.98.054007}
  {\bibfield  {journal} {\bibinfo  {journal} {Phys. Rev. D}\ }\textbf {\bibinfo
  {volume} {98}},\ \bibinfo {pages} {054007} (\bibinfo {year} {2018})},\
  \Eprint {https://arxiv.org/abs/1712.04558} {arXiv:1712.04558 [hep-ph]}
  \BibitemShut {NoStop}%
\bibitem [{\citenamefont {Cervera-Lierta}\ \emph {et~al.}(2017)\citenamefont
  {Cervera-Lierta}, \citenamefont {Latorre}, \citenamefont {Rojo},\ and\
  \citenamefont {Rottoli}}]{Cervera-Lierta:2017tdt}%
  \BibitemOpen
  \bibfield  {author} {\bibinfo {author} {\bibfnamefont {A.}~\bibnamefont
  {Cervera-Lierta}}, \bibinfo {author} {\bibfnamefont {J.~I.}\ \bibnamefont
  {Latorre}}, \bibinfo {author} {\bibfnamefont {J.}~\bibnamefont {Rojo}},\ and\
  \bibinfo {author} {\bibfnamefont {L.}~\bibnamefont {Rottoli}},\ }\bibfield
  {title} {\bibinfo {title} {{Maximal Entanglement in High Energy Physics}},\
  }\href {https://doi.org/10.21468/SciPostPhys.3.5.036} {\bibfield  {journal}
  {\bibinfo  {journal} {SciPost Phys.}\ }\textbf {\bibinfo {volume} {3}},\
  \bibinfo {pages} {036} (\bibinfo {year} {2017})},\ \Eprint
  {https://arxiv.org/abs/1703.02989} {arXiv:1703.02989 [hep-th]} \BibitemShut
  {NoStop}%
\bibitem [{\citenamefont {Berges}\ \emph {et~al.}(2019)\citenamefont {Berges},
  \citenamefont {Floerchinger},\ and\ \citenamefont
  {Venugopalan}}]{Berges:2018cny}%
  \BibitemOpen
  \bibfield  {author} {\bibinfo {author} {\bibfnamefont {J.}~\bibnamefont
  {Berges}}, \bibinfo {author} {\bibfnamefont {S.}~\bibnamefont
  {Floerchinger}},\ and\ \bibinfo {author} {\bibfnamefont {R.}~\bibnamefont
  {Venugopalan}},\ }\bibfield  {title} {\bibinfo {title} {{Entanglement and
  thermalization}},\ }\href {https://doi.org/10.1016/j.nuclphysa.2018.12.008}
  {\bibfield  {journal} {\bibinfo  {journal} {Nucl. Phys. A}\ }\textbf
  {\bibinfo {volume} {982}},\ \bibinfo {pages} {819} (\bibinfo {year}
  {2019})},\ \Eprint {https://arxiv.org/abs/1812.08120} {arXiv:1812.08120
  [hep-th]} \BibitemShut {NoStop}%
\bibitem [{\citenamefont {Beane}\ \emph {et~al.}(2019)\citenamefont {Beane},
  \citenamefont {Kaplan}, \citenamefont {Klco},\ and\ \citenamefont
  {Savage}}]{Beane:2018oxh}%
  \BibitemOpen
  \bibfield  {author} {\bibinfo {author} {\bibfnamefont {S.~R.}\ \bibnamefont
  {Beane}}, \bibinfo {author} {\bibfnamefont {D.~B.}\ \bibnamefont {Kaplan}},
  \bibinfo {author} {\bibfnamefont {N.}~\bibnamefont {Klco}},\ and\ \bibinfo
  {author} {\bibfnamefont {M.~J.}\ \bibnamefont {Savage}},\ }\bibfield  {title}
  {\bibinfo {title} {{Entanglement Suppression and Emergent Symmetries of
  Strong Interactions}},\ }\href
  {https://doi.org/10.1103/PhysRevLett.122.102001} {\bibfield  {journal}
  {\bibinfo  {journal} {Phys. Rev. Lett.}\ }\textbf {\bibinfo {volume} {122}},\
  \bibinfo {pages} {102001} (\bibinfo {year} {2019})},\ \Eprint
  {https://arxiv.org/abs/1812.03138} {arXiv:1812.03138 [nucl-th]} \BibitemShut
  {NoStop}%
\bibitem [{\citenamefont {Gorton}\ and\ \citenamefont
  {Johnson}(2019)}]{GortonJohnson2019a}%
  \BibitemOpen
  \bibfield  {author} {\bibinfo {author} {\bibfnamefont {O.}~\bibnamefont
  {Gorton}}\ and\ \bibinfo {author} {\bibfnamefont {C.~W.}\ \bibnamefont
  {Johnson}},\ }\bibfield  {title} {\bibinfo {title} {Entanglement entropy and
  proton-neutron interactions}} (\bibinfo {year} {2019}),\ \bibinfo {note}
  {{ESNT} workshop on proton-neutron pairing,
  http://esnt.cea.fr/Phocea/Page/index.php?id=84}\BibitemShut {NoStop}%
\bibitem [{\citenamefont {Beane}\ and\ \citenamefont
  {Ehlers}(2019)}]{Beane:2019loz}%
  \BibitemOpen
  \bibfield  {author} {\bibinfo {author} {\bibfnamefont {S.~R.}\ \bibnamefont
  {Beane}}\ and\ \bibinfo {author} {\bibfnamefont {P.}~\bibnamefont {Ehlers}},\
  }\bibfield  {title} {\bibinfo {title} {{Chiral symmetry breaking,
  entanglement, and the nucleon spin decomposition}},\ }\href
  {https://doi.org/10.1142/S0217732320500480} {\bibfield  {journal} {\bibinfo
  {journal} {Mod. Phys. Lett. A}\ }\textbf {\bibinfo {volume} {35}},\ \bibinfo
  {pages} {2050048} (\bibinfo {year} {2019})},\ \Eprint
  {https://arxiv.org/abs/1905.03295} {arXiv:1905.03295 [hep-ph]} \BibitemShut
  {NoStop}%
\bibitem [{\citenamefont {Tu}\ \emph {et~al.}(2020)\citenamefont {Tu},
  \citenamefont {Kharzeev},\ and\ \citenamefont {Ullrich}}]{Tu:2019ouv}%
  \BibitemOpen
  \bibfield  {author} {\bibinfo {author} {\bibfnamefont {Z.}~\bibnamefont
  {Tu}}, \bibinfo {author} {\bibfnamefont {D.~E.}\ \bibnamefont {Kharzeev}},\
  and\ \bibinfo {author} {\bibfnamefont {T.}~\bibnamefont {Ullrich}},\
  }\bibfield  {title} {\bibinfo {title} {{Einstein-Podolsky-Rosen Paradox and
  Quantum Entanglement at Subnucleonic Scales}},\ }\href
  {https://doi.org/10.1103/PhysRevLett.124.062001} {\bibfield  {journal}
  {\bibinfo  {journal} {Phys. Rev. Lett.}\ }\textbf {\bibinfo {volume} {124}},\
  \bibinfo {pages} {062001} (\bibinfo {year} {2020})},\ \Eprint
  {https://arxiv.org/abs/1904.11974} {arXiv:1904.11974 [hep-ph]} \BibitemShut
  {NoStop}%
\bibitem [{\citenamefont {Beane}\ and\ \citenamefont
  {Farrell}(2021)}]{Beane:2020wjl}%
  \BibitemOpen
  \bibfield  {author} {\bibinfo {author} {\bibfnamefont {S.~R.}\ \bibnamefont
  {Beane}}\ and\ \bibinfo {author} {\bibfnamefont {R.~C.}\ \bibnamefont
  {Farrell}},\ }\bibfield  {title} {\bibinfo {title} {{Geometry and
  entanglement in the scattering matrix}},\ }\href
  {https://doi.org/10.1016/j.aop.2021.168581} {\bibfield  {journal} {\bibinfo
  {journal} {Annals Phys.}\ }\textbf {\bibinfo {volume} {433}},\ \bibinfo
  {pages} {168581} (\bibinfo {year} {2021})},\ \Eprint
  {https://arxiv.org/abs/2011.01278} {arXiv:2011.01278 [hep-th]} \BibitemShut
  {NoStop}%
\bibitem [{\citenamefont {Beane}\ \emph {et~al.}(2021)\citenamefont {Beane},
  \citenamefont {Farrell},\ and\ \citenamefont {Varma}}]{Beane:2021zvo}%
  \BibitemOpen
  \bibfield  {author} {\bibinfo {author} {\bibfnamefont {S.~R.}\ \bibnamefont
  {Beane}}, \bibinfo {author} {\bibfnamefont {R.~C.}\ \bibnamefont {Farrell}},\
  and\ \bibinfo {author} {\bibfnamefont {M.}~\bibnamefont {Varma}},\ }\bibfield
   {title} {\bibinfo {title} {{Entanglement minimization in hadronic scattering
  with pions}},\ }\href {https://doi.org/10.1142/S0217751X21502055} {\bibfield
  {journal} {\bibinfo  {journal} {Int. J. Mod. Phys. A}\ }\textbf {\bibinfo
  {volume} {36}},\ \bibinfo {pages} {2150205} (\bibinfo {year} {2021})},\
  \Eprint {https://arxiv.org/abs/2108.00646} {arXiv:2108.00646 [hep-ph]}
  \BibitemShut {NoStop}%
\bibitem [{\citenamefont {Kharzeev}\ and\ \citenamefont
  {Levin}(2021)}]{Kharzeev:2021yyf}%
  \BibitemOpen
  \bibfield  {author} {\bibinfo {author} {\bibfnamefont {D.~E.}\ \bibnamefont
  {Kharzeev}}\ and\ \bibinfo {author} {\bibfnamefont {E.}~\bibnamefont
  {Levin}},\ }\bibfield  {title} {\bibinfo {title} {{Deep inelastic scattering
  as a probe of entanglement: Confronting experimental data}},\ }\href
  {https://doi.org/10.1103/PhysRevD.104.L031503} {\bibfield  {journal}
  {\bibinfo  {journal} {Phys. Rev. D}\ }\textbf {\bibinfo {volume} {104}},\
  \bibinfo {pages} {L031503} (\bibinfo {year} {2021})},\ \Eprint
  {https://arxiv.org/abs/2102.09773} {arXiv:2102.09773 [hep-ph]} \BibitemShut
  {NoStop}%
\bibitem [{\citenamefont {Robin}\ \emph {et~al.}(2021)\citenamefont {Robin},
  \citenamefont {Savage},\ and\ \citenamefont {Pillet}}]{Robin:2020aeh}%
  \BibitemOpen
  \bibfield  {author} {\bibinfo {author} {\bibfnamefont {C.}~\bibnamefont
  {Robin}}, \bibinfo {author} {\bibfnamefont {M.~J.}\ \bibnamefont {Savage}},\
  and\ \bibinfo {author} {\bibfnamefont {N.}~\bibnamefont {Pillet}},\
  }\bibfield  {title} {\bibinfo {title} {{Entanglement Rearrangement in
  Self-Consistent Nuclear Structure Calculations}},\ }\href
  {https://doi.org/10.1103/PhysRevC.103.034325} {\bibfield  {journal} {\bibinfo
   {journal} {Phys. Rev. C}\ }\textbf {\bibinfo {volume} {103}},\ \bibinfo
  {pages} {034325} (\bibinfo {year} {2021})},\ \Eprint
  {https://arxiv.org/abs/2007.09157} {arXiv:2007.09157 [nucl-th]} \BibitemShut
  {NoStop}%
\bibitem [{\citenamefont {Low}\ and\ \citenamefont
  {Mehen}(2021)}]{Low:2021ufv}%
  \BibitemOpen
  \bibfield  {author} {\bibinfo {author} {\bibfnamefont {I.}~\bibnamefont
  {Low}}\ and\ \bibinfo {author} {\bibfnamefont {T.}~\bibnamefont {Mehen}},\
  }\bibfield  {title} {\bibinfo {title} {{Symmetry from entanglement
  suppression}},\ }\href {https://doi.org/10.1103/PhysRevD.104.074014}
  {\bibfield  {journal} {\bibinfo  {journal} {Phys. Rev. D}\ }\textbf {\bibinfo
  {volume} {104}},\ \bibinfo {pages} {074014} (\bibinfo {year} {2021})},\
  \Eprint {https://arxiv.org/abs/2104.10835} {arXiv:2104.10835 [hep-th]}
  \BibitemShut {NoStop}%
\bibitem [{\citenamefont {Gong}\ \emph {et~al.}(2022)\citenamefont {Gong},
  \citenamefont {Parida}, \citenamefont {Tu},\ and\ \citenamefont
  {Venugopalan}}]{Gong:2021bcp}%
  \BibitemOpen
  \bibfield  {author} {\bibinfo {author} {\bibfnamefont {W.}~\bibnamefont
  {Gong}}, \bibinfo {author} {\bibfnamefont {G.}~\bibnamefont {Parida}},
  \bibinfo {author} {\bibfnamefont {Z.}~\bibnamefont {Tu}},\ and\ \bibinfo
  {author} {\bibfnamefont {R.}~\bibnamefont {Venugopalan}},\ }\bibfield
  {title} {\bibinfo {title} {{Measurement of Bell-type inequalities and quantum
  entanglement from \ensuremath{\Lambda}-hyperon spin correlations at high
  energy colliders}},\ }\href {https://doi.org/10.1103/PhysRevD.106.L031501}
  {\bibfield  {journal} {\bibinfo  {journal} {Phys. Rev. D}\ }\textbf {\bibinfo
  {volume} {106}},\ \bibinfo {pages} {L031501} (\bibinfo {year} {2022})},\
  \Eprint {https://arxiv.org/abs/2107.13007} {arXiv:2107.13007 [hep-ph]}
  \BibitemShut {NoStop}%
\bibitem [{\citenamefont {Roggero}(2021)}]{Roggero:2021asb}%
  \BibitemOpen
  \bibfield  {author} {\bibinfo {author} {\bibfnamefont {A.}~\bibnamefont
  {Roggero}},\ }\bibfield  {title} {\bibinfo {title} {{Entanglement and
  many-body effects in collective neutrino oscillations}},\ }\href
  {https://doi.org/10.1103/PhysRevD.104.103016} {\bibfield  {journal} {\bibinfo
   {journal} {Phys. Rev. D}\ }\textbf {\bibinfo {volume} {104}},\ \bibinfo
  {pages} {103016} (\bibinfo {year} {2021})},\ \Eprint
  {https://arxiv.org/abs/2102.10188} {arXiv:2102.10188 [hep-ph]} \BibitemShut
  {NoStop}%
\bibitem [{\citenamefont {Mueller}\ \emph {et~al.}(2022)\citenamefont
  {Mueller}, \citenamefont {Zache},\ and\ \citenamefont
  {Ott}}]{Mueller:2021gxd}%
  \BibitemOpen
  \bibfield  {author} {\bibinfo {author} {\bibfnamefont {N.}~\bibnamefont
  {Mueller}}, \bibinfo {author} {\bibfnamefont {T.~V.}\ \bibnamefont {Zache}},\
  and\ \bibinfo {author} {\bibfnamefont {R.}~\bibnamefont {Ott}},\ }\bibfield
  {title} {\bibinfo {title} {{Thermalization of Gauge Theories from their
  Entanglement Spectrum}},\ }\href
  {https://doi.org/10.1103/PhysRevLett.129.011601} {\bibfield  {journal}
  {\bibinfo  {journal} {Phys. Rev. Lett.}\ }\textbf {\bibinfo {volume} {129}},\
  \bibinfo {pages} {011601} (\bibinfo {year} {2022})},\ \Eprint
  {https://arxiv.org/abs/2107.11416} {arXiv:2107.11416 [quant-ph]} \BibitemShut
  {NoStop}%
\bibitem [{\citenamefont {Liu}\ \emph {et~al.}(2023)\citenamefont {Liu},
  \citenamefont {Low},\ and\ \citenamefont {Mehen}}]{Liu:2022grf}%
  \BibitemOpen
  \bibfield  {author} {\bibinfo {author} {\bibfnamefont {Q.}~\bibnamefont
  {Liu}}, \bibinfo {author} {\bibfnamefont {I.}~\bibnamefont {Low}},\ and\
  \bibinfo {author} {\bibfnamefont {T.}~\bibnamefont {Mehen}},\ }\bibfield
  {title} {\bibinfo {title} {{Minimal entanglement and emergent symmetries in
  low-energy QCD}},\ }\href {https://doi.org/10.1103/PhysRevC.107.025204}
  {\bibfield  {journal} {\bibinfo  {journal} {Phys. Rev. C}\ }\textbf {\bibinfo
  {volume} {107}},\ \bibinfo {pages} {025204} (\bibinfo {year} {2023})},\
  \Eprint {https://arxiv.org/abs/2210.12085} {arXiv:2210.12085 [quant-ph]}
  \BibitemShut {NoStop}%
\bibitem [{\citenamefont {Altman}\ \emph {et~al.}(2021)\citenamefont {Altman}
  \emph {et~al.}}]{Altman:2019vbv}%
  \BibitemOpen
  \bibfield  {author} {\bibinfo {author} {\bibfnamefont {E.}~\bibnamefont
  {Altman}} \emph {et~al.},\ }\bibfield  {title} {\bibinfo {title} {{Quantum
  Simulators: Architectures and Opportunities}},\ }\href
  {https://doi.org/10.1103/PRXQuantum.2.017003} {\bibfield  {journal} {\bibinfo
   {journal} {PRX Quantum}\ }\textbf {\bibinfo {volume} {2}},\ \bibinfo {pages}
  {017003} (\bibinfo {year} {2021})},\ \Eprint
  {https://arxiv.org/abs/1912.06938} {arXiv:1912.06938 [quant-ph]} \BibitemShut
  {NoStop}%
\bibitem [{\citenamefont {Reeh}\ and\ \citenamefont
  {Schlieder}(1961)}]{Reeh1961}%
  \BibitemOpen
  \bibfield  {author} {\bibinfo {author} {\bibfnamefont {H.}~\bibnamefont
  {Reeh}}\ and\ \bibinfo {author} {\bibfnamefont {S.}~\bibnamefont
  {Schlieder}},\ }\bibfield  {title} {\bibinfo {title} {Bemerkungen zur
  unit{\"a}r{\"a}quivalenz von lorentzinvarianten feldern},\ }\href
  {https://doi.org/10.1007/BF02787889} {\bibfield  {journal} {\bibinfo
  {journal} {Il Nuovo Cimento (1955-1965)}\ }\textbf {\bibinfo {volume} {22}},\
  \bibinfo {pages} {1051} (\bibinfo {year} {1961})}\BibitemShut {NoStop}%
\bibitem [{\citenamefont {Summers}\ and\ \citenamefont
  {Werner}(1985)}]{summers1985vacuum}%
  \BibitemOpen
  \bibfield  {author} {\bibinfo {author} {\bibfnamefont {S.~J.}\ \bibnamefont
  {Summers}}\ and\ \bibinfo {author} {\bibfnamefont {R.}~\bibnamefont
  {Werner}},\ }\bibfield  {title} {\bibinfo {title} {The vacuum violates bell's
  inequalities},\ }\href
  {https://doi.org/https://doi.org/10.1016/0375-9601(85)90093-3} {\bibfield
  {journal} {\bibinfo  {journal} {Physics Letters A}\ }\textbf {\bibinfo
  {volume} {110}},\ \bibinfo {pages} {257} (\bibinfo {year}
  {1985})}\BibitemShut {NoStop}%
\bibitem [{\citenamefont {Summers}\ and\ \citenamefont
  {Werner}(1987{\natexlab{a}})}]{summers1987bell1}%
  \BibitemOpen
  \bibfield  {author} {\bibinfo {author} {\bibfnamefont {S.~J.}\ \bibnamefont
  {Summers}}\ and\ \bibinfo {author} {\bibfnamefont {R.}~\bibnamefont
  {Werner}},\ }\bibfield  {title} {\bibinfo {title} {Bell's inequalities and
  quantum field theory. i. general setting},\ }\href
  {https://doi.org/10.1063/1.527733} {\bibfield  {journal} {\bibinfo  {journal}
  {Journal of Mathematical Physics}\ }\textbf {\bibinfo {volume} {28}},\
  \bibinfo {pages} {2440} (\bibinfo {year} {1987}{\natexlab{a}})},\ \Eprint
  {https://arxiv.org/abs/https://doi.org/10.1063/1.527733}
  {https://doi.org/10.1063/1.527733} \BibitemShut {NoStop}%
\bibitem [{\citenamefont {Summers}\ and\ \citenamefont
  {Werner}(1987{\natexlab{b}})}]{summers1987bell2}%
  \BibitemOpen
  \bibfield  {author} {\bibinfo {author} {\bibfnamefont {S.~J.}\ \bibnamefont
  {Summers}}\ and\ \bibinfo {author} {\bibfnamefont {R.}~\bibnamefont
  {Werner}},\ }\bibfield  {title} {\bibinfo {title} {Bell's inequalities and
  quantum field theory. ii. bell's inequalities are maximally violated in the
  vacuum},\ }\href {https://doi.org/10.1063/1.527734} {\bibfield  {journal}
  {\bibinfo  {journal} {Journal of Mathematical Physics}\ }\textbf {\bibinfo
  {volume} {28}},\ \bibinfo {pages} {2448} (\bibinfo {year}
  {1987}{\natexlab{b}})},\ \Eprint
  {https://arxiv.org/abs/https://doi.org/10.1063/1.527734}
  {https://doi.org/10.1063/1.527734} \BibitemShut {NoStop}%
\bibitem [{\citenamefont {Halvorson}\ and\ \citenamefont
  {Clifton}(2000)}]{Halvorson:1999pz}%
  \BibitemOpen
  \bibfield  {author} {\bibinfo {author} {\bibfnamefont {H.}~\bibnamefont
  {Halvorson}}\ and\ \bibinfo {author} {\bibfnamefont {R.}~\bibnamefont
  {Clifton}},\ }\bibfield  {title} {\bibinfo {title} {{Generic Bell correlation
  between arbitrary local algebras in quantum field theory}},\ }\href
  {https://doi.org/10.1063/1.533253} {\bibfield  {journal} {\bibinfo  {journal}
  {J. Math. Phys.}\ }\textbf {\bibinfo {volume} {41}},\ \bibinfo {pages} {1711}
  (\bibinfo {year} {2000})},\ \Eprint {https://arxiv.org/abs/math-ph/9909013}
  {arXiv:math-ph/9909013} \BibitemShut {NoStop}%
\bibitem [{\citenamefont {Witten}(2018)}]{Witten:2018zxz}%
  \BibitemOpen
  \bibfield  {author} {\bibinfo {author} {\bibfnamefont {E.}~\bibnamefont
  {Witten}},\ }\bibfield  {title} {\bibinfo {title} {{APS Medal for Exceptional
  Achievement in Research: Invited article on entanglement properties of
  quantum field theory}},\ }\href
  {https://doi.org/10.1103/RevModPhys.90.045003} {\bibfield  {journal}
  {\bibinfo  {journal} {Rev. Mod. Phys.}\ }\textbf {\bibinfo {volume} {90}},\
  \bibinfo {pages} {045003} (\bibinfo {year} {2018})},\ \Eprint
  {https://arxiv.org/abs/1803.04993} {arXiv:1803.04993 [hep-th]} \BibitemShut
  {NoStop}%
\bibitem [{\citenamefont {Valentini}(1991)}]{VALENTINI1991321}%
  \BibitemOpen
  \bibfield  {author} {\bibinfo {author} {\bibfnamefont {A.}~\bibnamefont
  {Valentini}},\ }\bibfield  {title} {\bibinfo {title} {Non-local correlations
  in quantum electrodynamics},\ }\href
  {https://doi.org/https://doi.org/10.1016/0375-9601(91)90952-5} {\bibfield
  {journal} {\bibinfo  {journal} {Physics Letters A}\ }\textbf {\bibinfo
  {volume} {153}},\ \bibinfo {pages} {321 } (\bibinfo {year}
  {1991})}\BibitemShut {NoStop}%
\bibitem [{\citenamefont {Reznik}(2003)}]{Reznik:2002fz}%
  \BibitemOpen
  \bibfield  {author} {\bibinfo {author} {\bibfnamefont {B.}~\bibnamefont
  {Reznik}},\ }\bibfield  {title} {\bibinfo {title} {{Entanglement from the
  vacuum}},\ }\href {https://doi.org/10.1023/A:1022875910744} {\bibfield
  {journal} {\bibinfo  {journal} {Found. Phys.}\ }\textbf {\bibinfo {volume}
  {33}},\ \bibinfo {pages} {167} (\bibinfo {year} {2003})},\ \Eprint
  {https://arxiv.org/abs/quant-ph/0212044} {arXiv:quant-ph/0212044}
  \BibitemShut {NoStop}%
\bibitem [{\citenamefont {Reznik}\ \emph {et~al.}(2005)\citenamefont {Reznik},
  \citenamefont {Retzker},\ and\ \citenamefont {Silman}}]{Reznik:2003mnx}%
  \BibitemOpen
  \bibfield  {author} {\bibinfo {author} {\bibfnamefont {B.}~\bibnamefont
  {Reznik}}, \bibinfo {author} {\bibfnamefont {A.}~\bibnamefont {Retzker}},\
  and\ \bibinfo {author} {\bibfnamefont {J.}~\bibnamefont {Silman}},\
  }\bibfield  {title} {\bibinfo {title} {{Violating Bell's inequalities in the
  vacuum}},\ }\href {https://doi.org/10.1103/PhysRevA.71.042104} {\bibfield
  {journal} {\bibinfo  {journal} {Phys. Rev. A}\ }\textbf {\bibinfo {volume}
  {71}},\ \bibinfo {pages} {042104} (\bibinfo {year} {2005})},\ \Eprint
  {https://arxiv.org/abs/quant-ph/0310058} {arXiv:quant-ph/0310058}
  \BibitemShut {NoStop}%
\bibitem [{\citenamefont {Retzker}\ \emph {et~al.}(2005)\citenamefont
  {Retzker}, \citenamefont {Cirac},\ and\ \citenamefont
  {Reznik}}]{Retzker_2005}%
  \BibitemOpen
  \bibfield  {author} {\bibinfo {author} {\bibfnamefont {A.}~\bibnamefont
  {Retzker}}, \bibinfo {author} {\bibfnamefont {J.~I.}\ \bibnamefont {Cirac}},\
  and\ \bibinfo {author} {\bibfnamefont {B.}~\bibnamefont {Reznik}},\
  }\bibfield  {title} {\bibinfo {title} {Detecting vacuum entanglement in a
  linear ion trap},\ }\bibfield  {journal} {\bibinfo  {journal} {Physical
  Review Letters}\ }\textbf {\bibinfo {volume} {94}},\ \href
  {https://doi.org/10.1103/physrevlett.94.050504}
  {10.1103/physrevlett.94.050504} (\bibinfo {year} {2005})\BibitemShut
  {NoStop}%
\bibitem [{\citenamefont {Klco}\ \emph {et~al.}(2023)\citenamefont {Klco},
  \citenamefont {Beck},\ and\ \citenamefont {Savage}}]{Klco:2021cxq}%
  \BibitemOpen
  \bibfield  {author} {\bibinfo {author} {\bibfnamefont {N.}~\bibnamefont
  {Klco}}, \bibinfo {author} {\bibfnamefont {D.~H.}\ \bibnamefont {Beck}},\
  and\ \bibinfo {author} {\bibfnamefont {M.~J.}\ \bibnamefont {Savage}},\
  }\bibfield  {title} {\bibinfo {title} {{Entanglement structures in quantum
  field theories: Negativity cores and bound entanglement in the vacuum}},\
  }\href {https://doi.org/10.1103/PhysRevA.107.012415} {\bibfield  {journal}
  {\bibinfo  {journal} {Phys. Rev. A}\ }\textbf {\bibinfo {volume} {107}},\
  \bibinfo {pages} {012415} (\bibinfo {year} {2023})},\ \Eprint
  {https://arxiv.org/abs/2110.10736} {arXiv:2110.10736 [quant-ph]} \BibitemShut
  {NoStop}%
\bibitem [{\citenamefont {Calabrese}\ and\ \citenamefont
  {Cardy}(2004)}]{Calabrese:2004eu}%
  \BibitemOpen
  \bibfield  {author} {\bibinfo {author} {\bibfnamefont {P.}~\bibnamefont
  {Calabrese}}\ and\ \bibinfo {author} {\bibfnamefont {J.~L.}\ \bibnamefont
  {Cardy}},\ }\bibfield  {title} {\bibinfo {title} {{Entanglement entropy and
  quantum field theory}},\ }\href
  {https://doi.org/10.1088/1742-5468/2004/06/P06002} {\bibfield  {journal}
  {\bibinfo  {journal} {J. Stat. Mech.}\ }\textbf {\bibinfo {volume} {0406}},\
  \bibinfo {pages} {P06002} (\bibinfo {year} {2004})},\ \Eprint
  {https://arxiv.org/abs/hep-th/0405152} {arXiv:hep-th/0405152} \BibitemShut
  {NoStop}%
\bibitem [{\citenamefont {Casini}\ and\ \citenamefont
  {Huerta}(2009{\natexlab{a}})}]{Casini:2008wt}%
  \BibitemOpen
  \bibfield  {author} {\bibinfo {author} {\bibfnamefont {H.}~\bibnamefont
  {Casini}}\ and\ \bibinfo {author} {\bibfnamefont {M.}~\bibnamefont
  {Huerta}},\ }\bibfield  {title} {\bibinfo {title} {{Remarks on the
  entanglement entropy for disconnected regions}},\ }\href
  {https://doi.org/10.1088/1126-6708/2009/03/048} {\bibfield  {journal}
  {\bibinfo  {journal} {JHEP}\ }\textbf {\bibinfo {volume} {03}},\ \bibinfo
  {pages} {048}},\ \Eprint {https://arxiv.org/abs/0812.1773} {arXiv:0812.1773
  [hep-th]} \BibitemShut {NoStop}%
\bibitem [{\citenamefont {Calabrese}\ and\ \citenamefont
  {Cardy}(2009)}]{Calabrese:2009qy}%
  \BibitemOpen
  \bibfield  {author} {\bibinfo {author} {\bibfnamefont {P.}~\bibnamefont
  {Calabrese}}\ and\ \bibinfo {author} {\bibfnamefont {J.}~\bibnamefont
  {Cardy}},\ }\bibfield  {title} {\bibinfo {title} {{Entanglement entropy and
  conformal field theory}},\ }\href
  {https://doi.org/10.1088/1751-8113/42/50/504005} {\bibfield  {journal}
  {\bibinfo  {journal} {J. Phys. A}\ }\textbf {\bibinfo {volume} {42}},\
  \bibinfo {pages} {504005} (\bibinfo {year} {2009})},\ \Eprint
  {https://arxiv.org/abs/0905.4013} {arXiv:0905.4013 [cond-mat.stat-mech]}
  \BibitemShut {NoStop}%
\bibitem [{\citenamefont {Casini}\ and\ \citenamefont
  {Huerta}(2009{\natexlab{b}})}]{Casini:2009sr}%
  \BibitemOpen
  \bibfield  {author} {\bibinfo {author} {\bibfnamefont {H.}~\bibnamefont
  {Casini}}\ and\ \bibinfo {author} {\bibfnamefont {M.}~\bibnamefont
  {Huerta}},\ }\bibfield  {title} {\bibinfo {title} {{Entanglement entropy in
  free quantum field theory}},\ }\href
  {https://doi.org/10.1088/1751-8113/42/50/504007} {\bibfield  {journal}
  {\bibinfo  {journal} {J. Phys. A}\ }\textbf {\bibinfo {volume} {42}},\
  \bibinfo {pages} {504007} (\bibinfo {year} {2009}{\natexlab{b}})},\ \Eprint
  {https://arxiv.org/abs/0905.2562} {arXiv:0905.2562 [hep-th]} \BibitemShut
  {NoStop}%
\bibitem [{\citenamefont {Calabrese}\ \emph {et~al.}(2009)\citenamefont
  {Calabrese}, \citenamefont {Cardy},\ and\ \citenamefont
  {Tonni}}]{Calabrese:2009ez}%
  \BibitemOpen
  \bibfield  {author} {\bibinfo {author} {\bibfnamefont {P.}~\bibnamefont
  {Calabrese}}, \bibinfo {author} {\bibfnamefont {J.}~\bibnamefont {Cardy}},\
  and\ \bibinfo {author} {\bibfnamefont {E.}~\bibnamefont {Tonni}},\ }\bibfield
   {title} {\bibinfo {title} {{Entanglement entropy of two disjoint intervals
  in conformal field theory}},\ }\href
  {https://doi.org/10.1088/1742-5468/2009/11/P11001} {\bibfield  {journal}
  {\bibinfo  {journal} {J. Stat. Mech.}\ }\textbf {\bibinfo {volume} {0911}},\
  \bibinfo {pages} {P11001} (\bibinfo {year} {2009})},\ \Eprint
  {https://arxiv.org/abs/0905.2069} {arXiv:0905.2069 [hep-th]} \BibitemShut
  {NoStop}%
\bibitem [{\citenamefont {Coser}\ \emph {et~al.}(2017)\citenamefont {Coser},
  \citenamefont {De~Nobili},\ and\ \citenamefont {Tonni}}]{Coser_2017}%
  \BibitemOpen
  \bibfield  {author} {\bibinfo {author} {\bibfnamefont {A.}~\bibnamefont
  {Coser}}, \bibinfo {author} {\bibfnamefont {C.}~\bibnamefont {De~Nobili}},\
  and\ \bibinfo {author} {\bibfnamefont {E.}~\bibnamefont {Tonni}},\ }\bibfield
   {title} {\bibinfo {title} {A contour for the entanglement entropies in
  harmonic lattices},\ }\href {https://doi.org/10.1088/1751-8121/aa7902}
  {\bibfield  {journal} {\bibinfo  {journal} {Journal of Physics A:
  Mathematical and Theoretical}\ }\textbf {\bibinfo {volume} {50}},\ \bibinfo
  {pages} {314001} (\bibinfo {year} {2017})}\BibitemShut {NoStop}%
\bibitem [{\citenamefont {Ruggiero}\ \emph {et~al.}(2018)\citenamefont
  {Ruggiero}, \citenamefont {Tonni},\ and\ \citenamefont
  {Calabrese}}]{Ruggiero:2018hyl}%
  \BibitemOpen
  \bibfield  {author} {\bibinfo {author} {\bibfnamefont {P.}~\bibnamefont
  {Ruggiero}}, \bibinfo {author} {\bibfnamefont {E.}~\bibnamefont {Tonni}},\
  and\ \bibinfo {author} {\bibfnamefont {P.}~\bibnamefont {Calabrese}},\
  }\bibfield  {title} {\bibinfo {title} {{Entanglement entropy of two disjoint
  intervals and the recursion formula for conformal blocks}},\ }\href
  {https://doi.org/10.1088/1742-5468/aae5a8} {\bibfield  {journal} {\bibinfo
  {journal} {J. Stat. Mech.}\ }\textbf {\bibinfo {volume} {1811}},\ \bibinfo
  {pages} {113101} (\bibinfo {year} {2018})},\ \Eprint
  {https://arxiv.org/abs/1805.05975} {arXiv:1805.05975 [cond-mat.stat-mech]}
  \BibitemShut {NoStop}%
\bibitem [{\citenamefont {Marcovitch}\ \emph {et~al.}(2009)\citenamefont
  {Marcovitch}, \citenamefont {Retzker}, \citenamefont {Plenio},\ and\
  \citenamefont {Reznik}}]{Marcovitch:2008sxc}%
  \BibitemOpen
  \bibfield  {author} {\bibinfo {author} {\bibfnamefont {S.}~\bibnamefont
  {Marcovitch}}, \bibinfo {author} {\bibfnamefont {A.}~\bibnamefont {Retzker}},
  \bibinfo {author} {\bibfnamefont {M.}~\bibnamefont {Plenio}},\ and\ \bibinfo
  {author} {\bibfnamefont {B.}~\bibnamefont {Reznik}},\ }\bibfield  {title}
  {\bibinfo {title} {{Critical and noncritical long-range entanglement in
  Klein-Gordon fields}},\ }\href {https://doi.org/10.1103/PhysRevA.80.012325}
  {\bibfield  {journal} {\bibinfo  {journal} {Phys. Rev. A}\ }\textbf {\bibinfo
  {volume} {80}},\ \bibinfo {pages} {012325} (\bibinfo {year} {2009})},\
  \Eprint {https://arxiv.org/abs/0811.1288} {arXiv:0811.1288 [quant-ph]}
  \BibitemShut {NoStop}%
\bibitem [{\citenamefont {Calabrese}\ \emph {et~al.}(2012)\citenamefont
  {Calabrese}, \citenamefont {Cardy},\ and\ \citenamefont
  {Tonni}}]{Calabrese:2012ew}%
  \BibitemOpen
  \bibfield  {author} {\bibinfo {author} {\bibfnamefont {P.}~\bibnamefont
  {Calabrese}}, \bibinfo {author} {\bibfnamefont {J.}~\bibnamefont {Cardy}},\
  and\ \bibinfo {author} {\bibfnamefont {E.}~\bibnamefont {Tonni}},\ }\bibfield
   {title} {\bibinfo {title} {{Entanglement negativity in quantum field
  theory}},\ }\href {https://doi.org/10.1103/PhysRevLett.109.130502} {\bibfield
   {journal} {\bibinfo  {journal} {Phys. Rev. Lett.}\ }\textbf {\bibinfo
  {volume} {109}},\ \bibinfo {pages} {130502} (\bibinfo {year} {2012})},\
  \Eprint {https://arxiv.org/abs/1206.3092} {arXiv:1206.3092
  [cond-mat.stat-mech]} \BibitemShut {NoStop}%
\bibitem [{\citenamefont {Calabrese}\ \emph {et~al.}(2013)\citenamefont
  {Calabrese}, \citenamefont {Cardy},\ and\ \citenamefont
  {Tonni}}]{Calabrese:2012nk}%
  \BibitemOpen
  \bibfield  {author} {\bibinfo {author} {\bibfnamefont {P.}~\bibnamefont
  {Calabrese}}, \bibinfo {author} {\bibfnamefont {J.}~\bibnamefont {Cardy}},\
  and\ \bibinfo {author} {\bibfnamefont {E.}~\bibnamefont {Tonni}},\ }\bibfield
   {title} {\bibinfo {title} {{Entanglement negativity in extended systems: A
  field theoretical approach}},\ }\href
  {https://doi.org/10.1088/1742-5468/2013/02/P02008} {\bibfield  {journal}
  {\bibinfo  {journal} {J. Stat. Mech.}\ }\textbf {\bibinfo {volume} {1302}},\
  \bibinfo {pages} {P02008} (\bibinfo {year} {2013})},\ \Eprint
  {https://arxiv.org/abs/1210.5359} {arXiv:1210.5359 [cond-mat.stat-mech]}
  \BibitemShut {NoStop}%
\bibitem [{\citenamefont {Klco}\ and\ \citenamefont
  {Savage}(2021{\natexlab{a}})}]{Klco:2020rga}%
  \BibitemOpen
  \bibfield  {author} {\bibinfo {author} {\bibfnamefont {N.}~\bibnamefont
  {Klco}}\ and\ \bibinfo {author} {\bibfnamefont {M.~J.}\ \bibnamefont
  {Savage}},\ }\bibfield  {title} {\bibinfo {title} {{Geometric quantum
  information structure in quantum fields and their lattice simulation}},\
  }\href {https://doi.org/10.1103/PhysRevD.103.065007} {\bibfield  {journal}
  {\bibinfo  {journal} {Phys. Rev. D}\ }\textbf {\bibinfo {volume} {103}},\
  \bibinfo {pages} {065007} (\bibinfo {year} {2021}{\natexlab{a}})},\ \Eprint
  {https://arxiv.org/abs/2008.03647} {arXiv:2008.03647 [quant-ph]} \BibitemShut
  {NoStop}%
\bibitem [{\citenamefont {Klco}\ and\ \citenamefont
  {Savage}(2021{\natexlab{b}})}]{Klco:2021biu}%
  \BibitemOpen
  \bibfield  {author} {\bibinfo {author} {\bibfnamefont {N.}~\bibnamefont
  {Klco}}\ and\ \bibinfo {author} {\bibfnamefont {M.~J.}\ \bibnamefont
  {Savage}},\ }\bibfield  {title} {\bibinfo {title} {{Entanglement Spheres and
  a UV-IR Connection in Effective Field Theories}},\ }\href
  {https://doi.org/10.1103/PhysRevLett.127.211602} {\bibfield  {journal}
  {\bibinfo  {journal} {Phys. Rev. Lett.}\ }\textbf {\bibinfo {volume} {127}},\
  \bibinfo {pages} {211602} (\bibinfo {year} {2021}{\natexlab{b}})},\ \Eprint
  {https://arxiv.org/abs/2103.14999} {arXiv:2103.14999 [hep-th]} \BibitemShut
  {NoStop}%
\bibitem [{\citenamefont {Plenio}(2005)}]{Plenio:2005cwa}%
  \BibitemOpen
  \bibfield  {author} {\bibinfo {author} {\bibfnamefont {M.~B.}\ \bibnamefont
  {Plenio}},\ }\bibfield  {title} {\bibinfo {title} {{Logarithmic Negativity: A
  Full Entanglement Monotone That is not Convex}},\ }\href
  {https://doi.org/10.1103/PhysRevLett.95.090503} {\bibfield  {journal}
  {\bibinfo  {journal} {Phys. Rev. Lett.}\ }\textbf {\bibinfo {volume} {95}},\
  \bibinfo {pages} {090503} (\bibinfo {year} {2005})},\ \Eprint
  {https://arxiv.org/abs/quant-ph/0505071} {arXiv:quant-ph/0505071}
  \BibitemShut {NoStop}%
\bibitem [{\citenamefont {Chitambar}\ \emph {et~al.}(2014)\citenamefont
  {Chitambar}, \citenamefont {Leung}, \citenamefont {Man{\v{c}}inska},
  \citenamefont {Ozols},\ and\ \citenamefont
  {Winter}}]{chitambar2014everything}%
  \BibitemOpen
  \bibfield  {author} {\bibinfo {author} {\bibfnamefont {E.}~\bibnamefont
  {Chitambar}}, \bibinfo {author} {\bibfnamefont {D.}~\bibnamefont {Leung}},
  \bibinfo {author} {\bibfnamefont {L.}~\bibnamefont {Man{\v{c}}inska}},
  \bibinfo {author} {\bibfnamefont {M.}~\bibnamefont {Ozols}},\ and\ \bibinfo
  {author} {\bibfnamefont {A.}~\bibnamefont {Winter}},\ }\bibfield  {title}
  {\bibinfo {title} {Everything you always wanted to know about locc (but were
  afraid to ask)},\ }\href@noop {} {\bibfield  {journal} {\bibinfo  {journal}
  {Communications in Mathematical Physics}\ }\textbf {\bibinfo {volume}
  {328}},\ \bibinfo {pages} {303} (\bibinfo {year} {2014})},\ \Eprint
  {https://arxiv.org/abs/1210.4583} {arXiv:1210.4583 [quant-ph]} \BibitemShut
  {NoStop}%
\bibitem [{\citenamefont {Simon}(2000)}]{Simon:2000zz}%
  \BibitemOpen
  \bibfield  {author} {\bibinfo {author} {\bibfnamefont {R.}~\bibnamefont
  {Simon}},\ }\bibfield  {title} {\bibinfo {title} {{Peres-Horodecki
  Separability Criterion for Continuous Variable Systems}},\ }\href
  {https://doi.org/10.1103/PhysRevLett.84.2726} {\bibfield  {journal} {\bibinfo
   {journal} {Phys. Rev. Lett.}\ }\textbf {\bibinfo {volume} {84}},\ \bibinfo
  {pages} {2726} (\bibinfo {year} {2000})},\ \Eprint
  {https://arxiv.org/abs/quant-ph/9909044} {arXiv:quant-ph/9909044}
  \BibitemShut {NoStop}%
\bibitem [{\citenamefont {Duan}\ \emph {et~al.}(2000)\citenamefont {Duan},
  \citenamefont {Giedke}, \citenamefont {Cirac},\ and\ \citenamefont
  {Zoller}}]{Duan_2000}%
  \BibitemOpen
  \bibfield  {author} {\bibinfo {author} {\bibfnamefont {L.-M.}\ \bibnamefont
  {Duan}}, \bibinfo {author} {\bibfnamefont {G.}~\bibnamefont {Giedke}},
  \bibinfo {author} {\bibfnamefont {J.~I.}\ \bibnamefont {Cirac}},\ and\
  \bibinfo {author} {\bibfnamefont {P.}~\bibnamefont {Zoller}},\ }\bibfield
  {title} {\bibinfo {title} {Inseparability criterion for continuous variable
  systems},\ }\href {https://doi.org/10.1103/physrevlett.84.2722} {\bibfield
  {journal} {\bibinfo  {journal} {Physical Review Letters}\ }\textbf {\bibinfo
  {volume} {84}},\ \bibinfo {pages} {2722–2725} (\bibinfo {year}
  {2000})}\BibitemShut {NoStop}%
\bibitem [{\citenamefont {Giedke}\ \emph {et~al.}(2001)\citenamefont {Giedke},
  \citenamefont {Duan}, \citenamefont {Cirac},\ and\ \citenamefont
  {Zoller}}]{giedke2001distillability}%
  \BibitemOpen
  \bibfield  {author} {\bibinfo {author} {\bibfnamefont {G.}~\bibnamefont
  {Giedke}}, \bibinfo {author} {\bibfnamefont {L.-M.}\ \bibnamefont {Duan}},
  \bibinfo {author} {\bibfnamefont {J.~I.}\ \bibnamefont {Cirac}},\ and\
  \bibinfo {author} {\bibfnamefont {P.}~\bibnamefont {Zoller}},\ }\bibfield
  {title} {\bibinfo {title} {Distillability criterion for all bipartite
  gaussian states},\ }\href
  {https://citeseerx.ist.psu.edu/viewdoc/download?doi=10.1.1.252.68&rep=rep1&type=pdf}
  {\bibfield  {journal} {\bibinfo  {journal} {Quantum Information and
  Computation}\ }\textbf {\bibinfo {volume} {1}},\ \bibinfo {pages} {79}
  (\bibinfo {year} {2001})},\ \Eprint {https://arxiv.org/abs/quant-ph/0104072}
  {arXiv:quant-ph/0104072 [quant-ph]} \BibitemShut {NoStop}%
\bibitem [{\citenamefont {{Giedke}}\ \emph {et~al.}(2001)\citenamefont
  {{Giedke}}, \citenamefont {{Kraus}}, \citenamefont {{Lewenstein}},\ and\
  \citenamefont {{Cirac}}}]{2001PhRvL..87p7904G}%
  \BibitemOpen
  \bibfield  {author} {\bibinfo {author} {\bibfnamefont {G.}~\bibnamefont
  {{Giedke}}}, \bibinfo {author} {\bibfnamefont {B.}~\bibnamefont {{Kraus}}},
  \bibinfo {author} {\bibfnamefont {M.}~\bibnamefont {{Lewenstein}}},\ and\
  \bibinfo {author} {\bibfnamefont {J.~I.}\ \bibnamefont {{Cirac}}},\
  }\bibfield  {title} {\bibinfo {title} {{Entanglement Criteria for All
  Bipartite Gaussian States}},\ }\href
  {https://doi.org/10.1103/PhysRevLett.87.167904} {\bibfield  {journal}
  {\bibinfo  {journal} {\prl}\ }\textbf {\bibinfo {volume} {87}},\ \bibinfo
  {eid} {quant-ph/0104050} (\bibinfo {year} {2001})},\ \Eprint
  {https://arxiv.org/abs/quant-ph/0104050} {arXiv:quant-ph/0104050 [quant-ph]}
  \BibitemShut {NoStop}%
\bibitem [{\citenamefont {{Botero}}\ and\ \citenamefont
  {{Reznik}}(2003)}]{2003PhRvA..67e2311B}%
  \BibitemOpen
  \bibfield  {author} {\bibinfo {author} {\bibfnamefont {A.}~\bibnamefont
  {{Botero}}}\ and\ \bibinfo {author} {\bibfnamefont {B.}~\bibnamefont
  {{Reznik}}},\ }\bibfield  {title} {\bibinfo {title} {{Modewise entanglement
  of Gaussian states}},\ }\href {https://doi.org/10.1103/PhysRevA.67.052311}
  {\bibfield  {journal} {\bibinfo  {journal} {\pra}\ }\textbf {\bibinfo
  {volume} {67}},\ \bibinfo {eid} {052311} (\bibinfo {year} {2003})},\ \Eprint
  {https://arxiv.org/abs/quant-ph/0209026} {arXiv:quant-ph/0209026 [quant-ph]}
  \BibitemShut {NoStop}%
\bibitem [{\citenamefont {Braunstein}\ and\ \citenamefont {van
  Loock}(2005)}]{Braunstein:2005zz}%
  \BibitemOpen
  \bibfield  {author} {\bibinfo {author} {\bibfnamefont {S.~L.}\ \bibnamefont
  {Braunstein}}\ and\ \bibinfo {author} {\bibfnamefont {P.}~\bibnamefont {van
  Loock}},\ }\bibfield  {title} {\bibinfo {title} {{Quantum information with
  continuous variables}},\ }\href {https://doi.org/10.1103/RevModPhys.77.513}
  {\bibfield  {journal} {\bibinfo  {journal} {Rev. Mod. Phys.}\ }\textbf
  {\bibinfo {volume} {77}},\ \bibinfo {pages} {513} (\bibinfo {year} {2005})},\
  \Eprint {https://arxiv.org/abs/quant-ph/0410100} {arXiv:quant-ph/0410100}
  \BibitemShut {NoStop}%
\bibitem [{\citenamefont {{Weedbrook}}\ \emph {et~al.}(2012)\citenamefont
  {{Weedbrook}}, \citenamefont {{Pirandola}}, \citenamefont
  {{Garc{\'\i}a-Patr{\'o}n}}, \citenamefont {{Cerf}}, \citenamefont {{Ralph}},
  \citenamefont {{Shapiro}},\ and\ \citenamefont
  {{Lloyd}}}]{2012RvMP...84..621W}%
  \BibitemOpen
  \bibfield  {author} {\bibinfo {author} {\bibfnamefont {C.}~\bibnamefont
  {{Weedbrook}}}, \bibinfo {author} {\bibfnamefont {S.}~\bibnamefont
  {{Pirandola}}}, \bibinfo {author} {\bibfnamefont {R.}~\bibnamefont
  {{Garc{\'\i}a-Patr{\'o}n}}}, \bibinfo {author} {\bibfnamefont {N.~J.}\
  \bibnamefont {{Cerf}}}, \bibinfo {author} {\bibfnamefont {T.~C.}\
  \bibnamefont {{Ralph}}}, \bibinfo {author} {\bibfnamefont {J.~H.}\
  \bibnamefont {{Shapiro}}},\ and\ \bibinfo {author} {\bibfnamefont
  {S.}~\bibnamefont {{Lloyd}}},\ }\bibfield  {title} {\bibinfo {title}
  {{Gaussian quantum information}},\ }\href
  {https://doi.org/10.1103/RevModPhys.84.621} {\bibfield  {journal} {\bibinfo
  {journal} {Reviews of Modern Physics}\ }\textbf {\bibinfo {volume} {84}},\
  \bibinfo {pages} {621} (\bibinfo {year} {2012})},\ \Eprint
  {https://arxiv.org/abs/1110.3234} {arXiv:1110.3234 [quant-ph]} \BibitemShut
  {NoStop}%
\bibitem [{\citenamefont {{Adesso}}\ \emph {et~al.}(2014)\citenamefont
  {{Adesso}}, \citenamefont {{Ragy}},\ and\ \citenamefont
  {{Lee}}}]{2014arXiv1401.4679A}%
  \BibitemOpen
  \bibfield  {author} {\bibinfo {author} {\bibfnamefont {G.}~\bibnamefont
  {{Adesso}}}, \bibinfo {author} {\bibfnamefont {S.}~\bibnamefont {{Ragy}}},\
  and\ \bibinfo {author} {\bibfnamefont {A.~R.}\ \bibnamefont {{Lee}}},\
  }\bibfield  {title} {\bibinfo {title} {{Continuous variable quantum
  information: Gaussian states and beyond}},\ }\href
  {https://doi.org/10.48550/arXiv.1401.4679} {\bibfield  {journal} {\bibinfo
  {journal} {Open Systems \& Information Dynamics}\ }\textbf {\bibinfo {volume}
  {21}},\ \bibinfo {eid} {arXiv:1401.4679} (\bibinfo {year} {2014})},\ \Eprint
  {https://arxiv.org/abs/1401.4679} {arXiv:1401.4679 [quant-ph]} \BibitemShut
  {NoStop}%
\bibitem [{\citenamefont {{Lami}}\ \emph {et~al.}(2018)\citenamefont {{Lami}},
  \citenamefont {{Serafini}},\ and\ \citenamefont
  {{Adesso}}}]{2016arXiv161205215L}%
  \BibitemOpen
  \bibfield  {author} {\bibinfo {author} {\bibfnamefont {L.}~\bibnamefont
  {{Lami}}}, \bibinfo {author} {\bibfnamefont {A.}~\bibnamefont {{Serafini}}},\
  and\ \bibinfo {author} {\bibfnamefont {G.}~\bibnamefont {{Adesso}}},\
  }\bibfield  {title} {\bibinfo {title} {{Gaussian entanglement revisited}},\
  }\href {https://doi.org/10.1088/1367-2630/aaa654} {\bibfield  {journal}
  {\bibinfo  {journal} {New Journal of Physics}\ }\textbf {\bibinfo {volume}
  {20}},\ \bibinfo {eid} {arXiv:1612.05215} (\bibinfo {year} {2018})},\ \Eprint
  {https://arxiv.org/abs/1612.05215} {arXiv:1612.05215 [quant-ph]} \BibitemShut
  {NoStop}%
\bibitem [{\citenamefont {Serafini}(2017)}]{Serafini2017}%
  \BibitemOpen
  \bibfield  {author} {\bibinfo {author} {\bibfnamefont {A.}~\bibnamefont
  {Serafini}},\ }\href {https://doi.org/10.1201/9781315118727} {\emph {\bibinfo
  {title} {Quantum Continuous Variables: A Primier of Theoretical Methods}}}\
  (\bibinfo  {publisher} {CRC Press},\ \bibinfo {year} {2017})\BibitemShut
  {NoStop}%
\bibitem [{\citenamefont {{Kaufman}}\ \emph {et~al.}(2016)\citenamefont
  {{Kaufman}}, \citenamefont {{Tai}}, \citenamefont {{Lukin}}, \citenamefont
  {{Rispoli}}, \citenamefont {{Schittko}}, \citenamefont {{Preiss}},\ and\
  \citenamefont {{Greiner}}}]{2016Sci...353..794K}%
  \BibitemOpen
  \bibfield  {author} {\bibinfo {author} {\bibfnamefont {A.~M.}\ \bibnamefont
  {{Kaufman}}}, \bibinfo {author} {\bibfnamefont {M.~E.}\ \bibnamefont
  {{Tai}}}, \bibinfo {author} {\bibfnamefont {A.}~\bibnamefont {{Lukin}}},
  \bibinfo {author} {\bibfnamefont {M.}~\bibnamefont {{Rispoli}}}, \bibinfo
  {author} {\bibfnamefont {R.}~\bibnamefont {{Schittko}}}, \bibinfo {author}
  {\bibfnamefont {P.~M.}\ \bibnamefont {{Preiss}}},\ and\ \bibinfo {author}
  {\bibfnamefont {M.}~\bibnamefont {{Greiner}}},\ }\bibfield  {title} {\bibinfo
  {title} {{Quantum thermalization through entanglement in an isolated
  many-body system}},\ }\href {https://doi.org/10.1126/science.aaf6725}
  {\bibfield  {journal} {\bibinfo  {journal} {Science}\ }\textbf {\bibinfo
  {volume} {353}},\ \bibinfo {pages} {794} (\bibinfo {year} {2016})},\ \Eprint
  {https://arxiv.org/abs/1603.04409} {arXiv:1603.04409 [quant-ph]} \BibitemShut
  {NoStop}%
\bibitem [{\citenamefont {Ho}\ and\ \citenamefont {Hsu}(2016)}]{Ho:2015rga}%
  \BibitemOpen
  \bibfield  {author} {\bibinfo {author} {\bibfnamefont {C.~M.}\ \bibnamefont
  {Ho}}\ and\ \bibinfo {author} {\bibfnamefont {S.~D.~H.}\ \bibnamefont
  {Hsu}},\ }\bibfield  {title} {\bibinfo {title} {{Entanglement and Fast
  Quantum Thermalization in Heavy Ion Collisions}},\ }\href
  {https://doi.org/10.1142/S0217732316501108} {\bibfield  {journal} {\bibinfo
  {journal} {Mod. Phys. Lett. A}\ }\textbf {\bibinfo {volume} {31}},\ \bibinfo
  {pages} {1650110} (\bibinfo {year} {2016})},\ \Eprint
  {https://arxiv.org/abs/1506.03696} {arXiv:1506.03696 [hep-th]} \BibitemShut
  {NoStop}%
\bibitem [{\citenamefont {Klco}\ and\ \citenamefont
  {Savage}(2020{\natexlab{a}})}]{Klco:2019yrb}%
  \BibitemOpen
  \bibfield  {author} {\bibinfo {author} {\bibfnamefont {N.}~\bibnamefont
  {Klco}}\ and\ \bibinfo {author} {\bibfnamefont {M.~J.}\ \bibnamefont
  {Savage}},\ }\bibfield  {title} {\bibinfo {title} {{Systematically
  Localizable Operators for Quantum Simulations of Quantum Field Theories}},\
  }\href {https://doi.org/10.1103/PhysRevA.102.012619} {\bibfield  {journal}
  {\bibinfo  {journal} {Phys. Rev. A}\ }\textbf {\bibinfo {volume} {102}},\
  \bibinfo {pages} {012619} (\bibinfo {year} {2020}{\natexlab{a}})},\ \Eprint
  {https://arxiv.org/abs/1912.03577} {arXiv:1912.03577 [quant-ph]} \BibitemShut
  {NoStop}%
\bibitem [{\citenamefont {Klco}\ and\ \citenamefont
  {Savage}(2020{\natexlab{b}})}]{Klco:2020aud}%
  \BibitemOpen
  \bibfield  {author} {\bibinfo {author} {\bibfnamefont {N.}~\bibnamefont
  {Klco}}\ and\ \bibinfo {author} {\bibfnamefont {M.~J.}\ \bibnamefont
  {Savage}},\ }\bibfield  {title} {\bibinfo {title} {{Fixed-point quantum
  circuits for quantum field theories}},\ }\href
  {https://doi.org/10.1103/PhysRevA.102.052422} {\bibfield  {journal} {\bibinfo
   {journal} {Phys. Rev. A}\ }\textbf {\bibinfo {volume} {102}},\ \bibinfo
  {pages} {052422} (\bibinfo {year} {2020}{\natexlab{b}})},\ \Eprint
  {https://arxiv.org/abs/2002.02018} {arXiv:2002.02018 [quant-ph]} \BibitemShut
  {NoStop}%
\bibitem [{\citenamefont {Horodecki}\ \emph {et~al.}(1996)\citenamefont
  {Horodecki}, \citenamefont {Horodecki},\ and\ \citenamefont
  {Horodecki}}]{Horodecki:1996nc}%
  \BibitemOpen
  \bibfield  {author} {\bibinfo {author} {\bibfnamefont {M.}~\bibnamefont
  {Horodecki}}, \bibinfo {author} {\bibfnamefont {P.}~\bibnamefont
  {Horodecki}},\ and\ \bibinfo {author} {\bibfnamefont {R.}~\bibnamefont
  {Horodecki}},\ }\bibfield  {title} {\bibinfo {title} {{On the necessary and
  sufficient conditions for separability of mixed quantum states}},\ }\href
  {https://doi.org/10.1016/S0375-9601(96)00706-2} {\bibfield  {journal}
  {\bibinfo  {journal} {Phys. Lett. A}\ }\textbf {\bibinfo {volume} {223}},\
  \bibinfo {pages} {1} (\bibinfo {year} {1996})},\ \Eprint
  {https://arxiv.org/abs/quant-ph/9605038} {arXiv:quant-ph/9605038}
  \BibitemShut {NoStop}%
\bibitem [{\citenamefont {Vidal}\ and\ \citenamefont
  {Werner}(2002)}]{Vidal:2002zz}%
  \BibitemOpen
  \bibfield  {author} {\bibinfo {author} {\bibfnamefont {G.}~\bibnamefont
  {Vidal}}\ and\ \bibinfo {author} {\bibfnamefont {R.}~\bibnamefont {Werner}},\
  }\bibfield  {title} {\bibinfo {title} {{Computable measure of
  entanglement}},\ }\href {https://doi.org/10.1103/PhysRevA.65.032314}
  {\bibfield  {journal} {\bibinfo  {journal} {Phys. Rev. A}\ }\textbf {\bibinfo
  {volume} {65}},\ \bibinfo {pages} {032314} (\bibinfo {year} {2002})},\
  \Eprint {https://arxiv.org/abs/quant-ph/0102117} {arXiv:quant-ph/0102117}
  \BibitemShut {NoStop}%
\bibitem [{\citenamefont {{Giedke}}\ \emph {et~al.}(2003)\citenamefont
  {{Giedke}}, \citenamefont {{Eisert}}, \citenamefont {{Cirac}},\ and\
  \citenamefont {{Plenio}}}]{2003quant.ph..1038G}%
  \BibitemOpen
  \bibfield  {author} {\bibinfo {author} {\bibfnamefont {G.}~\bibnamefont
  {{Giedke}}}, \bibinfo {author} {\bibfnamefont {J.}~\bibnamefont {{Eisert}}},
  \bibinfo {author} {\bibfnamefont {J.~I.}\ \bibnamefont {{Cirac}}},\ and\
  \bibinfo {author} {\bibfnamefont {M.~B.}\ \bibnamefont {{Plenio}}},\
  }\bibfield  {title} {\bibinfo {title} {{Entanglement transformations of pure
  Gaussian states}},\ }\href@noop {} {\bibfield  {journal} {\bibinfo  {journal}
  {arXiv e-prints}\ ,\ \bibinfo {eid} {quant-ph/0301038}} (\bibinfo {year}
  {2003})},\ \Eprint {https://arxiv.org/abs/quant-ph/0301038}
  {arXiv:quant-ph/0301038 [quant-ph]} \BibitemShut {NoStop}%
\bibitem [{\citenamefont {{Botero}}\ and\ \citenamefont
  {{Reznik}}(2004)}]{2004PhRvA..70e2329B}%
  \BibitemOpen
  \bibfield  {author} {\bibinfo {author} {\bibfnamefont {A.}~\bibnamefont
  {{Botero}}}\ and\ \bibinfo {author} {\bibfnamefont {B.}~\bibnamefont
  {{Reznik}}},\ }\bibfield  {title} {\bibinfo {title} {{Spatial structures and
  localization of vacuum entanglement in the linear harmonic chain}},\ }\href
  {https://doi.org/10.1103/PhysRevA.70.052329} {\bibfield  {journal} {\bibinfo
  {journal} {\pra}\ }\textbf {\bibinfo {volume} {70}},\ \bibinfo {eid} {052329}
  (\bibinfo {year} {2004})},\ \Eprint {https://arxiv.org/abs/quant-ph/0403233}
  {arXiv:quant-ph/0403233 [quant-ph]} \BibitemShut {NoStop}%
\bibitem [{\citenamefont {Miller}(1981)}]{inversesumref}%
  \BibitemOpen
  \bibfield  {author} {\bibinfo {author} {\bibfnamefont {K.}~\bibnamefont
  {Miller}},\ }\href
  {https://math.stackexchange.com/questions/17776/inverse-of-the-sum-of-matrices}
  {\bibinfo {title} {On the inverse of the sum of matrices}} (\bibinfo {year}
  {1981})\BibitemShut {NoStop}%
\bibitem [{\citenamefont {Williamson}(1936)}]{Williamson1936}%
  \BibitemOpen
  \bibfield  {author} {\bibinfo {author} {\bibfnamefont {J.}~\bibnamefont
  {Williamson}},\ }\bibfield  {title} {\bibinfo {title} {On the algebraic
  problem concerning the normal forms of linear dynamical systems},\ }\href
  {https://doi.org/10.2307/2371062} {\bibfield  {journal} {\bibinfo  {journal}
  {American Journal of Mathematics}\ }\textbf {\bibinfo {volume} {58}},\
  \bibinfo {pages} {141} (\bibinfo {year} {1936})},\ \bibinfo {note} {full
  publication date: Jan., 1936}\BibitemShut {NoStop}%
\bibitem [{\citenamefont {Audenaert}\ \emph {et~al.}(2002)\citenamefont
  {Audenaert}, \citenamefont {Eisert}, \citenamefont {Plenio},\ and\
  \citenamefont {Werner}}]{Audenaert:2002xfl}%
  \BibitemOpen
  \bibfield  {author} {\bibinfo {author} {\bibfnamefont {K.}~\bibnamefont
  {Audenaert}}, \bibinfo {author} {\bibfnamefont {J.}~\bibnamefont {Eisert}},
  \bibinfo {author} {\bibfnamefont {M.}~\bibnamefont {Plenio}},\ and\ \bibinfo
  {author} {\bibfnamefont {R.}~\bibnamefont {Werner}},\ }\bibfield  {title}
  {\bibinfo {title} {{Entanglement Properties of the Harmonic Chain}},\ }\href
  {https://doi.org/10.1103/PhysRevA.66.042327} {\bibfield  {journal} {\bibinfo
  {journal} {Phys. Rev. A}\ }\textbf {\bibinfo {volume} {66}},\ \bibinfo
  {pages} {042327} (\bibinfo {year} {2002})},\ \Eprint
  {https://arxiv.org/abs/quant-ph/0205025} {arXiv:quant-ph/0205025}
  \BibitemShut {NoStop}%
\bibitem [{\citenamefont {Kofler}\ \emph {et~al.}(2006)\citenamefont {Kofler},
  \citenamefont {Vedral}, \citenamefont {Kim},\ and\ \citenamefont
  {Brukner}}]{kofler2006entanglement}%
  \BibitemOpen
  \bibfield  {author} {\bibinfo {author} {\bibfnamefont {J.}~\bibnamefont
  {Kofler}}, \bibinfo {author} {\bibfnamefont {V.}~\bibnamefont {Vedral}},
  \bibinfo {author} {\bibfnamefont {M.~S.}\ \bibnamefont {Kim}},\ and\ \bibinfo
  {author} {\bibfnamefont {{\v{C}}.}~\bibnamefont {Brukner}},\ }\bibfield
  {title} {\bibinfo {title} {Entanglement between collective operators in a
  linear harmonic chain},\ }\href {https://doi.org/10.1103/physreva.73.052107}
  {\bibfield  {journal} {\bibinfo  {journal} {Physical Review A}\ }\textbf
  {\bibinfo {volume} {73}},\ \bibinfo {pages} {052107} (\bibinfo {year}
  {2006})}\BibitemShut {NoStop}%
\bibitem [{\citenamefont {Zych}\ \emph {et~al.}(2010)\citenamefont {Zych},
  \citenamefont {Costa}, \citenamefont {Kofler},\ and\ \citenamefont
  {Brukner}}]{Zych:2010yk}%
  \BibitemOpen
  \bibfield  {author} {\bibinfo {author} {\bibfnamefont {M.}~\bibnamefont
  {Zych}}, \bibinfo {author} {\bibfnamefont {F.}~\bibnamefont {Costa}},
  \bibinfo {author} {\bibfnamefont {J.}~\bibnamefont {Kofler}},\ and\ \bibinfo
  {author} {\bibfnamefont {C.}~\bibnamefont {Brukner}},\ }\bibfield  {title}
  {\bibinfo {title} {{Entanglement between smeared field operators in the
  Klein-Gordon vacuum}},\ }\href {https://doi.org/10.1103/PhysRevD.81.125019}
  {\bibfield  {journal} {\bibinfo  {journal} {Phys. Rev. D}\ }\textbf {\bibinfo
  {volume} {81}},\ \bibinfo {pages} {125019} (\bibinfo {year} {2010})},\
  \Eprint {https://arxiv.org/abs/1003.3354} {arXiv:1003.3354 [quant-ph]}
  \BibitemShut {NoStop}%
\bibitem [{\citenamefont {Mohammadi~Mozaffar}\ and\ \citenamefont
  {Mollabashi}(2017)}]{MohammadiMozaffar:2017nri}%
  \BibitemOpen
  \bibfield  {author} {\bibinfo {author} {\bibfnamefont {M.~R.}\ \bibnamefont
  {Mohammadi~Mozaffar}}\ and\ \bibinfo {author} {\bibfnamefont
  {A.}~\bibnamefont {Mollabashi}},\ }\bibfield  {title} {\bibinfo {title}
  {{Entanglement in Lifshitz-type Quantum Field Theories}},\ }\href
  {https://doi.org/10.1007/JHEP07(2017)120} {\bibfield  {journal} {\bibinfo
  {journal} {JHEP}\ }\textbf {\bibinfo {volume} {07}},\ \bibinfo {pages}
  {120}},\ \Eprint {https://arxiv.org/abs/1705.00483} {arXiv:1705.00483
  [hep-th]} \BibitemShut {NoStop}%
\bibitem [{\citenamefont {Di~Giulio}\ and\ \citenamefont
  {Tonni}(2020)}]{DiGiulio:2019cxv}%
  \BibitemOpen
  \bibfield  {author} {\bibinfo {author} {\bibfnamefont {G.}~\bibnamefont
  {Di~Giulio}}\ and\ \bibinfo {author} {\bibfnamefont {E.}~\bibnamefont
  {Tonni}},\ }\bibfield  {title} {\bibinfo {title} {{On entanglement
  hamiltonians of an interval in massless harmonic chains}},\ }\href
  {https://doi.org/10.1088/1742-5468/ab7129} {\bibfield  {journal} {\bibinfo
  {journal} {J. Stat. Mech.}\ }\textbf {\bibinfo {volume} {2003}},\ \bibinfo
  {pages} {033102} (\bibinfo {year} {2020})},\ \Eprint
  {https://arxiv.org/abs/1911.07188} {arXiv:1911.07188 [cond-mat.stat-mech]}
  \BibitemShut {NoStop}%
\bibitem [{\citenamefont {Cohen}(1998)}]{PhysRevLett.80.2493}%
  \BibitemOpen
  \bibfield  {author} {\bibinfo {author} {\bibfnamefont {O.}~\bibnamefont
  {Cohen}},\ }\bibfield  {title} {\bibinfo {title} {Unlocking hidden
  entanglement with classical information},\ }\href
  {https://doi.org/10.1103/PhysRevLett.80.2493} {\bibfield  {journal} {\bibinfo
   {journal} {Phys. Rev. Lett.}\ }\textbf {\bibinfo {volume} {80}},\ \bibinfo
  {pages} {2493} (\bibinfo {year} {1998})}\BibitemShut {NoStop}%
\bibitem [{\citenamefont {Sanpera}\ \emph {et~al.}(1998)\citenamefont
  {Sanpera}, \citenamefont {Tarrach},\ and\ \citenamefont
  {Vidal}}]{PhysRevA.58.826}%
  \BibitemOpen
  \bibfield  {author} {\bibinfo {author} {\bibfnamefont {A.}~\bibnamefont
  {Sanpera}}, \bibinfo {author} {\bibfnamefont {R.}~\bibnamefont {Tarrach}},\
  and\ \bibinfo {author} {\bibfnamefont {G.}~\bibnamefont {Vidal}},\ }\bibfield
   {title} {\bibinfo {title} {Local description of quantum inseparability},\
  }\href {https://doi.org/10.1103/PhysRevA.58.826} {\bibfield  {journal}
  {\bibinfo  {journal} {Phys. Rev. A}\ }\textbf {\bibinfo {volume} {58}},\
  \bibinfo {pages} {826} (\bibinfo {year} {1998})}\BibitemShut {NoStop}%
\bibitem [{\citenamefont {Vidal}\ and\ \citenamefont
  {Tarrach}(1999)}]{Vidal:1998ch}%
  \BibitemOpen
  \bibfield  {author} {\bibinfo {author} {\bibfnamefont {G.}~\bibnamefont
  {Vidal}}\ and\ \bibinfo {author} {\bibfnamefont {R.}~\bibnamefont
  {Tarrach}},\ }\bibfield  {title} {\bibinfo {title} {{Robustness of
  entanglement}},\ }\href {https://doi.org/10.1103/PhysRevA.59.141} {\bibfield
  {journal} {\bibinfo  {journal} {Phys. Rev. A}\ }\textbf {\bibinfo {volume}
  {59}},\ \bibinfo {pages} {141} (\bibinfo {year} {1999})},\ \Eprint
  {https://arxiv.org/abs/quant-ph/9806094} {arXiv:quant-ph/9806094}
  \BibitemShut {NoStop}%
\bibitem [{\citenamefont {Smolin}(2001)}]{PhysRevA.63.032306}%
  \BibitemOpen
  \bibfield  {author} {\bibinfo {author} {\bibfnamefont {J.~A.}\ \bibnamefont
  {Smolin}},\ }\bibfield  {title} {\bibinfo {title} {Four-party unlockable
  bound entangled state},\ }\href {https://doi.org/10.1103/PhysRevA.63.032306}
  {\bibfield  {journal} {\bibinfo  {journal} {Phys. Rev. A}\ }\textbf {\bibinfo
  {volume} {63}},\ \bibinfo {pages} {032306} (\bibinfo {year}
  {2001})}\BibitemShut {NoStop}%
\bibitem [{\citenamefont {Hiesmayr}(2021)}]{Hiesmayr2021}%
  \BibitemOpen
  \bibfield  {author} {\bibinfo {author} {\bibfnamefont {B.~C.}\ \bibnamefont
  {Hiesmayr}},\ }\bibfield  {title} {\bibinfo {title} {Free versus bound
  entanglement, a np-hard problem tackled by machine learning},\ }\href
  {https://doi.org/10.1038/s41598-021-98523-6} {\bibfield  {journal} {\bibinfo
  {journal} {Scientific Reports}\ }\textbf {\bibinfo {volume} {11}},\ \bibinfo
  {pages} {19739} (\bibinfo {year} {2021})}\BibitemShut {NoStop}%
\bibitem [{\citenamefont {{Wolfram Research{,} Inc.}}()}]{Mathematica}%
  \BibitemOpen
  \bibfield  {author} {\bibinfo {author} {\bibnamefont {{Wolfram Research{,}
  Inc.}}},\ }\href {https://www.wolfram.com/mathematica} {\bibinfo {title}
  {Mathematica, {V}ersion 11.1}},\ \bibinfo {note} {champaign, IL,
  2020}\BibitemShut {NoStop}%
\bibitem [{\citenamefont {Dur}\ \emph {et~al.}(2000)\citenamefont {Dur},
  \citenamefont {Vidal},\ and\ \citenamefont {Cirac}}]{Dur:2000zz}%
  \BibitemOpen
  \bibfield  {author} {\bibinfo {author} {\bibfnamefont {W.}~\bibnamefont
  {Dur}}, \bibinfo {author} {\bibfnamefont {G.}~\bibnamefont {Vidal}},\ and\
  \bibinfo {author} {\bibfnamefont {J.~I.}\ \bibnamefont {Cirac}},\ }\bibfield
  {title} {\bibinfo {title} {{Three qubits can be entangled in two inequivalent
  ways}},\ }\href {https://doi.org/10.1103/PhysRevA.62.062314} {\bibfield
  {journal} {\bibinfo  {journal} {Phys. Rev. A}\ }\textbf {\bibinfo {volume}
  {62}},\ \bibinfo {pages} {062314} (\bibinfo {year} {2000})},\ \Eprint
  {https://arxiv.org/abs/quant-ph/0005115} {arXiv:quant-ph/0005115}
  \BibitemShut {NoStop}%
\bibitem [{\citenamefont {{Giedke}}\ \emph {et~al.}(2000)\citenamefont
  {{Giedke}}, \citenamefont {{Duan}}, \citenamefont {{Cirac}},\ and\
  \citenamefont {{Zoller}}}]{2000quant.ph..7061G}%
  \BibitemOpen
  \bibfield  {author} {\bibinfo {author} {\bibfnamefont {G.}~\bibnamefont
  {{Giedke}}}, \bibinfo {author} {\bibfnamefont {L.-M.}\ \bibnamefont
  {{Duan}}}, \bibinfo {author} {\bibfnamefont {J.~I.}\ \bibnamefont
  {{Cirac}}},\ and\ \bibinfo {author} {\bibfnamefont {P.}~\bibnamefont
  {{Zoller}}},\ }\bibfield  {title} {\bibinfo {title} {{All inseparable
  two-mode Gaussian continuous variable states are distillable}},\ }\href@noop
  {} {\bibfield  {journal} {\bibinfo  {journal} {arXiv e-prints}\ ,\ \bibinfo
  {eid} {quant-ph/0007061}} (\bibinfo {year} {2000})},\ \Eprint
  {https://arxiv.org/abs/quant-ph/0007061} {arXiv:quant-ph/0007061 [quant-ph]}
  \BibitemShut {NoStop}%
\end{thebibliography}%

\onecolumngrid
\newpage
\appendix

\section{Examples: Underlying Entanglement}
\label{app:underlying}

In this section, we provide two simple examples of how partial measurement (with classical communication of the result) can attenuate classical correlations to allow extraction of entanglement that would otherwise be obscured from the remaining quantum system.
We begin with systems of three qubits and end with a four-site example in the latticized free scalar field.
These examples are intended to clarify the mechanism of volume measurement discussed in the main text.

The entanglement of three-qubit quantum states can be organized into two categories, GHZ and W, such that every state within a particular category is connected by local operations and classical communication~\cite{Dur:2000zz}.
The GHZ state is a three-qubit state described to have genuine three-party entanglement.  This statement is quantified through a maximal three-tangle, $\tau_3(GHZ) = 1$, and is qualitatively understood as the tracing of any one party leaving a maximally mixed, separable state in the remaining two Hilbert spaces.
However, these observations do not indicate the absence of underlying two-party entanglement.
In particular, it is well known that a local transformation of one qubit in the GHZ state provides an opportunity to extract a maximally entangled pair,
\begin{equation}
(\mathbb{I}_4 \otimes H) |GHZ\rangle = \frac{1}{\sqrt{2}} \left( \frac{|00\rangle + |11\rangle}{\sqrt{2}} |0\rangle + \frac{|00\rangle - |11\rangle}{\sqrt{2}} |1\rangle \right) \ \ \ ,
\end{equation}
where $H$ is the single-qubit Hadamard operator.
This simple observation indicates that a local operation and measurement on the third qubit, followed by classical communication to inform a local unitary, allows deterministic extraction of an entangled pair from a GHZ state.
Though the entanglement of the W state is optimally robust to the tracing of any one qubit, such that the two-qubit reduced density matrices exhibit the maximal entanglement that can be expected in a two-qubit subset of a three-qubit state~\cite{Dur:2000zz}, if the third party can be trusted collaboratively then the GHZ state is a theoretically more efficient two-party entanglement distribution resource.

The second example of obscured entanglement made visible by modification of classical correlations through partial measurement and LOCC will be in a six-site periodic lattice of the free scalar vacuum.
The vacuum state, Eq.~\eqref{eq:vacuumWF}, is
\begin{equation}
|\psi\rangle = \frac{\det \mathbf{K}^{1/4}}{\pi^{\frac{3}{2}}}  \int \text{d} \boldsymbol{\phi} \ e^{  -\frac{1}{2} \boldsymbol{\phi}^T \mathbf{K} \boldsymbol{\phi}   } |\boldsymbol{\phi} \rangle \ \ \ ,
\end{equation}
with $|\boldsymbol{\phi}\rangle = |\phi_1, \phi_2, \phi_3, \phi_4, \phi_5, \phi_6\rangle$.
In this system with a mass of $m = 0.3$, for example, the $(1\times 1)_{\rm mixed}$ negativity between two sites with the external volume traced vanishes at distance $\tilde{r} = 2$, 
\begin{equation}
\mathcal{N}_{1|2} = \mathcal{N}_{1|6} = 0.422 \qquad
\mathcal{N}_{1|3} = \mathcal{N}_{1,5} = 0.039 \qquad
\mathcal{N}_{1|4} = 0 \ \ \ .
\end{equation}
Calculating the reduced CM $\boldsymbol\sigma_{1,4}^{({\rm t})}$ determines that the two sites in isolation are in fact separable at distance $\tilde{r} = 2$, i.e., characterized by a mixed state compatible with a tensor product decomposition.
However, by examining a particular convex decomposition of pure states in the $\rho_{1,4}$ mixture (rather than the combined information of the reduced density matrix alone), the presence of underlying entanglement can be determined.
As discussed for Eq.~\eqref{eq:purestatepostmeasurement},
the ensemble of pure states composing the mixture tagged by volume measurement in the field basis is described by a single CM (along with a classical distribution of first moment displacements dependent on the measured $\phi_{2,3,5,6}$ configuration) of the form,
\begin{equation}
\boldsymbol{\sigma}^{({\rm m}, \phi)}_{1,4} =
\left(
\begin{array}{cccc}
 0.758 & 0. & 0.028 & 0. \\
 0. & 1.321 & 0. & -0.049 \\
 0.028 & 0. & 0.758 & 0. \\
 0. & -0.049 & 0. & 1.321 \\
\end{array}
\right) \ \ \ .
\end{equation}
This CM is calculated to have
$\mathcal{N}(\boldsymbol{\sigma}^{({\rm m}, \phi)}_{1,4}) = 0.054$.
Every pure-state ensemble present in this two-site mixture upon volume measurement carries the same non-zero entanglement.
However, this entanglement becomes entirely obscured by classical correlations when those measurement results are lost (volume trace).
As a general feature of long distance entanglement in the presence of finite bandwidth representations of the field, obscured entanglement is a physical mechanism behind recently quantified negativity and entanglement spheres~\cite{Klco:2020rga,Klco:2021biu}.
This example and the distillability of two-mode Gaussian negativity~\cite{2000quant.ph..7061G,giedke2001distillability} further underlines how distillable entanglement can be recovered 
from a separable subsystem (beyond the entanglement sphere) upon externally-controlled displacements corresponding to the classical correlations.

\section{CM Formalism for Gaussian Mixed States}
\label{app:cmmixedformalism}
Generically, a mixed quantum state may be expressed as a weighted (convex) combination of pure state density matrices,
\begin{equation}
\rho = \sum_{k} \lambda_k \rho_k \ \ \ ,
\end{equation}
with the probabilistic normalization condition that $\sum_k\lambda_k=1$.
For Gaussian states, whose covariance matrix formalism allows a polynomial framework, it is convenient to understand how an ensemble of quantum states described by pure CMs, $\{ \boldsymbol{\sigma} \}$ and associated first moment displacements, $\{\bar{q}\}$, are combined to form those of the mixed Gaussian state directly.
A result essential to the main text is provided below, while more thorough discussions may be found in e.g., Ref.~\cite{Serafini2017}.

Consider a mixture of $M$ pure ensembles.
If the CM were simply a matrix of expectation values of a single operator, then its value calculated in a mixture would be simply the weighted average of its values in each of the pure state ensembles composing the mixture.
Of course, the matrix elements of the CM are instead related to a difference of two-point and a product of one-point expectation values,
\begin{equation}
\sigma_{ij} =2 \langle (\hat{q}_i - \bar{q}_i) (\hat{q}_j - \bar{q}_j) \rangle = 2\Big[\langle \hat{q}_i \hat{q}_j \rangle - \langle \hat{q}_i \rangle \langle \hat{q}_j\rangle\Big] \ \ \ ,
\end{equation}
with $\mathbf{q} = \left\{ \hat{\phi}_1, \hat{\pi}_1, \cdots, \hat{\phi}_N, \hat{\pi}_N\right\}$ for the present scalar field application.
The pure states present in the mixed state are characterized by CM elements of the form
\begin{equation}
\sigma_{k,ij} = 2 \Big( \Tr \left[ \rho_{k} \hat{q}_i \hat{q}_j \right] - \Tr \left[ \rho_{k} \hat{q}_i \right] \Tr \left[ \rho_k \hat{q}_j \right] \Big)   \qquad k \in \{1, \cdots, M\} \ \ \ .
\end{equation}
For the composite mixed state, the CM matrix elements may be expanded as
\begin{align}
\sigma_{ij} &= 2 \Big( \Tr\left[ \rho \hat{q}_i \hat{q}_j \right] - \Tr \left[ \rho \hat{q}_i \right] \Tr \left[ \rho \hat{q}_j \right] \Big) \\
&= 2  \sum_{k} \lambda_k \Tr \left[\rho_k \hat{q}_i \hat{q}_j \right] - 2\left(\sum_{k} \lambda_k \Tr \left[ \rho_k \hat{q}_i\right] \right) \left(\sum_{k} \lambda_k \Tr \left[ \rho_k \hat{q}_j\right] \right) \\
&= \sum_{k} \lambda_k \sigma_{k,ij}  + 2 \sum_k \lambda_k \Tr \left[\rho_k \hat{q}_i\right] \Tr\left[\rho_k \hat{q}_j \right] - 2\left(\sum_{k} \lambda_k \Tr \left[ \rho_k \hat{q}_i\right] \right) \left(\sum_{k} \lambda_k \Tr \left[ \rho_k \hat{q}_j\right] \right)  \ \ \ ,
\end{align}
where the last step has been chosen in order to introduce the direct ensemble average of pure state CMs.
Clearly, if no displacements are present, the ensemble CM is simply the weighted average of the ensemble CMs.

For the scalar field application producing mixtures upon tracing of the lattice volume external to the patches, the $\phi \rightarrow -\phi$ symmetry assures that the last term vanishes.
Furthermore, as shown in Eq.~\eqref{eq:purestatepostmeasurement}, the volume measurement in the field basis causes ensembles to differ only in their displacements, allowing the first sum to simplify to a single universal CM of the underlying pure states.
The second term is positive semi-definite and will be identified as the classical noise $\mathbf{Y}$ in Eq.~\eqref{eq:sigmaplusYrelation}, leading to Eq.~\eqref{eq:Yexpression}.
Importantly, the presence of the non-vanishing second term emphasizes how a mixture of pure states all with the same CM can become described by a different CM modified by a classical distribution of first moment displacements.

\section{Examples: Delocalization by Classical Correlations}
\label{app:CCdelocalization}

In this section, examples are provided in a simple discrete-variable quantum system to accompany the discussion in Sec.~\ref{sec:delocalizing} of the main text.
When classical correlations are introduced in the form of coherent quantum noise,  the $(1_A\times 1_B)_{\rm pure}$ two-body entanglement structure of an underlying pure state is shown to be delocalized.
In particular, the introduction of correlated classical noise is shown to transform a system with $(1_A \times 1_B)_{\rm pure}$ structured entanglement into a form that exhibits features na\"ively regarded to be associated with many-body entanglement structure.
In the CV applications of the main text, analogous correlated classical noise arises from tracing field degrees of freedom outside a pair of detection patches in the free scalar field.
As such, the inset of Fig.~\ref{fig:scalarUnderlyintEntanglement} indicates that  transforming to a basis that produces $(1_A \times 1_B)_{\rm pure}$ pairs in an underlying pure state is not a sufficient criteria for achieving the $(1_A \times 1_B)_{\rm mixed}$ core-halo structure~\cite{Klco:2021cxq} of two-body entanglement in the mixed state upon tracing of the volume.
Though the following demonstrations are straightforward, it is intended that their explicit incorporation adds in the clarity of the manuscript.

Consider a four-qubit system of two Bell states stretched across a bipartitioning boundary between two spaces, $A$ and $B$,
\begin{equation}
|\psi\rangle  = \left(\frac{|00\rangle + |11\rangle}{\sqrt{2}}\right)_{AB}^{\otimes 2}  = |\Phi^+\rangle_{AB} \otimes |\Phi^+\rangle_{AB} \ \ \ .
\end{equation}
The logarithmic negativity, $\mathcal{N}_{A|B} = 2$, simply reflects the number of Bell pairs shared between the two Hilbert spaces.
Because the logarithmic negativity is additive among tensor products, $\mathcal{N}_{A|B}\left(\rho_{1,AB} \otimes \rho_{2,AB} \right)= \mathcal{N}_{A|B}(\rho_{1}) + \mathcal{N}_{A|B}(\rho_2)$ for arbitrary density matrices $\rho_{1,2}$,
the entirety of the $A|B$ logarithmic negativity of $|\psi\rangle$ (with $\rho_1 = \rho_2$) is trivially captured with a two-body structure, $\mathcal{N}_{A|B}(|\psi\rangle) = \sum_j \mathcal{N}_{(1_A \times 1_B)_j}(|\psi\rangle)$, summing over contributions from each of the two pairs.

Imagine next that a Gaussian-distributed set of single-qubit (non-entangling) unitary rotations is applied to the state, yielding a mixed-state density matrix of the form
\begin{equation}
 \rho(\boldsymbol{\Sigma}) = \sqrt{\frac{\det \boldsymbol{\Sigma}}{(2\pi)^4}} \int \dif \boldsymbol{\theta} \  R_x\left(\boldsymbol{\theta}\right) |\psi\rangle \langle \psi | R_x(-\boldsymbol{\theta}) \ e^{- \frac{1}{2}\boldsymbol{\theta}^T \mathbf{\Sigma} \boldsymbol{\theta}} \ \ \ ,
\end{equation}
where $R_x(\boldsymbol{\theta}) = R_x(\theta_1) \otimes R_x(\theta_2) \otimes R_x(\theta_3) \otimes R_x(\theta_4)$ and the single qubit rotation is parameterized as $R_x(\theta) = e^{-i \theta \sigma_x}$.
Physically, this type of quantum channel could arise from, for example, spatial or temporal inhomogeneities in environmental electromagnetic fields or fluctuations in the control pulses used to manipulate the qubits.
If the sampled distributions of rotation angles are independent, the accessible entanglement may be reduced, but the two-body structure will naturally remain.  For example, taking the distribution defined by $\boldsymbol{\Sigma}_1 = \mathbb{I}_4/{\sigma^2}$ with $\sigma = 0.1$, i.e., equivalent for each qubit, yields a total and two-body negativity that remains equal $\mathcal{N}_{A|B}\left(\rho(\boldsymbol{\Sigma}_1)\right) = \sum_j\mathcal{N}_{(1_A\times 1_B)_j}\left(\rho(\boldsymbol{\Sigma}_1)\right) = 1.94$.

Finally, if the sampled distributions of rotation angles are correlated, the reduced accessible entanglement may additionally be delocalized beyond the $(1 \times 1)$ structure.
For example, utilizing (for convenience) the covariance matrix of four neighboring sites in a free scalar field vacuum in the thermodynamic limit, $\boldsymbol{\Sigma}_2 = \mathbf{K}^\infty_{m = 1}$ from Eq.~\eqref{eq:KrhatTDL}, the total accessible entanglement $\mathcal{N}_{A|B}\left(\rho\left( \boldsymbol{\Sigma}_2\right)\right) = 0.183 \neq \sum_j\mathcal{N}_{(1_A \times 1_B)_j}\left(\rho\left( \boldsymbol{\Sigma}_2\right)\right) = 0.179$.
While it would require entangling operations in the context of pure state transformations, the correlations of a classical channel are sufficient to create a delocalization of accessible entanglement.

\section{Williamson Normal Form (Constructive)}
\label{app:williamson}

For any bonafide CM, the associated system may be transformed to the basis of normal modes through symplectic operations,
\begin{equation}
  \mathbf{D} = \mathbf{S} \boldsymbol{\sigma} \mathbf{S}^T  \qquad \mathbf{D} = \bigoplus_{j = 1}^n d_j \mathbb{I}_2 \ \ \ ,
\end{equation}
where the $d_j$'s are the (positive) symplectic eigenvalues of the CM.
As detailed in the following,
Ref.~\cite{Serafini2017} indicates a process for determining the diagonalizing symplectic operation through the Jordan decomposition ($\mathbf{L}.\mathbf{J}.\mathbf{L}^{-1}$) of the matrix product $i \boldsymbol{\Omega} \boldsymbol{\sigma}$.
In particular, the goal is for the CM to be decomposed as
\begin{equation}
  \boldsymbol{\sigma} = \mathbf{S}^{-1} \mathbf{D} \mathbf{S}^{-T} \quad \rightarrow \quad i \boldsymbol{\Omega} \boldsymbol{\sigma} = i \boldsymbol{\Omega} \mathbf{S}^{-1} \mathbf{D} \mathbf{S}^{-T} \ \ \ .
\end{equation}
If $\mathbf{S}$ is symplectic, $\mathbf{S}^T \boldsymbol{\Omega} = \boldsymbol{\Omega} \mathbf{S}^{-1}$, then the decomposition becomes
\begin{equation}
  i \boldsymbol{\Omega} \boldsymbol{\sigma} = i \mathbf{S}^T \boldsymbol{\Omega} \mathbf{D} \mathbf{S}^{-T} \ \ \ .
\end{equation}
The matrix of commutators, $\boldsymbol{\Omega}_N = \bigoplus_{j = 1}^N i \boldsymbol{\tau}_y$, is diagonalized in the Fock basis, achieved through left and right transformation by a unitary,
\begin{equation}
  u \boldsymbol{\Omega}_1 u^\dagger = -i \sigma_z \qquad u = \frac{1}{\sqrt{2}} \begin{pmatrix}
    1 & i \\
    1 & -i
  \end{pmatrix} \ \ \ ,
\end{equation}
such that
\begin{equation}
  i \boldsymbol{\Omega} \boldsymbol{\sigma} = \mathbf{S}^T \mathbf{U}^\dagger \left( \bigoplus^n \sigma_z \right) \mathbf{U} \mathbf{D}  \mathbf{S}^{-T} = \mathbf{S}^T \mathbf{U}^\dagger \left( \bigoplus^n_{j = 1} d_j\sigma_z \right) \mathbf{U}  \mathbf{S}^{-T}  \qquad \mathbf{U} = \bigoplus^n u \ \ \ ,
\end{equation}
where the commutation $\left[ \mathbf{U}, \mathbf{D}\right]$ has been utilized.
Note that the left and right factors are relative inverses, $\left( \mathbf{S}^T \mathbf{U}^\dagger \right) . \left( \mathbf{U} \mathbf{S}^{-T} \right) = \mathbb{I}$.
Thus, the symplectic transforming to the normal form of the CM can be identified from the Jordan decomposition of $i \boldsymbol{\Omega} \boldsymbol{\sigma}$.
In particular, if $\mathbf{L}$ is chosen in the Jordan decomposition such that the diagonal is ordered as above, then the normal form symplectic may be identified,
\begin{equation}
  i \boldsymbol{\Omega} \boldsymbol{\sigma} = \mathbf{L} \left( \bigoplus^n_{j = 1} d_i\sigma_z \right) \mathbf{L}^{-1} \qquad \rightarrow \qquad \mathbf{S} = (\mathbf{L}\mathbf{U})^T = \mathbf{U}^T \mathbf{L}^T \ \ \ .
\end{equation}
This structure may require fixing an ambiguity in the individual phases of the vectors comprising $\mathbf{L}$.
A further normalization condition on the vectors of $\mathbf{L}$ must be set in order for $\mathbf{S}$ to be symplectic, $\mathbf{S}^T \boldsymbol{\Omega} \mathbf{S} = \boldsymbol{\Omega}$,
\begin{equation}
    \mathbf{L} \mathbf{U} \boldsymbol{\Omega} \mathbf{U}^T \mathbf{L}^T = \boldsymbol{\Omega} \ \ \ .
\end{equation}
In the transformation of the symplectic matrix, $u \boldsymbol{\Omega}_1 u^T = \sigma_y$. Furthermore, the pair of vectors associated with each mode in $\mathbf{L}$ are related by a complex conjugate, $\mathbf{L}^T_{j,x} = \left(\mathbf{L}^T_{j,p}\right)^*$, thus allowing the relation $\left( \bigoplus \sigma_x\right) \mathbf{L}^T = \mathbf{L}^\dagger$.  The resulting constraint becomes,
\begin{equation}
  \boldsymbol{\Omega} = \mathbf{L} \left( \bigoplus \sigma_y\right) \mathbf{L}^T =-i \mathbf{L} \left( \bigoplus \sigma_z \sigma_x \right) \mathbf{L}^T = -i \mathbf{L} \left( \bigoplus \sigma_z \right) \mathbf{L}^\dagger
\end{equation}
and thus,
\begin{equation}
  i \mathbf{L}^{-1} \boldsymbol{\Omega} \mathbf{L}^{-\dagger} = \bigoplus\sigma_Z \ \ \ .
\end{equation}
In order to enforce this condition, the vectors of $\mathbf{L}$ may be normalized in the inverse space as
\begin{equation}
  L^{-1}_{j,x} = \frac{1}{\sqrt{\alpha_j}} L^{-1}_{j,x} \ \ , \ \ L^{-1}_{j,p} = \frac{1}{\sqrt{\alpha_j}} L^{-1}_{j,p} \qquad , \quad i \mathbf{L}^{-1} \Omega \mathbf{L}^{-\dagger} = \bigoplus_{j = 1}^n \alpha_j\sigma_z \ \ \ .
\end{equation}
After this normalization, the transformation constructed as $\mathbf{S} = \left(\mathbf{L}\mathbf{U}\right)^T$ will be symplectic and perform the CM symplectic diagonalization.

\FloatBarrier
\section{Tables of Numerical Values}
\FloatBarrier

In this section, tables of are provided of all numerical values presented in the figures of the main text.

\begin{table}
\scriptsize
\begin{tabular}{c|ccc}
\hline
\hline
$\tilde{r}/d$ & $\mathcal{N}_{A|B}(\boldsymbol{\sigma}^{({\rm m}, \phi)})$ & $\mathcal{N}_{A|B}(\boldsymbol{\sigma}^{({\rm m}, \pi)})$ & $\mathcal{N}_{A|B}(\boldsymbol{\sigma}^{({\rm t})})$ \\
\hline
\hline
$0.00$&$1.419$&$3.280$&$1.383$\\
$0.25$&$2.175\times 10^{-1}$&$2.609$&$1.297\times 10^{-1}$\\
$0.50$&$1.194\times 10^{-1}$&$2.411$&$4.999\times 10^{-2}$\\
$0.75$&$7.872\times 10^{-2}$&$2.298$&$2.519\times 10^{-2}$\\
$1.00$&$5.668\times 10^{-2}$&$2.219$&$1.342\times 10^{-2}$\\
$1.25$&$4.307\times 10^{-2}$&$2.160$&$6.793\times 10^{-3}$\\
$1.50$&$3.397\times 10^{-2}$&$2.114$&$3.047\times 10^{-3}$\\
$1.75$&$2.755\times 10^{-2}$&$2.075$&$1.262\times 10^{-3}$\\
$2.00$&$2.282\times 10^{-2}$&$2.042$&$5.828\times 10^{-4}$\\
$2.25$&$1.924\times 10^{-2}$&$2.013$&$3.099\times 10^{-4}$\\
$2.50$&$1.645\times 10^{-2}$&$1.988$&$1.752\times 10^{-4}$\\
$2.75$&$1.423\times 10^{-2}$&$1.966$&$9.788\times 10^{-5}$\\
$3.00$&$1.244\times 10^{-2}$&$1.946$&$5.036\times 10^{-5}$\\
$3.25$&$1.097\times 10^{-2}$&$1.927$&$2.229\times 10^{-5}$\\
$3.50$&$9.750\times 10^{-3}$&$1.911$&$8.616\times 10^{-6}$\\
$3.75$&$8.724\times 10^{-3}$&$1.895$&$3.582\times 10^{-6}$\\
$4.00$&$7.852\times 10^{-3}$&$1.881$&$1.806\times 10^{-6}$\\
$4.25$&$7.106\times 10^{-3}$&$1.868$&$1.025\times 10^{-6}$\\
$4.50$&$6.462\times 10^{-3}$&$1.855$&$6.053\times 10^{-7}$\\
$4.75$&$5.902\times 10^{-3}$&$1.844$&$3.495\times 10^{-7}$\\
$5.00$&$5.412\times 10^{-3}$&$1.833$&$1.844\times 10^{-7}$\\
$5.25$&$4.981\times 10^{-3}$&$1.822$&$8.202\times 10^{-8}$\\
$5.50$&$4.600\times 10^{-3}$&$1.812$&$3.004\times 10^{-8}$\\
$5.75$&$4.261\times 10^{-3}$&$1.803$&$1.072\times 10^{-8}$\\
$6.00$&$3.959\times 10^{-3}$&$1.794$&$4.790\times 10^{-9}$\\
$6.25$&$3.687\times 10^{-3}$&$1.786$&$2.607\times 10^{-9}$\\
$6.50$&$3.443\times 10^{-3}$&$1.778$&$1.557\times 10^{-9}$\\
\end{tabular}
\begin{tabular}{c|ccc}
$6.75$&$3.222\times 10^{-3}$&$1.770$&$9.533\times 10^{-10}$\\
$7.00$&$3.022\times 10^{-3}$&$1.763$&$5.675\times 10^{-10}$\\
$7.25$&$2.840\times 10^{-3}$&$1.756$&$3.075\times 10^{-10}$\\
$7.50$&$2.674\times 10^{-3}$&$1.749$&$1.378\times 10^{-10}$\\
$7.75$&$2.522\times 10^{-3}$&$1.742$&$4.839\times 10^{-11}$\\
$8.00$&$2.383\times 10^{-3}$&$1.736$&$1.487\times 10^{-11}$\\
$8.25$&$2.255\times 10^{-3}$&$1.730$&$5.474\times 10^{-12}$\\
$8.50$&$2.137\times 10^{-3}$&$1.724$&$2.724\times 10^{-12}$\\
$8.75$&$2.028\times 10^{-3}$&$1.718$&$1.590\times 10^{-12}$\\
$9.00$&$1.928\times 10^{-3}$&$1.713$&$9.918\times 10^{-13}$\\
$9.25$&$1.834\times 10^{-3}$&$1.707$&$6.261\times 10^{-13}$\\
$9.50$&$1.747\times 10^{-3}$&$1.702$&$3.817\times 10^{-13}$\\
$9.75$&$1.667\times 10^{-3}$&$1.697$&$2.104\times 10^{-13}$\\
$10.00$&$1.591\times 10^{-3}$&$1.692$&$9.334\times 10^{-14}$\\
$10.25$&$1.521\times 10^{-3}$&$1.688$&$3.035\times 10^{-14}$\\
$10.50$&$1.455\times 10^{-3}$&$1.683$&$8.195\times 10^{-15}$\\
$10.75$&$1.394\times 10^{-3}$&$1.679$&$2.303\times 10^{-15}$\\
$11.00$&$1.336\times 10^{-3}$&$1.674$&$9.918\times 10^{-16}$\\
$11.25$&$1.282\times 10^{-3}$&$1.670$&$5.528\times 10^{-16}$\\
$11.50$&$1.231\times 10^{-3}$&$1.666$&$3.435\times 10^{-16}$\\
$11.75$&$1.183\times 10^{-3}$&$1.662$&$2.223\times 10^{-16}$\\
$12.00$&$1.138\times 10^{-3}$&$1.658$&$1.437\times 10^{-16}$\\
$12.25$&$1.095\times 10^{-3}$&$1.654$&$8.882\times 10^{-17}$\\
$12.50$&$1.055\times 10^{-3}$&$1.650$&$4.890\times 10^{-17}$\\
$12.75$&$1.017\times 10^{-3}$&$1.647$&$2.044\times 10^{-17}$\\
$13.00$&$9.806\times 10^{-4}$&$1.643$&$5.500\times 10^{-18}$\\
$13.25$&$9.464\times 10^{-4}$&$1.640$&$1.271\times 10^{-18}$\\
\end{tabular}
\begin{tabular}{c|ccc}
$13.50$&$9.139\times 10^{-4}$&$1.636$&$2.366\times 10^{-19}$\\
$13.75$&$8.831\times 10^{-4}$&$1.633$&$8.067\times 10^{-20}$\\
$14.00$&$8.538\times 10^{-4}$&$1.630$&$4.235\times 10^{-20}$\\
$14.25$&$8.260\times 10^{-4}$&$1.626$&$2.599\times 10^{-20}$\\
$14.50$&$7.995\times 10^{-4}$&$1.623$&$1.700\times 10^{-20}$\\
$14.75$&$7.743\times 10^{-4}$&$1.620$&$1.133\times 10^{-20}$\\
$15.00$&$7.502\times 10^{-4}$&$1.617$&$7.430\times 10^{-21}$\\
$15.25$&$7.273\times 10^{-4}$&$1.614$&$4.587\times 10^{-21}$\\
$15.50$&$7.053\times 10^{-4}$&$1.611$&$2.437\times 10^{-21}$\\
$15.75$&$6.844\times 10^{-4}$&$1.608$&$8.272\times 10^{-22}$\\
$16.00$&$6.644\times 10^{-4}$&$1.605$&$1.175\times 10^{-22}$\\
$16.25$&$6.452\times 10^{-4}$&$1.603$&$2.131\times 10^{-23}$\\
$16.50$&$6.269\times 10^{-4}$&$1.600$&$0$\\
$16.75$&$6.093\times 10^{-4}$&$1.597$&$0$\\
$17.00$&$5.925\times 10^{-4}$&$1.595$&$0$\\
$17.25$&$5.763\times 10^{-4}$&$1.592$&$0$\\
$17.50$&$5.608\times 10^{-4}$&$1.590$&$0$\\
$17.75$&$5.459\times 10^{-4}$&$1.587$&$0$\\
$18.00$&$5.316\times 10^{-4}$&$1.585$&$0$\\
$18.25$&$5.179\times 10^{-4}$&$1.582$&$0$\\
$18.50$&$5.047\times 10^{-4}$&$1.580$&$0$\\
$18.75$&$4.920\times 10^{-4}$&$1.578$&$0$\\
$19.00$&$4.797\times 10^{-4}$&$1.575$&$0$\\
$19.25$&$4.679\times 10^{-4}$&$1.573$&$0$\\
$19.50$&$4.566\times 10^{-4}$&$1.571$&$0$\\
$19.75$&$4.456\times 10^{-4}$&$1.568$&$0$\\
$20.00$&$4.351\times 10^{-4}$&$1.566$&$0$\\
\hline
\hline
\end{tabular}
\caption{Numerical values for calculations presented by the main panel of Fig.~\ref{fig:scalarUnderlyintEntanglement}.}
\end{table}

\begin{table}
\renewcommand{\arraystretch}{1.4}
\begin{tabular}{c|cc|ccc}
\hline
\hline
$\tilde{r}/d$ & $\mathcal{N}_{A|B}(\boldsymbol{\sigma}^{({\rm m}, \phi)})$ & $\mathcal{N}_{A|B}^{2-\text{body}}(\boldsymbol{\sigma}^{({\rm m}, \phi)}_W)$ & $\mathcal{N}_{A|B}(\boldsymbol{\sigma}^{({\rm t}, \phi)})$ & $\mathcal{N}_{A|B}^{2-\text{body}}(\boldsymbol{\sigma}^{({\rm t}, \phi)}_W)$  & $\mathcal{N}_{A|B}^{2-\text{body}}(\boldsymbol{\sigma}^{({\rm t}, \phi)}_{\mathcal{N}})$\\
\hline
\hline
$0$&$1.419$&$1.419$&$1.383$&$1.295$&$1.383$\\
$\frac{1}{16}$&$5.287\times 10^{-1}$&$5.287\times 10^{-1}$&$4.933\times 10^{-1}$&$4.100\times 10^{-1}$&$4.933\times 10^{-1}$\\
$\frac{1}{8}$&$3.541\times 10^{-1}$&$3.541\times 10^{-1}$&$2.752\times 10^{-1}$&$2.171\times 10^{-1}$&$2.752\times 10^{-1}$\\
$\frac{3}{16}$&$2.695\times 10^{-1}$&$2.695\times 10^{-1}$&$1.820\times 10^{-1}$&$1.075\times 10^{-1}$&$1.820\times 10^{-1}$\\
$\frac{1}{4}$&$2.175\times 10^{-1}$&$2.175\times 10^{-1}$&$1.297\times 10^{-1}$&$3.137\times 10^{-2}$&$1.297\times 10^{-1}$\\
$\frac{5}{16}$&$1.818\times 10^{-1}$&$1.818\times 10^{-1}$&$9.736\times 10^{-2}$&$0$&$9.736\times 10^{-2}$\\
$\frac{3}{8}$&$1.556\times 10^{-1}$&$1.556\times 10^{-1}$&$7.598\times 10^{-2}$&$0$&$7.598\times 10^{-2}$\\
$\frac{7}{16}$&$1.354\times 10^{-1}$&$1.354\times 10^{-1}$&$6.101\times 10^{-2}$&$0$&$6.101\times 10^{-2}$\\
$\frac{1}{2}$&$1.194\times 10^{-1}$&$1.194\times 10^{-1}$&$4.999\times 10^{-2}$&$0$&$4.999\times 10^{-2}$\\
$\frac{9}{16}$&$1.064\times 10^{-1}$&$1.064\times 10^{-1}$&$4.157\times 10^{-2}$&$0$&$4.157\times 10^{-2}$\\
$\frac{5}{8}$&$9.556\times 10^{-2}$&$9.556\times 10^{-2}$&$3.493\times 10^{-2}$&$0$&$3.493\times 10^{-2}$\\
\hline
\hline
\end{tabular}
\caption{Numerical values for calculations presented by the inset of Fig.~\ref{fig:scalarUnderlyintEntanglement}.}
\end{table}

\begin{table}
$m \approx 0$ \\
\centering
\resizebox{\columnwidth}{!}{%
\begin{tabular}{cc|cccccccccccccccc}
\hline
\hline
&$\tilde{r}$ & \multicolumn{16}{c}{$\mathbf{G}\mathbf{H}^{\Gamma}$ ground state wavefunction (left patch)} \\
\hline
\hline
&$0$&$0.034$&$0.052$&$0.067$&$0.081$&$0.093$&$0.106$&$0.119$&$0.133$&$0.147$&$0.162$&$0.179$&$0.199$&$0.222$&$0.250$&$0.287$&$0.346$\\
&$5$&$0.059$&$0.089$&$0.112$&$0.131$&$0.149$&$0.164$&$0.178$&$0.191$&$0.202$&$0.212$&$0.219$&$0.224$&$0.223$&$0.216$&$0.197$&$0.150$\\
$\boldsymbol{\sigma}^{({\rm m},
\phi)}$ &$50$&$0.081$&$0.119$&$0.146$&$0.166$&$0.182$&$0.194$&$0.204$&$0.210$&$0.213$&$0.213$&$0.210$&$0.203$&$0.191$&$0.173$&$0.146$&$0.102$\\
&$150$&$0.087$&$0.127$&$0.154$&$0.174$&$0.189$&$0.200$&$0.207$&$0.212$&$0.213$&$0.211$&$0.206$&$0.197$&$0.183$&$0.164$&$0.137$&$0.095$\\
&$300$&$0.089$&$0.129$&$0.156$&$0.176$&$0.191$&$0.201$&$0.208$&$0.212$&$0.213$&$0.210$&$0.204$&$0.195$&$0.181$&$0.162$&$0.134$&$0.093$\\
\hline
&$0$&$0.028$&$0.047$&$0.062$&$0.077$&$0.090$&$0.103$&$0.117$&$0.131$&$0.146$&$0.161$&$0.179$&$0.199$&$0.222$&$0.251$&$0.290$&$0.350$\\
&$5$&$-0.028$&$0.021$&$0.061$&$0.096$&$0.127$&$0.155$&$0.180$&$0.203$&$0.222$&$0.237$&$0.248$&$0.251$&$0.245$&$0.223$&$0.172$&$0.056$\\
$\boldsymbol{\sigma}^{({\rm t},
\phi)}$&$50$&$0.021$&$-0.022$&$-0.159$&$-0.179$&$-0.109$&$0.007$&$0.131$&$0.237$&$0.302$&$0.313$&$0.261$&$0.149$&$-0.008$&$-0.165$&$-0.219$&$0.039$\\
&$150$&$-0.000$&$0.020$&$-0.135$&$0.259$&$0.002$&$-0.268$&$-0.134$&$0.188$&$0.278$&$0.008$&$-0.287$&$-0.130$&$0.311$&$-0.126$&$0.015$&$-0.000$\\
&$300$&$2.278\times 10^{-7}$&$-0.000$&$0.000$&$-0.007$&$0.037$&$-0.123$&$0.266$&$-0.392$&$0.399$&$-0.282$&$0.135$&$-0.043$&$0.008$&$-0.000$&$0.000$&$-3.040\times 10^{-7}$\\
\hline
\hline
\end{tabular}}
\caption{Numerical values for calculations presented by Fig.~\ref{fig:negGSwf} in the massless regime.}
\end{table}

\begin{table}
$m \neq 0$ \\
\centering
\resizebox{\columnwidth}{!}{%
\begin{tabular}{cc|cccccccccccccccc}
\hline
\hline
&$\tilde{r}$ & \multicolumn{16}{c}{$\mathbf{G}\mathbf{H}^{\Gamma}$ ground state wavefunction (left patch)} \\
\hline
\hline
&$0$&$0.001$&$0.002$&$0.004$&$0.005$&$0.008$&$0.011$&$0.015$&$0.022$&$0.031$&$0.044$&$0.063$&$0.092$&$0.137$&$0.208$&$0.327$&$0.561$\\
&$5$&$0.004$&$0.007$&$0.011$&$0.015$&$0.021$&$0.029$&$0.039$&$0.054$&$0.073$&$0.099$&$0.134$&$0.179$&$0.236$&$0.304$&$0.369$&$0.380$\\
$\boldsymbol{\sigma}^{({\rm m},
\phi)}$&$20$&$0.007$&$0.011$&$0.017$&$0.023$&$0.031$&$0.042$&$0.056$&$0.074$&$0.097$&$0.126$&$0.163$&$0.207$&$0.258$&$0.311$&$0.349$&$0.328$\\
&$40$&$0.008$&$0.013$&$0.019$&$0.026$&$0.035$&$0.047$&$0.062$&$0.081$&$0.105$&$0.134$&$0.171$&$0.214$&$0.262$&$0.310$&$0.342$&$0.315$\\
&$70$&$0.009$&$0.014$&$0.021$&$0.028$&$0.038$&$0.050$&$0.065$&$0.085$&$0.109$&$0.139$&$0.175$&$0.217$&$0.264$&$0.309$&$0.338$&$0.308$\\
\hline
&$0$&$0.001$&$0.002$&$0.004$&$0.005$&$0.008$&$0.011$&$0.015$&$0.022$&$0.031$&$0.044$&$0.063$&$0.092$&$0.137$&$0.208$&$0.327$&$0.561$\\
&$5$&$-0.005$&$0.002$&$0.011$&$0.022$&$0.034$&$0.049$&$0.068$&$0.093$&$0.124$&$0.164$&$0.212$&$0.268$&$0.323$&$0.355$&$0.297$&$-0.066$\\
$\boldsymbol{\sigma}^{({\rm t},
\phi)}$&$20$&$-0.002$&$0.053$&$-0.095$&$-0.114$&$-0.049$&$0.051$&$0.159$&$0.256$&$0.324$&$0.343$&$0.292$&$0.152$&$-0.058$&$-0.211$&$0.029$&$0.002$\\
&$40$&$-0.000$&$0.008$&$-0.074$&$0.200$&$-0.057$&$-0.248$&$-0.095$&$0.205$&$0.312$&$0.073$&$-0.279$&$-0.228$&$0.311$&$-0.097$&$0.009$&$-0.000$\\
&$70$&$1.054\times 10^{-7}$&$-0.000$&$0.000$&$-0.006$&$0.033$&$-0.119$&$0.269$&$-0.402$&$0.403$&$-0.270$&$0.119$&$-0.034$&$0.006$&$-0.000$&$0.000$&$-1.070\times 10^{-7}$\\
\hline
\hline
\end{tabular}}
\caption{Numerical values for calculations presented by Fig.~\ref{fig:negGSwf} in the massive regime.}
\end{table}

\end{document}